\documentclass[onecolumn,showpacs,pre,floatfix,amsmath,amssymb,notitlepage,superscriptaddress]{revtex4-1}

\usepackage{bm,bbm}
\usepackage{graphicx,epsfig,subfigure}
\usepackage{amsmath,amssymb}
\usepackage{amsthm}
\usepackage{color,xcolor}
\usepackage{url}
\usepackage[utf8]{inputenc}
\usepackage{enumitem}
\usepackage{setspace}
\usepackage{dcolumn}% Align table columns on decimal point
\usepackage{bm}% bold math
\usepackage{times}
\usepackage{enumerate}
\usepackage{footnote}
\input epsf
\usepackage[titletoc]{appendix}
\usepackage{lineno}

\usepackage{mathbbol}
\usepackage{bm}
\usepackage{bbm}
\usepackage{graphicx}
\usepackage{url}
\usepackage{xcolor}
\usepackage{enumitem}

\usepackage{hyperref}
\hypersetup{
    colorlinks=true,
    linkcolor=blue,
    filecolor=magenta,      
    urlcolor=cyan,
    pdftitle={Overleaf Example},
    pdfpagemode=FullScreen,
    }

\begin{document}

\title{Simplicial cascades  are orchestrated by the multidimensional geometry of neuronal complexes}
\author{Bengier Ülgen Kılıç}\email{bengieru@buffalo.edu}
\author{Dane Taylor}\email{danet@buffalo.edu}
\affiliation{Department of Mathematics, University at Buffalo, State University of New York, Buffalo, NY 14260, USA}

\date{\today} 

\clearpage

%\linenumbers
%%%%%%%%%%%%%%%%%%%%%%%%%%%%%%%%%%%%%%%%%%%%%%%%%%%%%%%%%%%%%%%%%%%%%%%%
%%%%%%%%%%%%%%%%%%%%%%%%%%%%%%%%%%%%%%%%%%%%%%%%%%%%%%%%%%%%%%%%%%%%%%%%

\begin{abstract} 
\begin{center}
    \textbf{Abstract}
\end{center}
Cascades arise in many contexts (e.g., neuronal avalanches, social contagions, and system failures). Despite evidence that propagations often involve higher-order dependencies,
%that cannot be decomposed into dyadic interactions, theory for 
cascade theory has largely focused on models with pairwise/dyadic interactions. Here, we develop a simplicial threshold model (STM) for nonlinear cascades over simplicial complexes that encode dyadic, triadic and higher-order interactions. We study STM cascades over ``small-world'' models that contain both short- and long-range $k$-simplices, exploring how spatio-temporal patterns manifest as a frustration between local and nonlocal propagations.
We show that higher-order coupling and nonlinear thresholding can coordinate to  robustly guide cascades along a simplicial-generalization of paths that we call $k$-dimensional ``geometrical channels''. 
%(i.e., which generalize paths) 
We also find   this coordination  to enhance the diversity and efficiency of cascades over a ``neuronal complex'', i.e., a simplicial-complex-based model for a neuronal network.
We support these findings with bifurcation theory and a data-driven approach based on latent geometry.
%that extends the related pursuits of graph embedding and nonlinear dimension reduction to simplicial complexes. 
%
%\drt{as well as enhance their diversity and efficiency. %of cascades over a ``neuronal complex'', i.e., a   neuronal network model that is based on a simplicial complex.
%which has important implications for memory capacity and efficiency of higher-order models of neuronal systems. 
Our findings and mathematical techniques provide fruitful directions for uncovering the multiscale, multidimensional mechanisms that orchestrate the spatio-temporal patterns of nonlinear cascades.

\end{abstract}

\maketitle

\clearpage

%%%%%%%%%%%%%%%%%%%%%%%%%%%%%%%%%%%%%%%%%%%%%%%%%%%%%%%%%%%%%%%%%%%%%%%%
\section{Introduction}\label{sec:intro}
%%%%%%%%%%%%%%%%%%%%%%%%%%%%%%%%%%%%%%%%%%%%%%%%%%%%%%%%%%%%%%%%%%%%%%%%

Cascading activity has been widely observed in  diverse types of real-world systems including networks of spiking neurons \cite{luczak,beggsetal,shew2011information}, 
the dissemination of information and opinions across social networks \cite{dirketal,centola2010spread,Watts5766,ruan2015kinetics}, epidemic spreading \cite{colizza2007modeling,masuda2013predicting,pastor2015epidemic}, failures within critical infrastructures \cite{BrummittE680,Buldetal,dobson2007complex}, and traffic jams \cite{Li669}. Models of such phenomena are often formulated as a spreading process in which a small, localized dynamical change  produces an avalanche of  effects across a network, and as such the mathematical models of these disparate applications are often closely related \cite{gleeson2013binary,porter2016dynamical}. Frequently, the network is spatially embedded \cite{barthelemy2011spatial} and there exist both short- and long-range edges \cite{roxin2004self,percha2005transition,watts1998collective,bassett2006small}, causing a cascade's spatio-temporal patterns to exhibit two competing phenomena \cite{taylor2015topological,centola2007cascade,centola2007complex,mahler2021analysis,marvel2013small}: wavefront propagation (WFP), where  spreading propagates locally across short-range edges; and  the appearance of new clusters (ANC), where it  propagates to distant locations across long-range edges. Whether a cascade predominantly propagates locally versus globally  informs experts on how to take appropriate  steps  toward  analysis, prediction,  control and/or sampling for various applications including advertisement-seeding strategies \cite{onnela2010spontaneous,bentley2021social}, mitigation and containment of epidemics \cite{hollingsworth2006will,epstein2007controlling,colizza2007modeling},
neuromodulation and stimulation \cite{gu2015controllability,medaglia2020personalizing}, contingency analysis for power grids \cite{dobson2007complex,hines2016cascading}, and management of supply chains \cite{pathak2007complexity,dolgui2018ripple,mari2015adaptivity}.

However, local WFP and non-local ANC also  depend on a cascade's precise propagation mechanism. In social networks, for example, people are often  reluctant to adopt a new belief/opinion unless several friends and family have already adopted it \cite{centola2010spread,ruan2015kinetics}, and such a threshold criterion causes social contagions to  preferably spread by local  WFP, and ANC   occurs less frequently \cite{taylor2015topological,centola2007cascade,centola2007complex,mahler2021analysis}. The integrate-and-fire mechanism of neurons is also  a threshold criterion \cite{kistler1997reduction}; however, neurons exhibit a   variety of other dynamical features (e.g., stochasticity, refractory periods, and inhibitory interactions   \cite{brette2007simulation}), thereby complicating the relation between neuronal threshold mechanisms and WFP/ANC. Importantly, it has been shown that the diversity of spatio-temporal patterns for neuronal cascades reflects a neuro-systems' memory capacity \cite{shew2013functional}, which helps explain certain cognitive impairments \cite{li2021collapse} and can be optimized by tuning the dynamics to criticality via a balancing of excitation/inhibition  \cite{beggsetal,larremore2011predicting}.
While considerable empirical and theoretical progress has been made regarding the origins and benefits of neuronal cascades having various properties (e.g., wide dynamic range), uncovering the mathematical and biological mechanisms responsible for orchestrating in real time how and where cascades propagate remains an open challenge.
An important step in this direction is to identify and understand structural/dynamical mechanisms that are plausible and  can potentially organize whether cascades can robustly spread locally along intended pathways despite the presence of structural and dynamical noise.

A promising direction is that recent research has highlighted that dyadic (i.e., pairwise) interactions encoded in graphs are insufficient representations for many  dynamical processes (e.g., circuit logic \cite{james2016information}, neuron responses \cite{Yu17514,maclean}, ecological networks \cite{mayfield}, power-grid failures \cite{ghasemi2021data}, supply chains \cite{dass2011holistic}, and group decision making \cite{lanchier2013stochastic,civilini2021evolutionary,noonan2021dynamics,patania}), which has inspired rapid growth in developing models and theory for dynamical processes over hypergraphs and simplicial complexes that encode dyadic, triadic, and higher-order combinatorial interactions.  Simplicial-complex models have been employed to study the macroscopic activity of brain regions \cite{petri2014homological,petri2021simplicial,giusti2016two}, and dynamical theory has been recently extended to many   higher-order systems including synchronization models  \cite{PhysRevLett.124.218301,PhysRevResearch.2.023281,gambuzzaetal,calmon2021topological}, social contagions \cite{petri2018simplicial,PhysRevResearch.2.023032,neuhauser2021opinion}, epidemic spreading  \cite{petrietal,barratetal,restrepoetal,PhysRevResearch.2.012049,higham2021epidemics}, random walks and diffusion \cite{mukherjee2016random,parzanchevski2017simplicial,estrada2018random,carletti2021random}, consensus \cite{yu2011distributed,neuhauser2020multibody} general models of ordinary differential equations \cite{ferrazetal,landry2021hypergraph}, and the optimization of higher-order dynamics \cite{skardal2021higher,ziegler2021balanced}.
Nevertheless, it has not been explored how  higher-order interactions  affect cascades' spatio-temporal WFP/ANC patterns nor the subsequent  implications for  neuronal avalanches.

Thus motivated, we extend a popular threshold model for cascades \cite{Watts5766} with binary dynamics \cite{gleeson2013binary} to develop  theoretical insights for the combined effects of thresholding and higher-order interactions on nonlinear cascades over simplicial complexes. We propose a simplicial threshold model (STM) for cascades in which  a vertex $v_i$ becomes active only when the aggregate activity across its simplicial  neighbors---which includes the states of adjacent edges, 2-simplices, and larger combinatorial sets of vertices---surpasses a threshold $T_i$. By assigning active/inactive states to vertices, edges, and higher-dimensional simplices, STM cascades provide a bridge between   modeling frameworks that exclusively describe  dynamics at the individual level (e.g., belief propagation and neuron firing) or at the  group level  (e.g., group decision making and the collective dynamics of cortical columns). It is natural to assume for some systems that groups influence individuals, and vice versa. However, such   interactions are inherently difficult to represent by graphs, due to the different dimensionality of individuals and groups. STM cascades assign states to $k$-simplices of various dimension $k$, thereby allowing simplicial cascades to nonlinearly propagate  in response to the states of individuals as well as groups of different sizes.

\begin{figure}[t!]
	\centering
	\includegraphics[width= .52\linewidth]{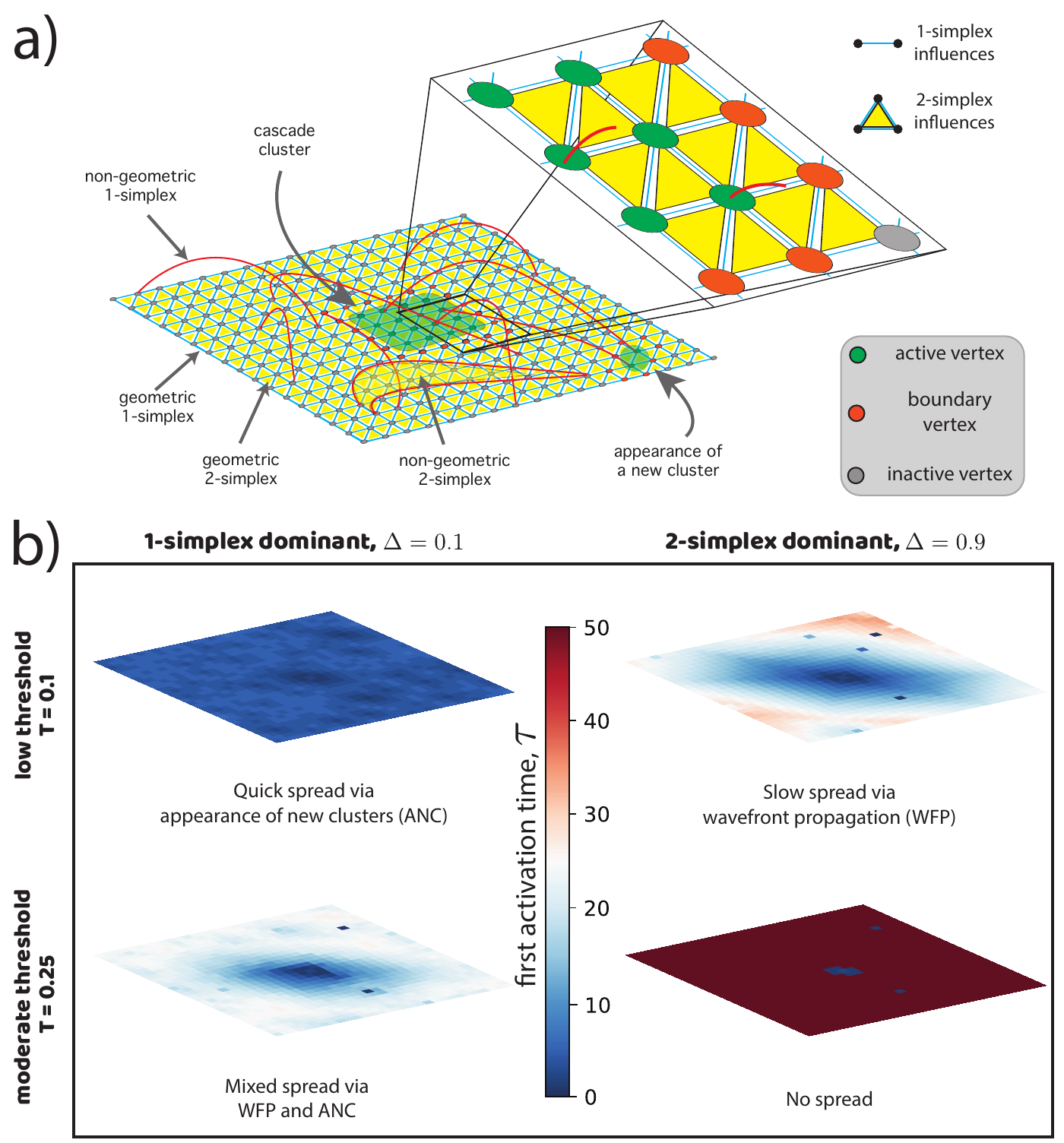}
	\caption{
		{\bf Local and non-local spreading patterns for a simplicial threshold model (STM) for cascades.}
		{\bf a)}~
		We initialize a STM cascade near the center of a  two-dimensional (2D) noisy geometric complex, which contains both short-range geometric $k$-simplices and long-range non-geometric  $k$-simplices. In this example, we study the clique complex \cite{kaczynski2004computational}  associated with a spatial graph in which vertices  are arranged in $30\times 30$ triangular lattice (yielding a ``geometrical substrate'' comprised of vertices and geometric  1- and 2-simplices) and non-geometric edges are added uniformly at random (yielding a   ``topological noise'' that manifest  by non-geometric $k$-simplices in the clique complex).STM cascades can propagate  by either local wavefront propagation (WFP)  over a geometrical substrate  or non-locally over non-geometric simplices to yield appearances of new clusters (ANC). Propagation to any boundary vertex $v_i$---which is inactive but has active simplicial neighbors (i.e., adjacent 1-simplices, 2-simplices, etc)---requires that the total activity  across its simplicial neighbors (which can be aggregated in different ways) surpasses a threshold $T$. 
		{\bf b)}~
		We study 2D STM cascades  that utilizes $k$-simplices with dimension $k\le \kappa=2$,  the relative interaction strength of 2-simplices versus 1-simplices is tuned by a parameter $\Delta\in[0,1]$ (see Eq.~\eqref{eq1}). We depict the influence of $T$ and $\Delta$ on cascades' spatio-temporal  patterns by visualizing the activation times $\tau_i$ at which each vertex $v_i$ first becomes active.  When  $T$ and $\Delta$ are both small (top-left subpanel), STM cascades rapidly progress via ANC, yielding a `splotchy' pattern. Increasing either $T$ or $\Delta$ suppresses ANC, thereby robustly guiding cascades along a geometrical substrate despite the presence of non-geometric $k$-simplices. (Observe in the bottom-right subpanel that STM cascades won't spread if $T$ and/or $\Delta$  are too large.) In summary, multidimensional interactions and  thresholding can coordinate to direct how and where  cascades spread, which has implications for neuronal avalanches and other spatio-temporal cascades. See Section `Simplicial cascades robustly follow geometrical substrates and channels' for further experiment details.
		}
	\label{fig1}
\end{figure}

We study  WFP and ANC phenomena for STM cascades over noisy geometric complexes that contain both short- and long-range simplices. Short-range simplices provide a ``geometrical substrate'' structure comprised of $k$-dimensional  simplices that are are ``lower adjacent'' by their $(k-1)$-dimensional faces \cite{kaczynski2004computational}. In contrast, long-range simplices impose a topological perturbation, or  `noise', to the geometrical substrate. Networks containing both short and long-range connections have been widely observed and analyzed in the context of neuronal networks and other applications \cite{roxin2004self,percha2005transition,watts1998collective,bassett2006small}. As shown in Fig.~\ref{fig1}, the    presence of short- and long-range  simplices, coupled with the nonlinear interplay between higher-order interactions and  threshold-based activations,  yields complicated   spatio-temporal patterns for STM cascades. We find that thresholding and higher-order interactions play a similar mechanistic function: 
they both inhibit long-range spreading,  which leads to less frequent ANC and  promotes local  WFP. However, their combination more robustly guides STM cascades along the geometrical substrate, thereby enabling cascades to reliably spread by WFP despite the presence of topological noise.
This  mechanism  for robustly organizing the spatio-temporal patterns of higher-order cascades has  significant implications for the aforementioned applications in which dyadic-interaction  models   insufficiently represent real-world cascades.

Our work is especially motivated by the study of cascading neuronal activity in brains \cite{shew2011information,larremore2011predicting,shew2013functional}, since it is well-known that such dynamics involves both thresholding and higher-order interactions \cite{Yu17514,reimann2017cliques}. Nevertheless, it has not been explored whether these two dynamical features coordinate in a nonlinear way to  benefit brain function or play a role in orchestrating the propagation paths for brain activity.
As an initial step in this direction, we study STM cascades over a simplicial complex model for a \emph{C. elegans} synapse network \cite{Choe-2004-connectivity,Kaiser-2006-placement} in which pairwise edges represent neuronal synapses, and we use higher-order $k$-simplices in a simplicial complex to encode $(k+1)$-dimensional  nonlinear dependencies (e.g.,  co-activations) among neurons \cite{reimann2017cliques}. We refer to this model as a ``neuronal complex,'' and our experiments reveal that higher-order interactions promote the diversity and energy efficiency of STM cascades over the \emph{C. elegans} neuronal complex.
%we   a  cascade model that is intentionally simple to provide, as a baseline,  fundamental insights into how higher-order interactions can possibly help  orchestrating the spatio-temporal propagation of cascades. We 
We emphasize that our findings for STM cascades are obtained for a model that   we define to be intentionally  simple so as to isolate and study nonlinear interplay between thresholding and higher-order coupling. Therefore,  it remains unknown whether similar phenomena  arise for biological networks of  neurons and if our findings/methodologies can extend to more bio-realistic neuron models  (e.g., Hodgkin–Huxley neurons \cite{brette2007simulation}). Nevertheless,  our findings  provide an important baseline of understanding for how  higher-order interactions can potentially help orchestrate the spatio-temporal propagation of cascades over a neuronal network.

We support these findings with  bifurcation theory to characterize WFP and ANC  for STM cascades spreading over $k$-dimensional channels, which we define as a geometrical substrate comprised of $k$-simplices that extends in one dimension, thereby generalizing  the  graph-based notion of a `path' to the setting of simplicial complexes. Our theory relies on a combinatorial analysis that considers the various possible dynamical responses for boundary vertices, which have  simplicial neighbors that are active, but they themselves are not yet active. We also introduce simplicial cascade maps that attribute simplicial complexes with a latent geometry in which pairwise distances reflect the time required for STM cascades to travel between vertices. Simplicial cascade maps are a simplicial-complex generalization of contagion maps \cite{taylor2015topological}, and they may similarly  be used to quantitatively study the competition between WFP and ANC using techniques from high-dimensional data analysis, nonlinear-dimension reduction, manifold learning, and topological data analysis. Our proposed mathematical tools and computational experiments reveal that the  multidimensional geometry of simplicial complexes can coordinate with the nonlinear propagation mechanism of  thresholding to robustly orchestrate higher-order cascades, which is a promising direction for uncovering the multiscale, multidimensional mechanisms that facilitate higher-order information processing in neuro-systems, and more broadly, that determine the spatio-temporal patterns of cascades across other social, biological, and technological systems.

%%%%%%%%%%%%%%%%%%%%%%%%%%%%%%%%%%%%%%%%%%%%%%%%%
%%%%%%%%%%%%%%%%%%%%%%%%%%%%%%%%%%%%%%%%%%%%%%%%%
\section{Results}\label{sec:results}
%%%%%%%%%%%%%%%%%%%%%%%%%%%%%%%%%%%%%%%%%%%%%%%%%
%%%%%%%%%%%%%%%%%%%%%%%%%%%%%%%%%%%%%%%%%%%%%%%%%

%%%%%%%%%%%%%%%%%%%%%%%%%%%%%%%%%%%%%%%%%%%%%%%%%
\subsection{Simplicial threshold model (STM) for cascades}\label{sec:stm}
%%%%%%%%%%%%%%%%%%%%%%%%%%%%%%%%%%%%%%%%%%%%%%%%%

We first briefly describe simplicial complexes  \cite{kaczynski2004computational}. Consider a set $\mathcal{C}_0=\{v_{1},\dots,v_{N}\}$ of $N$ vertices.
Each vertex $v_i\in \mathcal{C}_0$ is assigned a coordinate ${\bf y}^{(i)}\in \mathbb{R}^p$ in a $p$-dimensional ambient metric space. (We note in passing that vertices in ``abstract'' simplicial complexes do not have such coordinates; however, we will focus on the traditional definition herein.) We assume a Euclidean metric, although it may also be advantageous to explore other metric spaces \cite{krioukov2010hyperbolic,boguna2021network}.  
We can define a $k$-dimensional simplex $(v_0,\dots,v_{k})$, or simply $k$-simplex, by an unordered set of vertices $v_i$ with cardinality $k+1$. For example, a $0$-simplex is equivalent to a vertex $v_i$ and a 1-simplex is equivalent to an undirected, unweighted edge $(v_i,v_j)$.
Lastly, we define a  $K$-dimensional simplicial complex  $\{\mathcal{C}_k\}_{k=0}^K$ as the union of sets $\mathcal{C}_k$,  each of which contain   simplices of dimension $k$. 
For example, a 1-dimensional (1D) simplicial complex is a  graph $\{\mathcal{C}_k\}_{k=0}^1$, where $\mathcal{C}_0$ is a set of vertices having spatial coordinates and $\mathcal{C}_1$ is a set of undirected, unweighted edges. Intuitively, a 2-dimensional simplicial complex is a spatial graph with ``filled in'' triangles.
%Each vertex $v_i\in \mathcal{C}_0$ is assigned a  coordinate ${\bf y}^{(i)}\in \mathbb{R}^p$ in a $p$-dimensional ambient metric space. (Note that vertices in ``abstract'' simplicial complexes do not have such coordinates.) We assume a Euclidean metric, although it may be advantageous to explore other metric spaces \cite{krioukov2010hyperbolic,boguna2021network}. 
To define our cascade model, we further define notions of degree, or connectivity, among $k$-simplices. For each vertex $v_i \in \mathcal{C}_0$, we define $d_{i}^{k}$ as the number of $k$-simplices to which it is adjacent: $d_{i}^{1}$ is the 1-simplex degree of vertex $v_i$ (often called node degree for graphs), $d_i^{2}$ is its 2-simplex degree, and so on. We also define for each vertex $v_i$ the sets  $\mathcal{N}^k(i) = \{s \in \mathcal{C}_k | i\in s \}$ that contain its $k$-dimensional simplicial neighbors. It follows that $d^k_i = |\mathcal{N}^k(i)|$ for each vertex $v_i$ and simplex dimension $k$.

\begin{figure}[t]
	\centering
	\includegraphics[width= 1\linewidth]{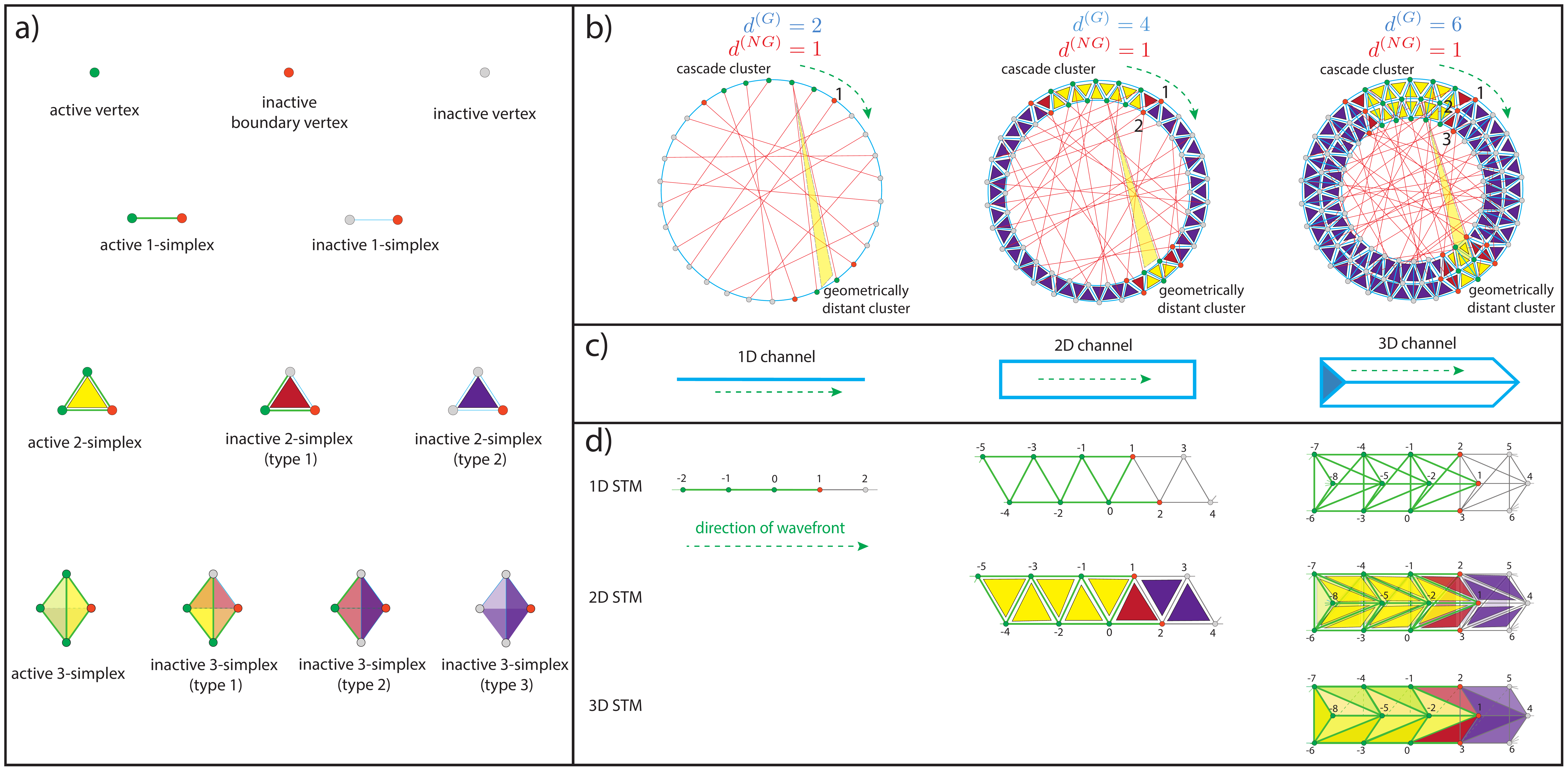}
	\caption{
		{\bf Wavefront propagation (WFP) for $\kappa$-dimensional STM cascades on noisy ring complexes.}
		{\bf a)}~
		Each $k$-simplex with $k\le\kappa$ is given  a binary state $x_i^k(t)\in\{0,1\}$ indicating whether it is inactive or active, respectively, at time $t=0,1,2,\dots$. Cascade propagation occurs when an inactive boundary vertex is adjacent to sufficiently many active $k$-simplices, in which case it (and possibly some of its adjacent $k$-simplex neighbors) will become active upon the next time step. There are different types of inactive $k$-simplices, depending on how many of their vertices are active.
		{\bf b)}~
		Noisy ring complexes (see Methods section `Generative model for noisy ring complexes') generalize noisy ring lattices \cite{taylor2015topological} and contain vertices that lie on a 1D ring manifold that is embedded in a 2D `ambient' space. Each vertex has  $d^{(NG)} = 1$ non-geometric edge (red lines) to a distant vertex and  $d^{(G)}$ geometric edges (blue lines) to nearby vertices with  $d^{(G)}\in\{ 2,4,6\}$ (left, middle, and right columns, respectively). Higher-dimensional simplices arise in the associated clique complexes and are similarly classified as geometric/non-geometric. To simplify our illustrations, we place vertices   alongside the manifold when $d^{(G)}>2$, and we do not visualize $3$-simplices.
		{\bf c)}~ 
		Geometric $k$-simplices with $k\le K$ compose a $K$-dimensional geometrical substrate. For noisy ring complexes, $K=d^{(G)}/2$ and the substrate is a $K$-dimensional channel---which is a non-intersecting sequence of lower-adjacent $K$-simplices. Channels generalize the graph-theoretic notion of a `path'.
		{\bf d)}~
		STM cascades with different dimension $\kappa\le K$ can propagate by WFP along a $K$-dimensional channel. Note that an STM cascade does not utilize all available $k$-simplices when $\kappa<K$.
		}
	\label{fig2}
\end{figure}

We now define STM cascades in which all $k$-simplices of dimension $k\le \kappa$  are given binary dynamical states $x_i^{k}(t) \in\{0,1\}$, i.e., inactive vs active, where index $i$ enumerates the simplices of dimension $k$ and  $t\ge0$ is time. For 2-dimensional (2D) STM cascades (i.e., $\kappa=2$),    the states  of vertices, 1-simplices, and 2-simplices are  given by $\{x_i^{0}(t)\}$, $\{x_i^{1}(t)\}$, and $\{x_i^{2}(t)\}$, respectively. Parameter $\kappa$ is called the STM cascade's dimension, and it may differ from that of the simplicial complex {as long as $\kappa \leq K$}. For $k>0$, the states of $k$-simplices  are directly determined  by the states of vertices; a $k$-simplex $(v_0,\dots,v_{k})$ is active only when $k$ of the vertices are active. For example, an edge $(v_i,v_j)$ is active if at least one vertex $v_i$ or $v_j$ is active, a 2-simplex is active if at least two
vertices are active, and so on. See Fig.~\ref{fig2}a) for a  visualization of  states for vertices, 1-simplices, 2-simplices, and 3-simplices. We present these examples from the perspective of  a boundary vertex, which we define as a vertex that is inactive but has at least one active simplicial neighbor.

The vertices' states evolve via a   discrete-time process that we define  for general $\kappa$ in  Methods section `STM cascades'. Here, we present a simplified dynamics for 2D STM cascades, and our later simulations will also focus on $\kappa=2$. At time step $t+1$, the state $x_i^0(t)$ of each vertex $v_i$  possibly changes according to a threshold criterion
\begin{equation}\label{eq:xx}
    x_{i}^0({t+1}) = \left\{ \begin{array}{rcl}
         1, &  &\text{ if either } x_{i}^0({t})=1 \text{ or } R_{i}({t}) > T_i \, ,\\
         0, & &\text{ if } x_{i}^0({t})=0 \text{ and } R_{i}({t}) \le T_i\, ,
    \end{array}\right. 
\end{equation}
where $T_i$ is an activation threshold intrinsic to vertex $v_{i}$ and
\begin{equation}\label{eq1}
R_{i}({t}) =  (1-\Delta)f_{i}^1({t}) +  \Delta~f_{i}^2({t})
\end{equation}
is a weighted average of cascade activity across the simplicial neighbors of vertex $v_i$. Parameter $\Delta$ tunes the relative influence of 2-simplices and
$f_{i}^1(t) = \frac{1}{d^1_i}\sum_{j \in\mathcal{N}_{A}^1(i,t)} x_j^1(t)$ and
$f_{i}^2(t) = \frac{1}{d^2_i}\sum_{j \in\mathcal{N}_{A}^2(i,t)} x_j^2(t)$ are the fractions of adjacent $1$-  and $2$-simplices that are active at time $t$. One can also interpret $R_{i}({t})$ as $v_i$'s ``simplicial exposure'' to a cascade at time $t$. When vertices change their states, we allow $k$-simplices with $k>0$ to update their states instantaneously, and we leave open the investigation of more complicated dependencies such as delayed state changes for higher-dimensional $k$-simplices. Also, note that the limit $\Delta\to0$ yields a 1D STM cascade, which  is equivalent to the Watts threshold model \cite{Watts5766} for cascades over graphs.

To narrow the scope of our  experiments, herein we initialize all STM cascades at time $t=0$ using cluster seeding (see Methods section `Cluster seeding'), in which case we select a vertex and set all of its adjacent vertices to be active, while  all other vertices are inactive. Thresholding can potentially prevent localized initial conditions from propagating into large-scale cascades, and cluster seeding helps overcome this dynamical barrier \cite{gleeson2007seed,taylor2015topological}. Our  experiments are also   simplified by assuming an identical threshold  $T_i=T$ for each vertex $v_i$. This allows us to explore the cooperative effects of thresholding and higher-order interactions for 2D STM cascades by varying only two   parameters: threshold $T$ and 2-simplex influence $\Delta$.

Notably, we also define and study a stochastic variant of STM cascades  in Supplementary Note  `Stochastic Simplicial Threshold Model'  in which the vertices' states change via a nonlinear stochastic process instead of the deterministic nonlinear dynamics defined by Eqs.~\eqref{eq:xx}--\eqref{eq1}.

%%%%%%%%%%%%%%%%%%%%%%%%%%%%%%%%%%%%%%%%%%%%%%%%%
\subsection{Noisy geometric complexes, geometrical substrates and channels}\label{sec:2D}
%%%%%%%%%%%%%%%%%%%%%%%%%%%%%%%%%%%%%%%%%%%%%%%%%

We study the spatio-temporal patterns of STM cascades over noisy geometric complexes, which contain both short- and long-range simplices and are a generalization of noisy geometric networks \cite{taylor2015topological}. Short- and long-range interactions have been observed in a wide variety of applications (e.g., face-to-face and online interactions in social networks) and are known to play an important structural/dynamical role for neuronal activity \cite{roxin2004self,percha2005transition}. It's also worth noting that noisy geometric networks  exhibit the small-world property \cite{watts1998collective} under certain parameter choices, and noisy geometric complexes will likely exhibit a simplicial analogue to this property \cite{klamt2009hypergraphs,bolle2006thermodynamics}.

To explore WFP and ANC in an analytically tractable setting,
we assume that the  vertices $\mathcal{C}_0$ lie along a manifold within an ambient space $\mathbb{R}^p$, and that all $k$-simplices are one of two types: geometric simplices that connect vertices that are nearby on the manifold, and long-range non-geometric simplices that connect distant vertices.  Each $k$-simplex is considered to be geometric if and only if all of its associated faces are geometric. After categorizing $k$-simplices as geometric or non-geometric, we further refine the notion of $k$-simplex degrees. Specifically, we let  $d^{k,G}_i$ and  $d^{k,NG}_i$ denote geometric and non-geometric $k$-simplex degrees, respectively, of a vertex $v_i$ so that $d^k_i = d^{k,G}_i + d^{k,NG}_i$. We provide visualizations of synthetic examples of noisy geometric complexes in Fig.~\ref{fig1} and Fig.~\ref{fig2}b), where the vertices lie on a 2D plane and a 1D ring manifold, respectively. In these synthetic  models, we construct geometric edges by connecting each vertex to several of its nearest neighbors, and we create non-geometric edges uniformly at random between pairs of vertices that do not yet have an edge. In either case, we construct noisy geometric complexes by considering the associated clique complexes for these vertices and edges. See Methods section `Generative model for noisy ring complexes' for further details about this construction.

We define the subgraph (or sub-complex) restricted to geometric edges (or simplices) as a geometrical substrate, and  propagation along a substrate is called WFP, by definition. Importantly, a substrate's geometry and dimensionality can in principle differ from that of the manifold and that of the full simplicial complex that contains both geometric and non-geometric edges. For example, in Fig.~\ref{fig2}b) we depict three noisy ring complexes  in which the vertices have different geometric degrees: $d^{(G)}\in\{2,4,6\}$. In all cases, the vertices lie on a 1D ring manifold that is embedded in a 2D ambient space; however, as shown in Fig.~\ref{fig2}c), the resulting $K$-dimensional geometrical substrates have different dimensions with $K=d^{(G)}/2$. Because each substrate extends in 1 dimension along the 1D ring manifold, each   is a $K$-dimensional geometrical channel, which we define as a non-intersecting sequence of lower-adjacent $K$-simplices (i.e., each subsequent $K$-simplex intersects with the preceding $K$-simplex by a $(K-1)$-simplex that is a face to both $K$-simplices \cite{kaczynski2004computational}). A channel is a higher-dimensional generalization of a ``non-intersecting path'' in a graph in which wavefronts travel the fastest, and it is closely  related to the graph-based concepts of $k$-clique rolling \cite{derenyi2005clique} and complex paths \cite{guilbeault2021topological}.

Before continuing, we highlight that it is important to understand the different types of `dimension' that have been introduced. We assume that a noisy geometric complex lies on a manifold of some dimension and is  within a $p$-dimensional metric space. The dimension $K$ of a simplicial complex refers to the maximum dimension of its $k$-simplices. Within a given simplicial complex, there can exist a geometrical substrate of some possibly smaller dimension $k$. Finally a STM cascade has its own dimension, $\kappa$, which is the largest $k$-simplex dimension that is utilized by the nonlinear dynamics. In principle, all of these dimensions can differ.

%%%%%%%%%%%%%%%%%%%%%%%%%%%%%%%%%%%%%%%%%%%%%%%%%
\subsection{Simplicial cascades robustly follow geometrical substrates and channels} \label{sec:impact}
%%%%%%%%%%%%%%%%%%%%%%%%%%%%%%%%%%%%%%%%%%%%%%%%%

We study the coordinated effects of thresholding and  higher-order interactions on WFP and ANC, and it is helpful to first provide precise definitions of these two phenomena that manifest as a frustration between local and non-local connections in a noisy geometric complex. 
We characterize a propagation to a vertex as  WFP if at the time of propagation, it is adjacent to at least one active vertex via a geometric $k$-simplex. 
In contrast, a  propagation to a vertex is called ANC if and only if that propagation occurs   solely due to its  adjacency to non-geometric active $k$-simplices, and all of its adjacent geometric $k$-simplices are inactive at the time of propagation.
%
%In contrast, a propagation to a vertex is characterized as ANC if and only if all of its geometric neighbors are inactive.
%
%For  1-, 2-, and higher-dimensional $k$-simplices, the propagation type of WFP vs ANC is inherited by the propagation type for its vertices. 
%
%that  are characterized as WFP or ANC based on the vertices whose changed states cause the $k$-simplex activation. That is, propagation to a $k$-simplex is called ANC if and only if it's state has changed in response only to ANC propagations to its associated vertices. In this way, if $k$-simplices are activated in response to proximity to both geometric and non-geometric $k$-simplices, then the propagations are still considered to be WFP. That is, `exposure' to active $k$-simplices can accelerate WFP, but it is still considered to be WFP.

Thresholding and  higher-order interactions can both suppress non-local ANC across non-geometric edges, which promotes simplicial cascades to locally propagate via WFP along a geometrical substrate. %
This is visualized in Fig.~\ref{fig1}, where we study 2D STM cascades over a 2D  noisy geometric complex (The manifold, simplicial complex, geometrical substrate, and STM cascades   all have the same dimension in this simple example.). We initialized the STM cascades with cluster seeding at a center vertex so that they could potentially spread outward via WFP along the 2D manifold (which is discretized by the geometrical substrate). 
In Fig.~\ref{fig1}b), we visualize the activation times $\tau_i$ (i.e., when each vertex $v_i$ first becomes active), showing results for STM cascades with four choices for the parameters $T$ and $\Delta$. We study 2D STM cascades over a 2D noisy geometric complex in which the vertices are positioned in a $30\times 30$ triangular lattice, and each vertex $v_i$ has  $d_i^{1,G}=6$ geometric edges to nearest neighbors (although vertices on the outside have fewer) as well as $d_i^{1,NG}=1$  non-geometric edge, which  are added uniformly at random between pairs of vertices. We then study the resulting clique complex. Observe for small $T$ and $\Delta$ (top-left subpanel) that STM cascades rapidly spread and predominantly exhibit ANC, which results in the  `splotchy'   pattern. In contrast, when either $T$ or $\Delta$ is increased, the simplicial cascade predominantly exhibits WFP, and not ANC, which slows propagation and enables the cascade to more reliably follow along the geometrical substrate (i.e., thereby overcoming the presence of long-range ``topological noise''). 
Finally, observe that if $T$ and $\Delta$ are too large (bottom-right subpanel), then the initial seed cluster does not lead to a cascade. 

This finding extends existing knowledge about the effects of short- and long-range connections on cascades. It is well-known that long-range edges allow traditional pairwise-progressing  cascades to rapidly spread  via the mechanism of ANC. This concept is most apparent in the context of epidemic spreading, and as a response, banning international airline travel  is often a first response to prevent long-range transmissions for epidemics \cite{hollingsworth2006will,epstein2007controlling,colizza2007modeling}. However, ANC is also   suppressed when the cascade's propagation mechanism requires a vertex's neighboring activity (i.e., `exposure') to surpass a threshold $T$ \cite{centola2007cascade,centola2007complex,centola2010spread,taylor2015topological,mahler2021analysis}. We find  that higher-order interactions can be as, if not more, effective at suppressing non-local ANC. Moreover, these two mechanisms can coordinate to more robustly guide cascades along a geometrical substrate despite the presence of topological (i.e., non-geometric) noise. In the next sections, we explore the potential benefits of this  structural/dynamical coordination as a multiscale/multidimensional mechanism to orchestrate neuronal avalanches.

\begin{figure*}[t]
	\centering
	\includegraphics[width= 1\linewidth]{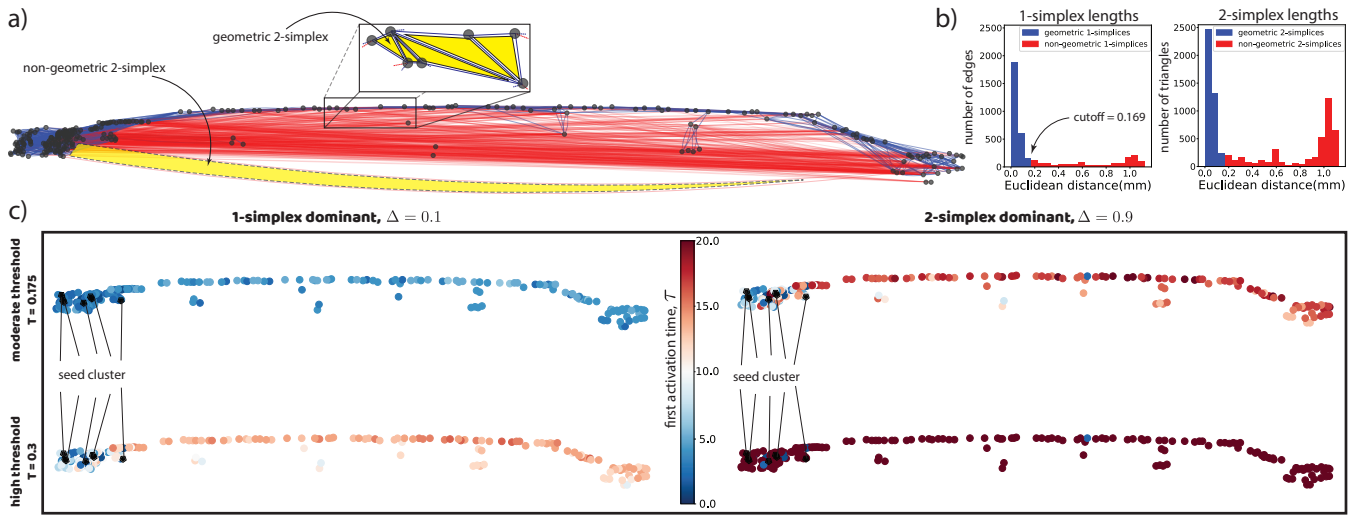}
	\caption{
		{\bf STM cascades on a \emph{C. elegans} neuronal complex.}
		{\bf a)}~2D visualization of experimentally measured  locations and synapse connections between neurons in nematode \emph{C. elegans} \cite{Choe-2004-connectivity, Kaiser-2006-placement}. We model higher-order nonlinear dynamical dependencies among sets of $(k+1)$ neurons using $k$-simplices in the associated clique complex for which we ignore edge directions.
		{\bf b)}~Histograms depict the distribution of lengths for 1- and 2-simplices, where we define the length of a 2-simplex as the maximum length over its faces.
		We distinguish geometric and non-geometric $1$-simplices by selecting a cutoff distance of 0.169 mm, and we characterize a 2-simplex as  geometric if and only if all of its faces are geometric.
		%Also, note that histograms indicate the undirected \emph{C. elegans} synapse network on which we ran our experiments.}
		{\bf c)}~Vertex colors depict their first-activation times $\tau_i$ for a 2D STM cascade that is initialized with the indicated seed cluster. Observe that $T$ and $\Delta$ affect the spatio-temporal pattern of activations (i.e., WFP and ANC) similarly to what was shown in Fig.~\ref{fig1}a).
	}
	\label{fig3}
\end{figure*}

%%%%%%%%%%%%%%%%%%%%%%%%%%%%%%%%%%%%%%%%%%%%%%%%%
\subsection{STM cascades on a \emph{C. elegans} neuronal complex} \label{contagionsondata2}
%%%%%%%%%%%%%%%%%%%%%%%%%%%%%%%%%%%%%%%%%%%%%%%%%

We observe similar cooperative effects of thresholding and higher-order interactions for STM cascades on a neuronal complex, which we define as a simplicial complex model that represents the higher-order nonlinear interdependencies between neurons. We study simplicial cascades over a neuronal complex representation for the neural circuitry and dynamics for nematode \emph{C. elegans} \cite{Choe-2004-connectivity, Kaiser-2006-placement}. In this example,  vertices   represent neurons' somas (i.e., cell bodies),  edges represent experimentally observed  synapses, and we use higher-order simplices to encode potential higher-order nonlinear dynamical relationships (e.g., co-activations) between combinatorial sets of neurons \cite{reimann2017cliques}. Notably, we   simulate STM cascades  on an undirected \emph{C. elegans} synapse network since our model and theory doesn't involve directed $k$-simplices. 

In Fig.~\ref{fig3}a), we visualize the \emph{C. elegans} neuronal complex. The locations of vertices reflect experimental measurements for the somas' centers. The length of each edge gives the distance between  somas, which we use as an estimate for the combined lengths of the axon and dendrite involved in each synapse. Geometric and non-geometric edges are indicated by blue and red lines, respectively. For simplicity, we do not visualize   higher-dimensional simplices. We provide a histogram of edge lengths in Fig.~\ref{fig3}b), and observe that most edges are short-range, but there are also many long-range connections. We  heuristically classify edges as geometric/non-geometric depending on whether   edge lengths are less than or greater than a ``cutoff'' distance of 0.169 mm. Note that this choice of threshold has no effect on the dynamics of STM cascades. Finally, we construct a neuronal complex by considering the graph's associated clique complex \cite{kaczynski2004computational}.

Observe that the \emph{C. elegans} neuronal complex  approximately lies on a 1D manifold that is embedded 2D, which occurs  due to the elongated shape of a nematode worm. Thus, we are interested in understanding the extent to which simplicial cascades locally propagate by WFP along the 1D manifold versus  non-local ANC. To provide insight, in Fig.~\ref{fig3}c) we visualize first activation times $\tau_i$ for 2D STM cascades with different parameters $T$ and $\Delta$. These subpanels recapitulate our visualizations in Fig.~\ref{fig1}b):  thresholding and higher-order interactions both suppress non-local ANC and can cooperatively promote WFP along a geometrical substrate or channel. (We will support this quantitatively below.)

While our knowledge of neuronal cascades has grown immensely in recent years \cite{luczak,beggsetal,shew2011information,shew2013functional,larremore2011predicting}, the mathematical mechanisms  responsible for directing where and how  cascades propagate have remained elusive. The coordination of higher-order nonlinear thresholding and the multidimensional geometry of simplicial complexes is a plausible structural/dynamical mechanism that can help self-organize neuronal cascades.  
%In brains, for example, thresholding and higher-order interactions can potentially be modulated in real time  via dendritic synapse location (by amplifying the signal) \cite{dendritic_integration} or via spike-timing dependence of synaptic plasticity \cite{STDP} (by integrating time-scales) to help direct cascades. 
Of course, the combined effects of other dynamical features (e.g., refractory periods, inhibition, and stochasticity \cite{brette2007simulation}) should also be explored. In particular, neurons are known to exhibit alternating states of  polarization/depolarization. In contrast, the STM cascades that we study here involve irreversible state transitions as a way to a baseline of understanding for the interplay between thresholding and higher-order interactions in the absence of the confounding effects of other dynamical features. 
As an initial step toward generalising STM cascades, we provide extended experiments in Supplementary Note `Stochastic Simplicial Threshold Model.' 
%to begin this pursuit. 
Nevertheless, our work highlights this emerging field as a promising direction for unveiling the multiscale mechanisms that orchestrate higher-order information processing within, but not limited to, neuronal systems.
%%complex cognitive systems, including 
%biological, social and technological systems. 

\begin{figure*}[t]
	\centering
	\includegraphics[width= 1\linewidth]{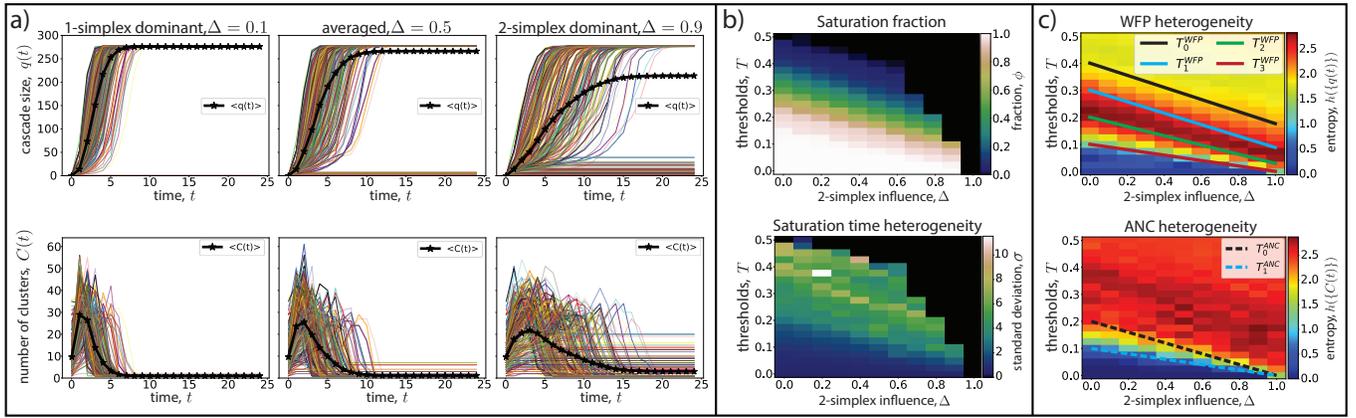}
	\caption{
		{\bf Higher-order interactions enhance diversity
		%memory capacity 
		and efficiency for STM cascade patterns on a \emph{C. elegans} neuronal complex.}
		{\bf a)}~
		Cascade size $q(t)$ (top) and the number $C(t)$ of spatially disjoint  clusters (bottom) versus time $t$  for 2D STM cascades with threshold $T=0.1$ and $\Delta\in\{0.1,0.5,0.9\}$. Different curves represent different initial conditions, and black curves give their means.
		{\bf b)}~
		For STM cascades with different  $T$ and $\Delta$, colors indicate (top) the fraction $\phi$ of initial conditions in which a cascade saturates the neuronal complex (i.e., spreads everywhere) and (bottom) the standard deviation $\sigma$ of the times at which saturations occur. Black regions indicate $(T,\Delta)$ values for which no cascades saturate the network. Cascades are most heterogeneous when $T$ and $\Delta$ are neither too small or large.
		{\bf c)}~
		Colors indicate the heterogeneity of WFP and ANC properties by showing   (top) $h(\{q(t)\})$ and (bottom) $h(\{C(t)\})$, where  $h(\cdot)$ denotes the discrete Shannon entropy of a set of cascades with different initial conditions (see Methods section `Entropy Calculation').  We focus on time $t=5$, since these measures are most reflective of WFP and ANC at early times. Lines indicate bifurcation theory that we will develop for STM cascades over noisy ring lattices, but as can be seen, the theory is also  qualitatively predictive for this neuronal complex.
		%we describe in section `Bifurcation theory for STM Cascades over geometrical channels'. 
		Observe that increasing $\Delta$ causes the spatio-temporal patterns' changes (i.e., bifurcations)  to occur for smaller $T$ values, which lowers the energy consumption per neuron activation.
	}
	\label{fig7}
\end{figure*}

%%%%%%%%%%%%%%%%%%%%%%%%%%%%%%%%%%%%%%%%%%%%%
\subsection{Higher-order interactions enhance patterns' diversity and efficiency}\label{sec:express}
%%%%%%%%%%%%%%%%%%%%%%%%%%%%%%%%%%%%%%%%%%%%%

Higher-order interactions promote  heterogeneity for   STM cascades' spatio-temporal patterns, which has important implications in the context of neuronal cascades. Specifically, neuronal networks  that exhibit   more `expressive' activity patterns have broader memory capacity \cite{shew2011information,shew2013functional}, which has been shown to occur for neuronal networks that are  tuned near `criticality'---i.e., a dynamical phase transition. At the same time, there is extensive empirical evidence that neuron interactions are higher-order \cite{Yu17514,reimann2017cliques}, yet mathematical theory development for neuronal cascades  has  largely remained limited to dyadic-interaction models (see, e.g., \cite{larremore2011predicting}).

Motivated by these insights, here we study the diversity and efficiency for STM cascades over the \emph{C. elegans} neuronal complex.
In Fig.~\ref{fig7}, we study how   parameters     $T$ and  $\Delta$ effect the heterogeneity of STM cascades.
%on the \emph{C. elegans} neuronal complex. 
In Fig.~\ref{fig7}a), we study WFP (top) and ANC (bottom) properties by plotting the cascade size $q(t)$ and the number of clusters $C(t)$, respectively, as   STM cascades propagate. The left, center, and right columns show results for STM cascades that are 1-simplex dominant ($\Delta=0.1$), averaged ($\Delta=0.5$) and 2-simplex dominant ($\Delta=0.9$), respectively. In each subpanel, different curves represent different initial conditions, whereby we select different vertices to initiate cluster seeding. Black curves indicate the means across initial conditions. Observe that some STM cascades  spread to the entire neuronal complex and are said to saturate the network, whereas others do not. Also,   early on, the numbers of clusters increase due to ANC, but they can later decrease as cascade clusters grow and merge. Moreover,   there is significant heterogeneity across the different cascades' {initializations}, which arises due to the heterogeneous connectivity of neurons within the neuronal complex. This heterogeneity becomes more prominent as $\Delta$ (the 2-simplex influence) increases.

In Fig.~\ref{fig7}b), we further study cascade heterogeneity for different initial conditions and different choices for $T$ and $\Delta$. We  plot (top) the fraction $\phi$ of cascades that saturate the  network  (i.e., when all vertices become active) and (bottom)  the standard deviation $\sigma$ for the times at which saturations occur. The black-colored regions highlight that no STM cascades saturate the network if $T$ and/or $\Delta$ are too large. Observe that the cascades' saturation fractions and times are most heterogeneous when $T$ and $\Delta$ are neither too small nor too large. This suggests thresholding and higher-order interactions may also play a `critical' role for helping tune  neuronal networks to exhibit maximal cascade pattern diversity (which is called `wide dynamic range' when considered from a multiscale perspective).

In Fig.~\ref{fig7}c), we  focus on $q(t)$ and $C (t)$ when $t=5$, which is an early time in which these values provide empirical quantitative measures for WFP and ANC, respectively. (At larger times $t$, it is difficult to distinguish WFP and ANC propagations since the cascades are so large.) For different $T$ and $\Delta$, we study the heterogeneity of these values  by computing the Shannon entropy of   (top) $h(\{q(t)\})$ and (bottom) $h(\{C(t)\})$ across the  different initial conditions. See Methods section `Entropy Calculation' for details. Observe that the entropy of cascade sizes is largest when $T$ and $\Delta$  are neither too small or too large, which is similar to our finding in Fig.~\ref{fig7}b).   When considering $h(\{C(t)\})$, we do not observe a similar  peak for intermediate values of $T$ and $\Delta$; however the changes in entropy  for   $C(t)$ and $q(t)$ occur at approximately the same values of $T $ and $\Delta$, since the spatio-temporal patterns (i.e., ANC and WFP) undergo changes at these particular parameter choices. 

We further highlight in   Figs.~\ref{fig7}b) and \ref{fig7}c) that as $\Delta$ increases, the dynamical changes can be observed to occur at smaller values of $T$. This has important implications for the efficiency of STM cascades. Specifically, we define the `activation energy' of a vertex to equal the minimum fraction of active vertices that are required in order for that vertex to become active. For our model, a vertex's activation energy monotonically decreases as $T$ decreases, since fewer neighboring activations are be required to overcome a smaller threshold barrier. In other words, a small threshold would allow a cascade to  propagate efficiently, with each vertex's activation requiring the activation of only a small number of other vertex activations. 

However, it can also be important that a threshold $T$ allows  cascades with different initial conditions to produce heterogeneous cascade patterns. Our experimental results   in Fig.~\ref{fig7}b) and \ref{fig7}c) have shown that increasing the 2-simplex influence  $\Delta$ shifts the phase transitions for dynamical behavior to occur for smaller $T$ values. In other words, the introduction of higher-order interactions for these experiments allows cascades with similarly complex  patterns to occur for smaller $T$ values (i.e., when their activation energies are smaller). 

As a concrete example, consider our visualization of $h(\{q(t)\})$ in in Fig.~\ref{fig7}c), which is given by the Shannon entropy of cascade sizes at  time $t=5$ across all initial conditions with cluster seeding. For each $\Delta$, we can consider the threshold $T$ at which $h(\{q(t)\})$ are most heterogeneous. For $\Delta\approx0$, $h(\{q(t)\})$ obtains its maximum near $T=0.25$, but for $\Delta\approx1$, $h(\{q(t)\})$  obtains its maximum near $T=0.1$ (i.e., when vertices have a smaller activation energy). In both cases, the maximum is approximately $h(\{q(5)\})\approx 2.8$. In this way,  the presence of higher-order interactions allows  cascade patterns with similarly complexity to be produced more efficiently.

%neuronal systems. 
%When the threshold $T$ is smaller, it requires fewer neighboring activations to induce a vertex %neuron  to activate. Thus, small-threshold systems inherently consume less energy, per activation. Since  larger $\Delta$ allows for heterogeneous cascade patterns to arise for smaller  $T$, this suggests that higher-order interactions can  improve the efficiency for systems can that exhibit diverse  cascade patterns over neuronal complexes.

Finally, by considering  STM cascades across the $(T,\Delta)$ parameter space, we can   systematically investigate the complementary effects of thresholding and higher-order interactions. We will develop bifurcation theory in the next section to guide this exploration, which is represented by the solid and dashed lines in Fig.~\ref{fig7}c). Importantly, our theory will be developed for STM cascades over the family of noisy ring complexes that we presented in Methods section `Generative model for noisy ring complexes', and as such, it is not guaranteed to be predictive for other simplicial complexes. That said, one can remarkably observe in Fig.~\ref{fig7}c) that this theory is   qualitatively predictive for \emph{C. elegans} neuronal complex.
%the behavior of STM cascades over empirical and synthetic models for noisy geometric complexes outside of this assumption. } 
%\buk{it is not clear which assumption here}

\begin{figure}[t]
	\centering
	\includegraphics[width=.6 \linewidth]{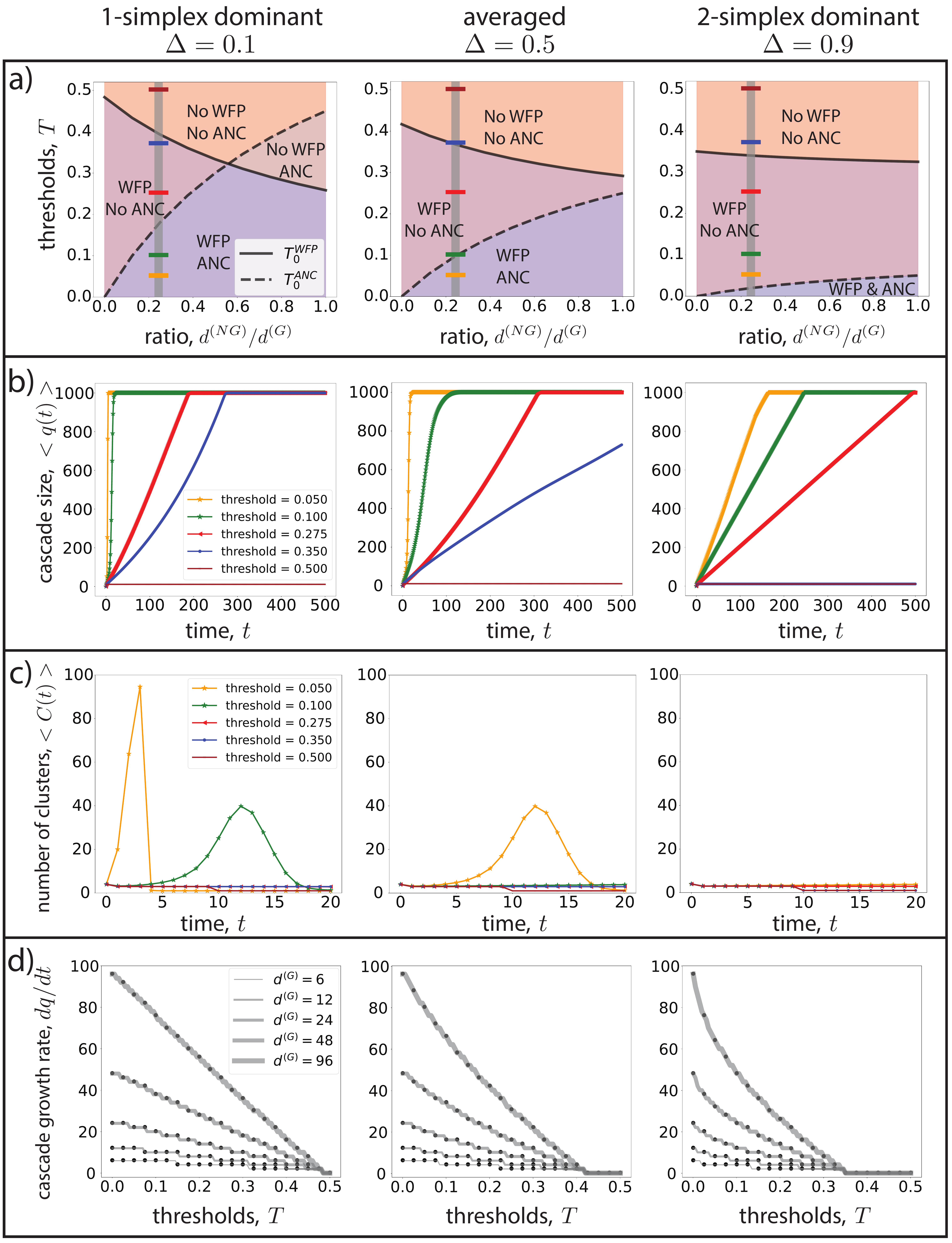}
	\caption{
		{\bf Bifurcation theory characterizes WFP and ANC over $K$-dimensional geometrical channels}. We consider 2D STM cascades over a noisy ring complex (recall Fig.~\ref{fig2}) for various $T$ and either (left) $\Delta=0.1$, (center) $\Delta=0.5$, or (right) $\Delta=0.9$.
		{\bf a)}~
		Bifurcation diagrams depict the critical thresholds $T_0^{WFP}$ and $T_0^{ANC}$ given by Eqs.~\eqref{eq:WFP_eqs} and \eqref{eq:ANC_eqs}, respectively, for different $T$, $d^{(NG)}$ and $d^{(G)}$.  We find four regimes that are characterized  by the absence/presence of WFP and ANC. Observe that increasing $\Delta$ suppresses ANC, and  the regime that exhibits ANC with no WFP disappears under higher-order coupling with $\Delta>0.1$.
		Vertical gray lines and horizontal colored marks identify the values  $ d^{(NG)}/d^{(G)}=0.25$ and $T\in\{0.05,0.1,0.275,0.35,0.5\}$, and in panels b) and c) we show for these values that the spatio-temporal patterns of  STM cascades are as predicted. 
		{\bf b)}~
		Colored curves indicate the sizes $q(t)$ of STM cascades versus time $t$, averaged across all possible initial conditions with cluster seeding. 
		{\bf c)}~
		Colored curves indicate the average number $C(t)$ of cascade clusters,   and one can observe a peak only when ANC occurs. Three scenarios give rise to WFP and ANC: $(\Delta,T) \in \{(0.1,0.05),(0.1,0.1),(0.5,0.05)\}$. Four scenarios give rise to no spreading: $(\Delta,T) \in \{(0.1,0.5),(0.5,0.5),(0.9,0.35),(0.9,0.5)\}$. The other selected values of $\Delta$ and $T$    yield WFP and no ANC, in which case $q(t)$  grows linearly, $dq/dt =2(j+1)$ for  $T \in[T^{WFP}_{j+1}, T^{WFP}_{j})$.
		{\bf d)}~
		Black symbols and gray curves indicate observed  and predicted values, respectively, of cascade growth rates, $dq/dt$, for STM cascades exhibiting WFP and no ANC for a noisy ring complex with $d^{(NG)} = 0$ and $d^{(G)} \in \{6,12,24,48,96\}$ (i.e., channel dimensions $K \in \{3,6,12,24,48\}$). Combining high-dimensional channels with higher-order interactions allows cascade growth rates to have a nonlinear sensitivity to changes for the threshold $T$.
	}
	\label{fig4}
\end{figure}

%%%%%%%%%%%%%%%%%%%%%%%%%%%%%%%%%%%%%%%%%%%%%
\subsection{Bifurcation theory for STM Cascades over geometrical channels}\label{sec:bif}
%%%%%%%%%%%%%%%%%%%%%%%%%%%%%%%%%%%%%%%%%%%%%

We analyze WFP and ANC for STM cascades over  a family of simplicial complexes in which $N$ vertices lie along a 1D manifold as shown in Fig.~\ref{fig2}b), and for which the vertices' degrees lack heterogeneity (although our experiments highlight that the theory can be qualitatively predictive beyond this assumption).  See Methods section `Generative model for noisy ring complexes' for their formation, which  generalizes the noisy ring lattices that are studied in \cite{taylor2015topological}, wherein the authors developed bifurcation theory to predict WFP and ANC properties for a threshold-based cascade model that is restricted to dyadic interactions. In the Methods section `Combinatorial analysis for bifurcation theory' we  describe bifurcation theory that characterizes  STM cascades over  noisy ring complexes. We present general theory for $\kappa$-dimensional STM cascades, and we summarize here bifurcation theory for 2D STM cascades. Our theory assumes large $N$ and is based on a combinatorial analysis for the different possible state changes for boundary vertices that have active simplicial neighbors, but they themselves are not yet active.  We focus on the early stage of cascades in which they are just beginning to spread, and we summarize our results below.
%\buk{Below, we summarize our results that hold for simplicial complexes obtained by k-regular graphs. As we show in Supplementary Material Sections `Effects of 1-simplex degree heterogeneity on STM cascades' and `Effects of 2-simplex degree heterogeneity on STM cascades', these results also hold for small perturbations to the k-regular graph families. Moreover, we find that the introduction of significant irregularity into a geometric substrate can significantly reduce WFP; however, the introduction of irregularity for non-geometric $k$-simplices has comparatively little effect (unless the cascade size is large, in which case it can increase the rate of ANC).}
%\todo{It seemed weird to discuss the extended results/experiments before introducing the results/experiments. so I moved the reference to SUpplementary notes to the end.}
%\buk{OK, but above paragraph seems to be changed, we perhaps need to highlight that in blue? Particularly, I was referencing this paragraph under the comment 1G.}

Our primary findings are two sequences of critical thresholds that characterize WFP and ANC and which depend on the STM parameter $\Delta$ and degrees $d^{(G)}$, $d^{(NG)}$, $d_{i}^1$, and $d_{i}^{2}$. The qualitative properties of WFP are determined by critical thresholds
\begin{equation}\label{eq:WFP_eqs}
    T^{WFP}_{j}  = (1-\Delta)\frac{s_j }{d_{i}^{1}} + \Delta\frac{1}{d_{i}^{2}}\binom{s_j }{2},
\end{equation}
where $s_j  = {d^{(G)}}/{2}-j$ is the number of active geometric 1-simplex neighbors and $ j \in \{0,1,.., {d^{(G)}}/{2}\}$. 
The first and second terms in Eq.~\eqref{eq:WFP_eqs} represent 1-simplex and 2-simplex influences, respectively. Here, we highlight that $\binom{s_j }{2}$ equals the number of geometric 2-simplex neighbors that are active, since we assume that for every pair of active  geometric 1-simplex neighbors of $v_i$, there   exists an associated geometric 2-simplex neighbor of $v_i$. This is true for the geometric substrate for which we develop theory, which is a clique complex associated with geometric edges that are arranged in a $k$-regular ring lattice.
While the thresholds $T^{WFP}_{j}$ may differ for  vertices $v_i$ that have different $k$-simplex degrees, they are the same for simplicial complexes that are ``$k$-simplex degree-regular'' and have identical local connectivity in the geometric substrate. Moreover, Eq.~\eqref{eq:WFP_eqs} has assumed that non-geometric $k$-simplices are inactive, which occurs with very high probability for small cascades in which $q(t)/N\ll1$.
%Moreover,  unless otherwise noted, we focus on thresholds associated with the median $k$-simplex degrees in our experiments. 
The resulting critical  thresholds identify  ranges $T \in[T^{WFP}_{j+1}, T^{WFP}_{j})$ such that the speed of WFP   is identical for any threshold $T$  within a given range. Within each range, the WFP speed is $j+1$. For noisy ring complexes, WFP progresses in the clockwise and counter-clockwise directions, given the cascade growth $q(t)\approx (2j+2)t$ for small $t$. There is no WFP when $T>T^{WFP}_{0}$.

Similarly, the qualitative properties of ANC are determined by critical thresholds 
\begin{equation}\label{eq:ANC_eqs}
    T^{ANC}_{j}  = (1-\Delta) \frac{d^{(NG)}-j}{d_{i}^{1}},
\end{equation}
where $j\in \{0,1,..,d^{(NG)}\}$ 
and $d^{(NG)}-j$ represents the number of adjacent non-geometric 1-simplices that are active. Note  that  there is not a second term in the right-hand side of Eq.~\eqref{eq:ANC_eqs},  since our theory assumes that the non-geometric 2-simplex neighbors of a vertex $v_i$   are inactive at early cascade times (i.e., small $t$). Such an event  occurs with vanishing probability when $q(t)/N$ is small. Also note that Eq.~\eqref{eq:ANC_eqs} assumes all geometric neighbors are inactive, which is required by the definition of ANC. It follows that the  probability of   ANC occurrences
%
%An ANC event occurs when a non-local boundary vertex $v_i$  becomes active, which requires a certain level of activity across its non-geometric simplicial neighbors (i.e., adjacent 1-simplices and 2-simplices that are long-range). 
%the probability of that happening
is the same for any   $T \in [T^{ANC}_{j+1},T^{ANC}_{j})$, and it is different for any two $T$ values in different regions. Notably, there is no ANC if $T>T^{ANC}_{0}$.

In Fig.~\ref{fig4}a), we show  bifurcation diagrams that characterize   WFP and ANC  for different choices of $T$ and the ratio $d^{(NG)}/d^{(G)}$.  Solid and dashed black  lines indicate $T^{WFP}_{0}$ and $T^{ANC}_{0}$, respectively. Different columns depict bifurcation diagrams for different  STM cascades that are either: (left, $\Delta=0.1$) 1-simplex dominant; (center, $\Delta=0.5$) averaged; or  (right, $\Delta=0.9$) 2-simplex dominant. The vertical gray  lines  and  horizontal colored marks indicate choices for the  ratio $d^{(NG)}/d^{(G)}$  and $T$ that are further studied in Figs.~\ref{fig4}b) and \ref{fig4}c). Observe in Fig.~\ref{fig4}a) that as $\Delta$ increases, the region of parameter space  exhibiting WFP and no ANC expands, whereas the region exhibiting WFP and ANC shrinks. Notably, the region exhibiting ANC and no WFP  vanishes altogether for $\Delta>0.1$. In other word, as STM cascades  are more strongly influenced by higher-order interactions, they exhibit an increase in WFP and a decrease in ANC; they more robustly propagate via WFP along a geometrical channel/substrate, and they are less impacted by the  `topological noise' that is imposed  by  the presence of long-range, non-geometric $k$-simplices.

In Figs.~\ref{fig4}b) and \ref{fig4}c), we plot the cascade size $q(t)$ and number $C(t)$ of clusters, respectively, as a function time $t$. These  are averaged across all possible initial conditions with cluster seeding. As before, the left, center and right columns depict the choices $\Delta\in\{0.1,0.5,0.9\}$. In each panel, we show several curves for different thresholds  $T\in\{0.05,0.1,0.275,0.37,0.5\}$. All panels reflect results for noisy ring complexes with $d^{(G)} = 8$ and $d^{(NG)} = 2$ (i.e., $d^{(NG)}/d^{(G)} = 0.25$). Our selection for these parameter choices was guided by the bifurcation diagrams in Fig.~\ref{fig4}a). We chose these particular values to highlight the impact of $\Delta$ and $T$ on WFP and ANC properties. In particular, cascades exhibiting WFP and no ANC will have linear growth for $q(t)$ and  the number of clusters $C(t)$ does not increase. (Note that it would be quadratic growth for WFP on the 2D manifold shown in Fig.~\ref{fig1}, cubic growth for 3D manifolds, and so on.) On the other hand, cascades exhibiting WFP and  ANC will have very rapid growth for $q(t)$ and an initial spike for the number of clusters $C(t)$. $C(t)$ can later decrease as clusters merge together. Finally, cascades do not spread if they neither exhibit  WFP nor ANC.

One can observe in Figs.~\ref{fig4}b) and \ref{fig4}c) that  the qualitative features of WFP and ANC occur for  different choices of $T$ and $\Delta$  exactly as predicted by our bifurcation theory. First, there is no spreading when $T=0.5$ and $\Delta\in\{0.1,0.5\}$ (left and center columns), or when $T\in\{0.35,0.5\}$ and $\Delta=0.9$ (right column), since $T>T_0^{WFP}$ and $T>T_0^{ANC}$ in these cases and there is neither WFP nor ANC. Second, there is a sharp rise in the number of clusters and rapid, super-linear growth only when $T\in\{0.05,0.1\}$ and $\Delta=0.1$ (left column) and when $T=0.05$ with $\Delta=0.5$ (center column), since $T<T_0^{WFP}$ and $T<T_0^{ANC}$ in these cases and both WFP and ANC occur. Third, for all other values of $T$ and $\Delta$, the curves exhibit linear growth  when $t$ is small, since $T<T_0^{WFP}$ and $T>T_0^{ANC}$ and there is WFP but no ANC. (The growth rate of spreading can be faster at later times $t$, since our bifurcation theory focuses on the nature of spreading dynamics at early stages of the cascades.)

In Fig.~\ref{fig4}d), we study how the speed of WFP along a geometric channel  is affected by the threshold $T$, 2-simplex influence parameter $\Delta$, and the channel dimension $K= d^{(G)} /2$. Black symbols and gray curves indicate observed and predicted values of cascade growth size, $dq/dt$, for $d^{(NG)}=0$ and different choices of $d^{(G)}\in\{6,12,24,48,96\}$. First, observe that our prediction $dq/dt=2(j+1)$ for  $T \in[T^{WFP}_{j+1}, T^{WFP}_{j})$ is very accurate for the different parameter values. Second, observe that $ {dq}/{dt}$ generally increases with the channel dimension $K$. Lastly, observe in the right column of  Fig.~\ref{fig4}d) that by introducing higher-order interactions (i.e., large $\Delta$), cascade growth rates $ {dq}/{dt}$ have a nonlinear sensitivity to changes of the threshold $T$. Such a nonlinear response could benefit the directing of cascade propagation   via mechanisms that   modulate  activation thresholds (e.g., neurochemical modulations).

\begin{figure*}[t]
	\centering
	\includegraphics[width= 1\linewidth]{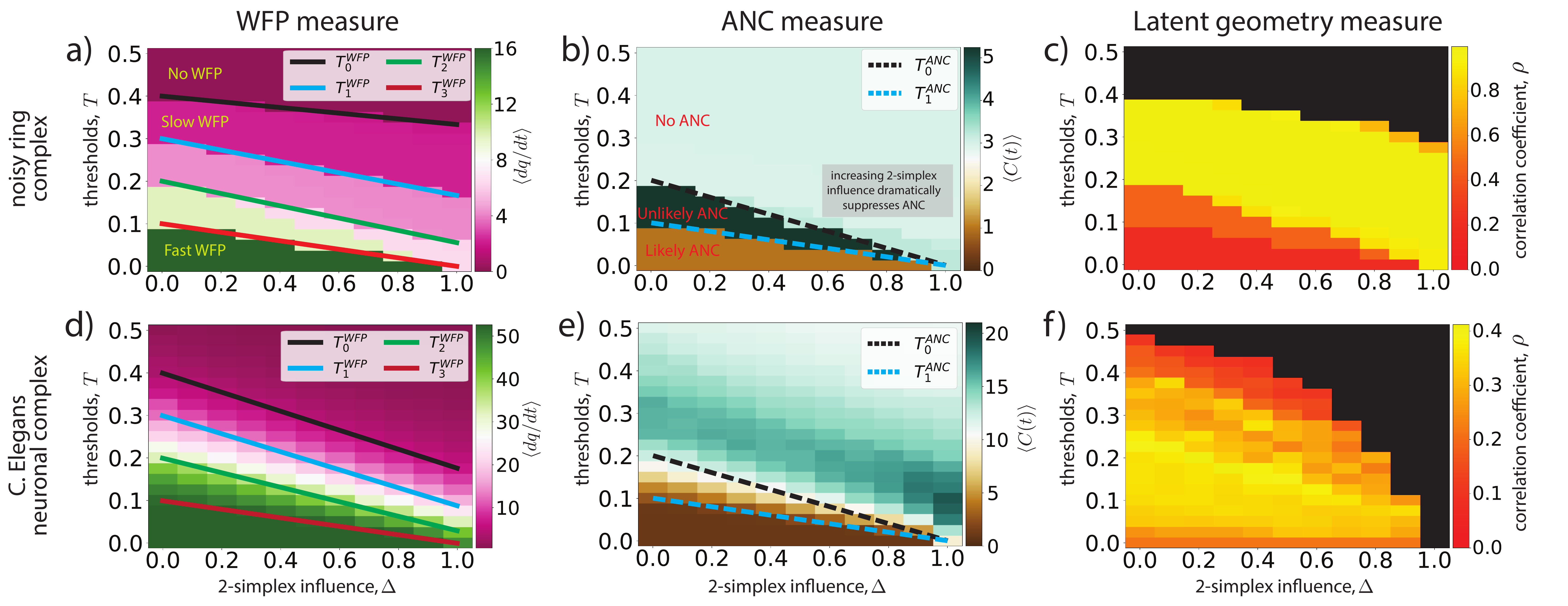}
	\caption{
		{\bf Empirical measurements for WFP and ANC rates predicted by critical thresholds $T_j^{WFP}$ and $T_j^{ANC}$.}
		We study 2D STM cascades with various $T$ and $\Delta$ over (top row) a noisy ring complex with $N=1000$ vertices, $d^{(G)} = 8$, and $d^{(NG)}=2$ and
		(bottom row) the \emph{C. elegans} neuronal complex.
		{\bf a)}~
		An empirical measure for WFP speed, $\frac{dq}{dt}$, which we compute at $t=5$ and average across all initial conditions with cluster seeding. Observe that 	$\frac{dq}{dt}$ undergoes changes  at the critical thresholds $T_j^{WFP}$ given by Eq.~\eqref{eq:WFP_eqs}, which vary with   $\Delta$. Within each region  $T\in [T_{j+1}^{WFP},T_{j}^{WFP})$, observe that the growth rate is close to our predicted rate of $2j+2$.
		{\bf b)}~
		An empirical measure for ANC, $C(t)$, which we  compute at $t=5$ and average across   initial conditions. Observe that  $C(t)$ undergoes changes that are accurately predicted by critical thresholds $T_j^{ANC}$ given in Eq.~\eqref{eq:ANC_eqs}. That is, there are three regions
		$T\in [T_{j+1}^{ANC},T_{j}^{ANC})$, and ANC events   occur at  approximately the same rate within each region.
		{\bf c)}~
		A Pearson correlation coefficient $\rho$ quantifies the extent to which STM cascades predominantly follow along the manifold via WFP. It is computed by comparing pairwise-distances  between vertices $v_i$ to $v_{i'}$ in the original 2D ambient space containing the ring manifold  to pairwise-distances $|| \bm{\tau}^{(i)}- \bm{\tau}^{(j)} ||_2 $ between a nonlinear embedding of $N$ vertices $v_i$ and $v_j$ using STM Cascade Maps $ \{i\} \mapsto \{\bm{\tau}^{(i)}\}\in\mathbb{R}^J  $  for which distances  reflect the time required for STM cascades to travel between vertices (see Methods section `Simplicial cascade maps').
		{\bf d)--f)}~
		Similar information as in panels a)--c), except for the \emph{C. elegans} neuronal complex.
	}
	\label{fig5}
\end{figure*}

In Fig.~\ref{fig5}, we study how increasing either $T$ or $\Delta$ generically slows the spread of STM cascades, and in particular, it slows the rates of both WFP and ANC behaviors. This is predicted by the other critical thresholds given in Eqs.~\eqref{eq:WFP_eqs} and \eqref{eq:ANC_eqs} for different values of $j\ge 0$. In Fig.~\ref{fig5}a), we use color to depict the average rate of change for $q(t)$ at time $t=5$, which is an empirical measure for WFP speed. We predict linear growth for $q(t)$ at a rate of ${dq}/{dt}\approx 2j+2$ for $T\in[T^{WFP}_{j+1},T^{WFP}_{j})$, which is very close to what we empirically observe. Observe that as $\Delta$ increases, the ranges associated with larger $j$ broaden, whereas the ranges associated with smaller $j$ narrow. This can be understood by examining the right-hand side of Eq.~\eqref{eq:WFP_eqs} and noting that the first term is linear, whereas the second term is combinatorial. Hence, as STM cascades are more strongly influenced by 2-simplex interactions, slower WFP becomes a more dominant phenomenon across the $T$-parameter space.

In Fig.~\ref{fig5}b), we use color to depict the average number $C(t)$ of clusters  at time $t=5$, which is an empirical measure for the rate of ANC. Our bifurcation theory predicts three ranges $T\in[T^{ANC}_{j+1},T^{ANC}_{j})$, and as expected, the observed number of clusters is similar within these ranges and different across them. Importantly,   increasing $\Delta$ causes all of the thresholds $T_j^{ANC}$ to approach 0. Thus, for any fixed $T$, increasing $\Delta$ will cause ANC events to vanish altogether. So while thresholding and higher-order interactions   play a similar mechanistic role in that they both suppress ANC and allow WFP, higher-order interactions achieve this much more effectively. 

In Figs.~\ref{fig5}d) and \ref{fig5}e), we depict similar information as Figs.~\ref{fig5}a) and \ref{fig5}b), except it is computed for the \emph{C. elegans} neuronal complex, rather than a noisy ring complex for which the bifurcation theory was developed. See Methods section `Critical Regimes for \emph{C. elegans}' for further information. Despite being outside the assumptions  of  our bifurcation theory,  Eqs.~\eqref{eq:WFP_eqs} and \eqref{eq:ANC_eqs} surprisingly predictive for the  qualitative behavior of
%reasonable job of predicting the qualitative behavior of 
WFP and ANC for the \emph{C. elegans} neuronal complex (that is,   spatio-temporal pattern changes still occur near the bifurcation lines). Also, observe that the transitions in Figs.~\ref{fig5}d) and \ref{fig5}e) are not as abrupt as those shown in Figs.~\ref{fig5}a) and \ref{fig5}b), since the neuronal complex has heterogeneous 1- and 2-simplex degrees, which is known to blur bifurcation  \cite{taylor2015topological}. Nevertheless, the theory accurately predicts the general trend for how increasing $\Delta$ leads to a suppression of ANC, thereby  promoting WFP.

In Supplementary Note ``Effects of 1-Simplex Degree Heterogeneity on STM Cascades'', we numerically study how heterogeneity added to the geometric and/or non-geometric 1-simplex degrees affects bifurcations that occur for WFP and ANC for 2D STM cascades over noisy ring complexes. Our main finding is that our analytically derived bifurcations remain qualitatively accurate even when a small  amount of degree heterogeneity is introduced. We also find that the introduction of heterogeneity for non-geometric edges decreases the range of $T$ for which WFP is predominantly exhibited over ANC. Interestingly, increasing the influence of 2-simplices (i.e., increasing $\Delta$) counterbalances this effect.  That is, higher-order interactions help simplicial cascades become more robust to the noise imposed by degree heterogeneity (see  Supplementary Figure 6, lower row).
In Supplementary Material Section `Effects of 2-simplex degree heterogeneity on STM cascades,'
we further  extend this study, finding that the introduction of significant irregularity into a geometric substrate can significantly reduce WFP; however, the introduction of many non-geometric $2$-simplices has comparatively little effect on STM cascades for early   times in which $q(t)$ is small.

%%%%%%%%%%%%%%%%%%%%%%%%%%%%%%%%%%%%%%%%%%%%%
\subsection{Latent geometry of simplicial cascades quantifies WFP vs ANC}\label{sec:simpmaps}
%%%%%%%%%%%%%%%%%%%%%%%%%%%%%%%%%%%%%%%%%%%%%

It was proposed in \cite{taylor2015topological} to quantitatively study   competition between WFP and ANC using techniques from high-dimensional data analysis, nonlinear dimension reduction, manifold learning, and topological data analysis. The approach relied on constructing ``contagion maps''  in which a set of vertices $\mathcal{C}_0$ in a graph are nonlinearly embedded in a Euclidean metric space so that the distances between vertices reflect the time required for contagions to traverse between them. Contagion maps are similar to  other nonlinear embeddings that are based on diffusion  \cite{coifman2006diffusion} and shortest-path distance \cite{tenenbaum2000global}, but in contrast, they provide insights about the dynamics of thresholded cascades (as opposed to the dynamics of heat diffusion, for example). We generalize this approach by attributing the vertices in a simplicial complex with  a latent geometry so that pairwise distances between   vertices reflect the time required for STM cascades to traverse between them.   See Methods section `Simplicial cascade maps' for  details on this construction.  Each cascade map uses $J$ different initial conditions with cluster seeding to yield a point cloud  $\{v_i\}\mapsto \{\bm{ \tau}^{(i)}\}\in \mathbb{R}^J$. For each,  we compute the Pearson correlation coefficient $\rho$ between pairwise distances $||\bm{ \tau}^{(i)}-\bm{\tau}^{(j)}||_2$ in the latent embedding and pairwise distances  between   vertices in the original ambient space (e.g., locations on a ring manifold in 2D or the empirically observed  locations of somas for \emph{C. elegans}). See Supplementary Note  ``Visualizations of STM Cascade Maps'' for visualizations of these point clouds and further discussion.

In Figs.~\ref{fig5}c) and \ref{fig5}f), we use the high-dimensional geometry of simplicial cascade maps to quantitatively study the competing phenomena for WFP and ANC for a noisy ring complex and the \emph{C. elegans} neuronal complex, respectively. We use color to visualize $\rho$ for different simplicial cascade maps  using STM cascades with different choices for $\Delta$ and $T$. Larger values of $\rho$ indicate parameter choices in which cascades exhibit a prevalence  of WFP versus ANC, whereas smaller $\rho$ indicate the opposite. Observe in both Figs.~\ref{fig5}c) and \ref{fig5}f) that larger $\rho$ values occur for an intermediate regime in which  $T$ and $\Delta$ are neither too larger nor too small. In this regime, the geometry of  simplicial cascade maps best matches the original 2D geometry, which occurs because   STM cascades predominantly exhibit WFP along the geometrical substrate and are not disrupted by ANC across   long-range simplices (i.e., the topological `noise'). By comparing the panels in Fig.~\ref{fig5}c) to those in Figs.~\ref{fig5}a) and \ref{fig5}b), observe that the regions of larger $\rho$ coincide with regions in which there is slow WFP and unlikely ANC, as is predicted by our bifurcation theory. Finally, observe that the $\rho$ values are generally larger for the noisy ring complex than for the \emph{C. elegans} complex. This likely occurs because the  noisy ring complexes that we study   have no degree heterogeneity, whereas the \emph{C. elegans} neuronal complex does have heterogeneous $k$-simplex degrees. Also, we note that  the \emph{C. elegans} neuronal complex contains many more non-geometric 2-simplices than that of the noisy ring complexes.
That said, our extended experiments in Supplementary Material Section `Effects of 2-simplex degree heterogeneity on STM cascades' suggest that it is the irregularity of geometric $k$-simplices---not the non-geometric $k$-simplices---that has the greatest impact on WFP and ANC.

%%%%%%%%%%%%%%%%%%%%%%%%%%%%%%%%%%%%%%%%%%%%%%%%%%%%%%%%%%%%
%%%%%%%%%%%%%%%%%%%%%%%%%%%%%%%%%%%%%%%%%%%%%%%%%%%%%%%%%%%%
\section{Discussion}\label{sec:discussion}
%%%%%%%%%%%%%%%%%%%%%%%%%%%%%%%%%%%%%%%%%%%%%%%%%%%%%%%%%%%%
%%%%%%%%%%%%%%%%%%%%%%%%%%%%%%%%%%%%%%%%%%%%%%%%%%%%%%%%%%%%
 
Nonlinear cascades arise in diverse types of social, biological, physical and technological systems, many of which are  insufficiently represented by cascade models that are restricted to pairwise (i.e., dyadic) interactions \cite{james2016information,Yu17514,maclean,mayfield,ghasemi2021data,dass2011holistic,lanchier2013stochastic,civilini2021evolutionary,noonan2021dynamics,patania}. Thus motivated, we have proposed a simplicial threshold model (STM) for  cascades over simplicial complexes that encode dyadic, triadic, and higher-order interactions. Our work complements recent higher-order models for epidemic spreading \cite{bodo2016sis,petrietal,barratetal,chowdaryetal,restrepoetal,PhysRevResearch.2.012049,higham2021epidemics}, social contagions \cite{PhysRevResearch.2.023032,neuhauser2021opinion}, and consensus \cite{yu2011distributed,neuhauser2020multibody,neuhauser2021consensus,sahasrabuddhe2021modelling}, and in particular, the effects of higher-order interactions on spatio-temporal patterns (i.e., WFP vs ANC) and the implications for neuronal avalanches have not yet been explored. By assigning the states  of active/inactive to individual vertices as well as groups of vertices, STM cascades provide a new modeling framework that can help bridge individual-based threshold models (e.g.,   social contagions and neuron interactions) with group-based threshold models (e.g.,  group decision making and interacting neuron groups such as cortical columns or structural communities). In particular, simplicial cascades allow for the modeling of ``multidimensional cascades'' in which the states of individuals influence the states of groups, and vice versa, and such interactions  cannot be appropriately represented by graph-based modeling. Herein, the dynamical states of higher-dimensional simplices are inherited by their associated vertices' states, and it would be interesting in future work to explore more complicated dependencies such as allowing time lags between when a vertex becomes active and when its adjacent higher-dimensional simplices subsequently become active. Such multidimensional models remain an exciting open avenue for research.

By studying STM  cascades over ``noisy geometric complexes''---a family of spatially embedded simplicial complexes that contain both short- and long-range $k$-simplices---our work  reveals the interplay between higher-order dynamical nonlinearity and the multidimensional geometry of simplicial complexes to be  a promising  direction for research into how complex systems  organize the spatio-temporal patterns of cascade dynamics. We  have shown that the coordination of higher-order coupling and thresholding allows STM cascades to robustly  suppress the appearance of new clusters (ANC), yielding local  wavefront propagation (WFP) along a geometrical substrate. STM cascades can propagate along $k$-dimensional geometrical channels (i.e., a sequence of ``lower adjacent'' $k$-simplices) despite the presence of  long-range simplices (which introduce a ``topological noise'' to the geometry). While \cite{taylor2015topological,mahler2021analysis} presents bifurcation theory describing how thresholding impacts WFP and ANC on noisy geometric networks containing short- and long-range edges,  no prior work has explored the effects of  higher-order coupling on WFP and ANC. This is problematic, since understanding whether a cascade predominantly spreads   locally or non-locally significantly impacts the steps that one takes, e.g., to predict and control cascades \cite{onnela2010spontaneous,bentley2021social,hollingsworth2006will,epstein2007controlling,colizza2007modeling,gu2015controllability,medaglia2020personalizing,dobson2007complex,hines2016cascading,pathak2007complexity,dolgui2018ripple,mari2015adaptivity}. Our bifurcation theory for STM cascades over geometrical channels (see Fig.~\ref{fig2} and Eqs.~\eqref{eq:WFP_eqs} and \eqref{eq:ANC_eqs}) was shown to accurately predict how WFP and ANC  change depending on parameters of the cascade (i.e., threshold $T$ and a parameter $\Delta$ that tunes the relative strength of 2-simplex interactions) and parameters of the noisy geometric complex (i.e., the $k$-simplex degrees, which measure the number of number of  geometric edges, $d^{(G)}$,  non-geometric edges, $d^{(NG)}$, and   2-simplices, $d^{2}$, that are adjacent to a vertex). 
This theory characterizes the absence/presence of WFP and ANC and their respective rates, and it  provides a solid theoretical foundation to support the exploration of WFP and ANC for higher-order cascades in a variety applied settings (e.g., neuronal avalanches,   cascading failures, and so on).

Our work   provides  important insights for higher-order information processing in neuronal networks  and other complex systems. Higher-order dependencies are widely observed for neuronal activity \cite{Yu17514,reimann2017cliques}, yet theory development for  neuronal cascades is largely restricted to pairwise-interaction models \cite{larremore2011predicting}.  Thus motivated, we  studied STM cascades over a `neuronal complex' that represents the structural and higher-order nonlinear dynamical dependencies among neurons in nematode \emph{C. elegans}. We have shown that thresholding and higher-order interactions  can collectively orchestrate the spatio-temporal patterns of STM cascades that spread across the multidimensional geometry of a neuronal complex, which we predict to be an important  mathematical mechanism that can potentially help brains direct   neuronal cascades and optimize the diversity
%memory capacity
and efficiency of cascades' spatio-temporal patterns (see Fig.~\ref{fig7}).  Given the importance of efficiency in brains, simplicial-complex modeling is expected to also lead to new perspectives for other types of efficiency, such as wiring efficiency \cite{sporns}. Moreover, we have   shown (see Fig.~\ref{fig2}d)) that the combination of higher-order coupling with high-dimensional channels allows the growth-rates of STM cascades to be nonlinearly sensitive to changes in $T$, which may benefit the directing of multiscale cascades via the (e.g., neurochemical) modulation of activation thresholds $T_i$. Moreover, the sizes and durations of neuronal avalanches are known to exhibit wide dynamical range \cite{shew2011information,larremore2011predicting,shew2013functional}, and we have shown that higher-order interactions can provide a mechanism for growth rates to have similar heavy-tailed heterogeneity (which we pose as a measurable hypothesis for the neuroscience community).

It is also worth noting that we have proposed an intentionally simple model for higher-order cascades with the goal of gaining concrete, analytically tractable insights. Future work should investigate the combined effects of other dynamical properties of neurons (e.g., alternating states of activity/inactivity, refractory periods, inhibition, directed edges, and stochasticity  \cite{brette2007simulation}) and other dynamical behaviors such as local/non-local  patterns for synchronized neuron firings  (which may benefit from recent advances in synchronization theory for  higher-order systems \cite{PhysRevLett.124.218301, PhysRevResearch.2.023281,gambuzzaetal,calmon2021topological,skardal2021higher}). In this same vein, future research should also investigate biological processes that could possibly mediate the coordination of higher-order interactions and thresholding, particularly by incorporating empirical neuronal data. Thus motivated,  we introduce and study a stochastic variant of STM cascades in  Supplementary Note `Stochastic   Simplicial Threshold Model'. We show that our results for deterministic STM cascades remain qualitatively similar as long as the propagation mechanism remains dominated by thresholding and not stochasticity. 

%\todo{mention directed edges}

Finally, we have introduced a technique called `simplicial cascade maps' that  embed a simplicial complex in a latent metric space. This nonlinear embedding extends contagion maps \cite{taylor2015topological}, which are recovered under the assumption of 1D STM cascades, and both mappings embed vertices so that the distance between vertices reflects how long cascades take to traverse from one vertex to another. Simplicial cascade maps generalize the well-developed field of graph embedding to the context of simplicial complexes, and we have used them to quantitatively study the extent to which STM cascades follow geometrical channels within a simplicial complex, i.e., as opposed to exhibiting non-local ANC phenomena. Although it is not our focus herein, simplicial cascade maps are expected to support  higher-order generalizations of  methodology development for manifold learning, topological data analysis, and nonlinear dimension reduction. Notably, STM cascades can robustly follow geometrical substrate despite the presence of topological noise, which is a property that can benefit these data-science pursuits when they are applied to noisy data.

%%%%%%%%%%%%%%%%%%%%%%%%%%%%%%%%%%%%%%%%%%%%%%%%%%%
%%%%%%%%%%%%%%%%%%%%%%%%%%%%%%%%%%%%%%%%%%%%%%%%%%%
\section{Methods}\label{sec:methods}
%%%%%%%%%%%%%%%%%%%%%%%%%%%%%%%%%%%%%%%%%%%%%%%%%%%
%%%%%%%%%%%%%%%%%%%%%%%%%%%%%%%%%%%%%%%%%%%%%%%%%%%

%................................................
\subsection{STM cascades}\label{sec:STM2}
%````````````````````````````````````````````````

In the text above, we focused on the case of 2D STM cascades. We  now define a general version for  STM cascades  of dimension $\kappa\ge 1$. At time step $t+1$, the state $x_i^0(t)$ of each vertex $v_i$ possibly changes according to the threshold criterion given by Eq.~\eqref{eq1} except that we now define
the simplicial exposure to be
\begin{equation}\label{eq1a}
    R_{i}^{t} = \sum_{k=1}^\kappa \alpha_k f_{i}^k(t),
\end{equation}
where $f_{i}^k(t) = \frac{1}{d^k_i}\sum_{j \in\mathcal{N}_{A}^k(i)} x_j^k(t)$ is the fraction of vertex $v_i$'s neighboring $k$-simplices that are active and $\{\alpha_k\}$ are non-negative weights that satisfy $1=\sum_k \alpha_k$. The choice $\alpha_1=(1-\Delta)$, $\alpha_2=\Delta$ and $\kappa = 2$ recovers the model for 2D STM cascades that we studied above. In   Supplementary Note `Stochastic Stochastic Simplicial Threshold Model', we formulate and study a stochastic generalization of this model.

%................................................
\subsection{Cluster seeding}\label{sec:seeding}
%````````````````````````````````````````````````

We initialize an STM cascade at a vertex $v_{i}$ with cluster seeding, which we define as follows. Let $\mathcal{N}^{1}(i)\subset \mathcal{C}_0$ denote the set of vertices that are adjacent to $v_i$ through $1$-simplices. We set $x_{j}^{0}(0) = 1$ for any  $ j\in \mathcal{N}^{1}(i)$ at time $t=0$ and  $x_{j'}^{0}(0) = 0$  for any $  j'\not\in \mathcal{N}^{1}(i)$. Thus, the size of an STM cascade at time $t=0$ is $q(0) = d_i^1$, which can possibly vary depending on the vertex degrees. Note that the seed vertex $v_{i}$ itself is not in the set $\mathcal{N}^{1}(i)$, since we  assume no self loops. Therefore $x_{i}^{0}(0)$ is inactive at $t=0$, but it will very likely become active at time $t=1$ (excluding the situation of pathologically large   $\Delta$ and $T$).

%................................................
\subsection{Generative model for noisy ring complexes}\label{sec:Ring}
%````````````````````````````````````````````````

We construct noisy ring complexes by considering the clique complexes associated with noisy ring lattices \cite{taylor2015topological}. First, we place $N$ vertices $v_i$ at angles $\theta_i = 2\pi (i/N)$ for $i\in\{1,\dots, N\}$. We then create   geometric edges by connecting each vertex to its $d^{(G)}$ nearest neighbors. We assume $d^{(G)}$ to be an even number so that ${d^{(G)}}/{2}$ edges go in either direction along the 1D manifold.  Next, we create non-geometric edges uniformly at random between the vertices so that each vertex has exactly $d^{(NG)}$ non-geometric edges. We generate non-geometric edges using the configuration model, except   we introduce a re-sampling procedure to avoid adding an edge that already exists. The resulting  graph is a noisy ring lattice, and we construct its associated clique complex to yield a noisy ring complex. (Recall that a clique complex is a simplicial complex that is derived from a graph, and there is a one-to-one correspondence between each clique  involving $(k+1)$ vertices in the graph and each $k$-simplex in the simplicial complex.) Finally, each $k$-simplex is then defined to be geometric or non-geometric, depending on whether it involves one or more non-geometric edge. This generative model yields noisy ring complexes that are  specified by three parameters: $N$, $d^{(G)}$ and $d^{(NG)}$.

Noisy ring complexes are particularly amenable to theory development because they  are degree  regular with respect to the 1-simplex degrees; each vertex $v_i$ is adjacent to exactly $d_i^{1}=  d^{(G)} + d^{(NG)}$ 1-simplices, where $d^{(G)}$ and $d^{(NG)}$ are the geometric and non-geometric 1-simplex degrees, respectively.  The degrees $d_i^{k}$ of higher-order  simplices  are not degree regular; however, the geometric degrees $d_i^{k,G}$  for   $k\ge1$ are identical across vertices due to the symmetry of the geometrical substrate (i.e., the `sub' simplicial complex that includes only geometric simplices).

While STM cascades can be studied over any simplicial complex, we focus herein on clique complexes, which helps facilitate the identification of adjacencies among $k$-simplices. If ${\bf A}$ is a graph's adjacency matrix so that $A_{ij} =1$ if $(v_i,v_j)\in \mathcal{C}_1$ and $A_{ij} =0$ otherwise, then an entry $B_{ij}$  in matrix ${\bf B} = {\bf A}^2 * {\bf A}$  encodes the number of 2-simplices  that are shared by vertices $v_i$ and $v_j$. (Here, $*$ denotes the Haddamard, or `entrywise', product.) In this work, we make use of matrices  ${\bf A}$ and ${\bf B}$ when numerically implementing 2D STM cascades over  clique complexes.

%................................................
\subsection{Entropy calculation}\label{sec:entropy}
%````````````````````````````````````````````````

We use Shannon entropy in Fig.~\ref{fig7}c) to quantify the diversity of spatio-temporal patterns of 2D STM cascades on a \emph{C. elegans} neuronal complex, and we compute it as follows. In each panel of Fig.~\ref{fig7}a), we plot (top) cascade size $q(t)$ and (bottom) the number of spatially distant cascade clusters $C(t)$, and different curves indicate $q(t)$ and $C(t)$ for  different initial conditions with cluster seeding. Focusing on $t=5$, we consider the sets $\{q(5)\}$ and $\{C(5)\}$ and approximate their probability distributions by constructing histograms with  $20$ bins. Letting $p_i$ denote the fraction of entries that fall into the $i$-th bin, we compute the associated discrete Shannon entropy 

\begin{equation}\label{f}
h  = -\sum_{i=1}^{20}p_{i}\log{p_{i}}.
\end{equation}

We note that   our choice for the number of bins does effect the total entropy; however, we find that it has little effect on the qualitative behavior for how heterogeneity changes across  the $(T,\Delta)$ parameter space, which is our main interest for Fig.~\ref{fig7}c).

%................................................
\subsection{Combinatorial analysis for bifurcation theory}\label{sec:Unfoldingof}
%````````````````````````````````````````````````

We now present the derivation of our bifurcation theory given in Eqs.~\eqref{eq:WFP_eqs} and \eqref{eq:ANC_eqs} for 2D STM cascades over noisy ring complexes. Recall for this model that $N$ nodes are positioned along a the unit circle and are spaced apart by an angle $\delta = 2\pi /N$. Therefore, neighboring vertices are positioned apart by angles $1\delta$, $2\delta$, and so on. Also, recall that each vertex has exactly $d^{(G)}$ geometric edges to nearest-neighbor vertices and $d^{(NG)}$ non-geometric edges to other vertices, which are added uniformly at random. This generative model for noisy geometric complexes helps us to develop theory for ANC and WFP, but as we shall show, it also has important implications for such phenomena.

We first describe ANC in the limit of large $N$ when the cascade size $q(t)$ is small. By definition, an ANC event occurs when a cascade propagates to a vertex  $v_i$ that is far from a cascade cluster, implying that all of its geometric $k$-simplices are inactive. It follows that the fractions of active adjacent $k$-simplices can only take on the following values
\begin{equation}\label{eq:fki}
    f^k_i \in \left\{0,\frac{1}{d_i^k},\frac{2}{d_i^k},\dots,\frac{d_i^{k,NG}}{d_i^k}\right\},
\end{equation}
depending on the number of active non-geometric $k$-simplices. For STM cascades over noisy ring complexes that are generated via the model that we describe in Methods section `Generative model for noisy ring complexes', we find that ANC events occur predominantly due to influences by non-geometric 1-simplices. In contrast, we find non-geometric 2-simplices to have a negligible effect on ANC in the limit of large  $N$, small $q(t)$, and  fixed $d^{(G)}$ and $d^{(NG)}$, which implies $f^2_i \approx 0 $ under these assumptions. 
%\todo{THIS SHOULD EMPHASIS AND USE SMALL q.}
Specifically, non-geometric 2-simplices (and higher-dimensional simplices) are    rare, because non-geometric edges are added uniformly at random. Consider a  vertex $v_i$ that is distant from a cascade cluster, and suppose that it has one non-geometric edge to an active vertex $v_j$. That edge is the face of a 2-simplex only if  $v_i$ has a second non-geometric edge to a third vertex that is already adjacent to $v_j$. This occurs with   probability   $1 - [(N-1 - d^1_i)/N]^{(d^{(NG)}-1)} \sim \mathcal{O}( {N}^{-1}) $, which approaches zero with increasing $N$. This result uses  that there are $N-1 - d^1_i$ possible vertices that $v_i$ can  connect to  without creating a non-geometric 2-simplex $v_j$. Since non-geometric edges are created uniformly at random, each of the remaining non-geometric edges for $v_i$ don't  create a 2-simplex with probability $[(N-1 - d^1_i)/N]$. Moreover, $[(N-1 - d^1_i)/N]^{(d^{(NG)}-1)}$ gives the probability that none of them do. Subtracting this probability by 1 gives the probability that there is at least one non-geometric 2-simplex between $v_i$ and $v_j$ (that is, given that they are already connected by a non-geometric edge). Therefore, while non-geometric 2-simplices (and higher-dimensional simplices) do arise in our generative model for noisy ring complexes, they are rare and have little effect on ANC for large systems.

To obtain the critical thresholds given in Eq.~\eqref{eq:ANC_eqs}, we approximate   $R_i(t)\approx (1-\Delta) f^1_i$ and observe that
\begin{equation}\label{eq:f1i}
    f^1_i \in \left\{0,\frac{1}{d^{(G)}+d^{(NG)}},\dots,\frac{d^{(NG)}}{d^{(G)}+d^{(NG)}}\right\} ,
\end{equation}
which uses that the 1-simplices are degree regular. If one considers a variable threshold $T$, then the probability that ANC events occur  will significantly change as $T$ surpasses the different $R_i(t)$ values corresponding to different $f^1_i$. For example, there are no ANC events when $T> (1-\Delta)\frac{d^{(NG)}}{d^{(G)}+d^{(NG)}}$.

Notably, our bifurcation theory  for ANC naturally  extends to $\kappa$-dimensional STM cascades in which $R_i(t) =\sum_{k=1}^\kappa \alpha_k f_i^k(t)$. In this case,   non-geometric $k$-simplices with $k>1$ also have little effect on ANC, and   the critical thresholds   are identical to those   in Eq.~\eqref{eq:ANC_eqs} with the variable substitution $(1-\Delta)\mapsto \alpha_1$.

\begin{figure*}[t]
	\centering
	\includegraphics[width= 1\linewidth]{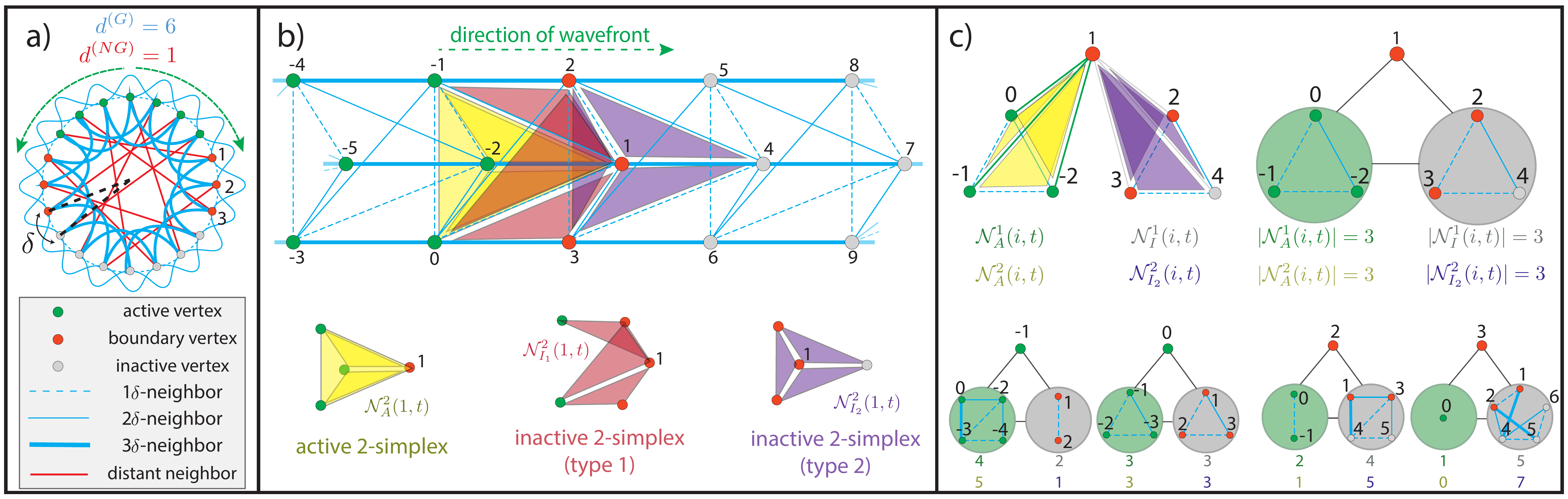}
	\caption{
		{\bf Bifurcation theory   obtained by examining the connections between boundary vertices and  active/inactive $1$- and $2$-simplices.}
		{\bf a)}~
		Visualization of a noisy ring complex with $d^{(G)} = 6$ and $d^{(NG)}=1$ with $N$ vertices that are spaced apart by an angle $\delta=2\pi/N$. Linestyles highlight that edges  connect neighbors with different proximity, and we label   vertices so that vertex   $v_1$ is  positioned  $1\delta$ to the   right of the wavefront, $v_2$ is at position $2\delta$, and so on. The speed of WFP is determined by the number of   boundary vertices  $\{v_1,v_2,v_3\}$ that become active upon the next time step.
		{\bf b)}~
		Visualization of active and inactive $k$-simplices for the  boundary vertex $v_1$ that is closest to the wavefront. For each $v_i$, we define  a set $\mathcal{N}_{A}^{k}(i,t)$ of adjacent $k$-simplices that are active and  sets $\mathcal{N}_{I_1}^{k}(i,t)$ and $\mathcal{N}_{I_2}^{k}(i,t)$ of inactive 2-simplices that are type 1 and 2, respectively. (Recall Fig.~\ref{fig2}.)
		{\bf c)}~
		By identifying a set of neighboring active vertices (green shaded regions) for a boundary vertex $v_i$, one can compute the number of adjacent 1- and 2-simplices that are active by    counting the number of those vertices and the number of edges among them, respectively. One can identify the number of inactive 1-simplices and type-2 inactive 2-simplices in a similar way (gray shaded regions). This approach is depicted for vertex $v_1$ (top) as well as other nearby vertices (bottom), and the associated  numbers are indicated for each.
	}
	\label{fig6}
\end{figure*}

We next develop bifurcation theory for WFP dynamics, and in this case,  higher-dimensional simplices   have a significant effect. Our analysis stems from considering boundary vertices that are  not yet  active but have  geometric simplicial neighbors---i.e., adjacent geometric 1-simplices, geometric 2-simplices, etc.---that are active. The propagation speed of a wavefront  along a geometrical channel is determined by the number of boundary vertices that become active upon each time step. For example, in Fig.~\ref{fig6}a) we visualize a noisy ring complex with $d^{(G)}=6 $ so that there are $d^{(G)}/2=3 $ boundary vertices   $\{v_1,v_2,v_3\}$ for the clockwise-progressing wavefront. (Recall that each vertex connects to $d^{(G)}/2$  nearest-neighbor vertices in either direction along the ring manifold.) Therefore, the speed of a wavefront is either 1, 2, or 3, depending on how many of them become active at each  time step. Note that the cascade exposure $R_i(t)$ defined in Eq.~\eqref{eq1} will be different for each boundary vertex, and because they are enumerated  closest-to-farthest from the wavefront, one has $f_1^k(t)\ge f_2^k(t) \ge f_3^k(t)$ 
and $R_1(t)\ge R_2(t) \ge R_3(t)$. Therefore, as   a threshold $T_i$ increases, the criterion $R_i(t)>T_i$ defined in Eq.~\eqref{eq:xx} will first fail for $v_3$, then $v_2$, and finally $v_1$. The wavefront shown in Fig.~\ref{fig6} won't propagate for any threshold that is larger than $R_1(t)$.

The $R_i(t)$ values of boundary vertices  reveal critical threshold values for WFP, and we identify them for noisy ring complexes by considering how each  $v_i$ is adjacent to geometric $k$-simplices that are either active or inactive. We may assume that the non-geometric $k$-simplices are inactive in the limit of large $N$ and small cascades size  $q(t)$ [technically, a non-geometric $k$-simplex is active with probability that is at most  $\mathcal{O}(q(t)/N)$], and so  we initially focus on   $d^{(NG)}=0$. We will later allow for nonzero $d^{(NG)}$ when we   compute the fractions $f_i^k(t)$. To this end, we define for each $v_i$ the sets $\mathcal{N}^k(i)$ of adjacent $k$-simplices, which we   partition   into sets $\mathcal{N}_A^k(i,t)$ and $\mathcal{N}_I^k(i,t)$ of  adjacent $k$-simplices that are active and inactive, respectively, at time $t$. Note that   $\mathcal{N}^k(i) = \mathcal{N}_A^k(i,t)\cup \mathcal{N}_I^k(i,t)$  and $d^k_i = |\mathcal{N}^k(i)|$ is the degree of $v_i$ with respect to $k$-simplices.   With these  definitions, the fractions of active $k$-simplices are given by
\begin{equation}
    f_i^k(t) = \frac{|\mathcal{N}_A^k(i,t)|}{d_i^k}.
\end{equation}
 
In Fig.~\ref{fig6}b), we  visualize a wavefront propagating along a geometrical channel for the noisy ring complex shown in Fig.~\ref{fig6}a). Vertices are positioned so that we may more easily identify whether $2$-simplices are active or inactive. Focusing on  the boundary vertex $v_1$ that is closest to the wavefront and has the largest exposure $R_i(t)$, we illustrate its set of adjacent 2-simplices that are active. Because  $v_1$ is adjacent to $|\mathcal{N}_A^1(1,t)|=3$ active 1-simplices and $|\mathcal{N}_A^2(1,t)|=3$ active 2-simplices, it follows that 
$f_1^1(t) = \frac{3}{d_i^1} $ and $f_1^2(t) = \frac{3}{d_i^2} $. (Note that the denominators include both geometric and non-geometric $k$-simplices.) We also visualize in Fig.~\ref{fig6}b) the inactive 2-simplices that are adjacent to boundary vertex $v_1$. Recall from  Fig.~\ref{fig2} that there are  two types of inactive 2-simplices, depending on whether a 2-simplex contains only one active vertex (type 1) or no active vertices (type 2). We   let $\mathcal{N}^2_{I_{1}}(i,t)$ and $\mathcal{N}^2_{I_{2}}(i,t)$ denote the sets of   inactive 2-simplices of types 1 and 2, respectively, and depict them for $v_1$. Observe that  $|\mathcal{N}_{I}^1(1,t)|= |\mathcal{N}_{I_1}^2(1,t)|=|\mathcal{N}_{I_2}^2(1,t)|=3$.

In Fig.~\ref{fig6}c), we highlight that one can easily compute the number of active  1- and 2-simplices that are adjacent to  a boundary vertex $v_i$ using three steps. First, we identify   the set $\{v_j|(i,j)\in \mathcal{N}_A^1(i,t)\}$ of  active vertices that   are connected to $v_i$ by   active 1-simplices (see green shaded regions). Second, we count the number of  vertices  in that set, which yields $|\mathcal{N}_A^1(i,t)|$ since there is a one-to-one correspondence between these vertices and the active 1-simplices that are adjacent to $v_i$. Third, we count the number of edges among those vertices, which yields $|\mathcal{N}_A^2(i,t)|$ since there is a one-to-one correspondence between  those edges and active 2-simplices. We can also  calculate the number of  type-2 inactive 2-simplices in a similar way. That is, we first identify  the set $\{v_j|(i,j)\in \mathcal{N}_I^1(i,t)\}$ of inactive vertices that  are connected to $v_i$ by inactive 1-simplices (see gray shaded regions in Fig.~\ref{fig6}c)). We then count how many vertices are in the set (which yields $|\mathcal{N}_I^1(i,t)|$) and the number of edges among those vertices (which yields $|\mathcal{N}_{I_2}^2(i,t)|$). The upper part of Fig.~\ref{fig6}c) illustrates this approach for $v_1$, and we do not visualize type-1 inactive 2-simplices, because they are more difficult to compute directly but can be found after the other sets are determined: $|\mathcal{N}_{I_{1}}^{2}(i,t)| = d_{i}^{2} - |\mathcal{N}_{A}^{2}(i,t)| - |\mathcal{N}_{I_{2}}^{2}(i,t)|$. The lower part of Fig.~\ref{fig6}c) illustrates this approach for the other two boundary vertices $\{v_2,v_3\}$ as well as two   vertices $\{v_{-1},v_0\}$ that are already active, since they are to the left of the wavefront.

Importantly, because each vertex has exactly $d^{(G)}/2$ 1-simplices going in either side along the ring manifold left,  there is always a clique of edges among vertices in  a set $\{v_j|(i,j)\in \mathcal{N}_A^1(i,t)\}$ for the boundary vertices. (This is not true for active vertices, such as $v_{-1}$, as shown in the lower part of Fig.~\ref{fig6}c).) Therefore, if a boundary vertex has $s_j$ active 1-simplices, then it must also have $\binom{s_j}{2}$ active 2-simplices. It follows that the different possible $f_i^1$ values for a boundary vertex $v_i$ are given by
\begin{equation}\label{eq:f1ii}
    f^1_i \in \left\{0,\frac{1}{d^1_i},\frac{2}{d^1_i},\dots,\frac{d^{(G)/2}}{d^1_i}\right\},
\end{equation}
and the corresponding $f_i^2$ values are
\begin{equation}\label{eq:f2i}
    f^2_i \in \left\{0,\frac{1}{d^2_i}\binom{1}{2},\frac{1}{d^2_i}\binom{2}{2},\dots,\frac{1}{d^2_i}\binom{d^{(G)}/2}{2}\right\} .
\end{equation}
We enumerate these possibilities by $j$ and use the definition $R_i(t) = (1-\Delta)f^1_i + \Delta f^2_i$ to obtain the critical threshold values for WFP given by Eq.~\eqref{eq:WFP_eqs}. 
For $\kappa$-dimensional STM cascades, setting $T=R_i(t)$ yields a more  general set of bifurcation lines:
\begin{equation}\label{eq:WFP_eqs2}
    T^{WFP}_{j}  = \sum_{k=1}^\kappa \alpha_k \frac{1}{d_{i}^{2}} \binom{s_j^{(G)}}{k}.
\end{equation}
In either case,  $(j+1)$  boundary vertices will become active upon each time step when  $T\in [T_{j+1}^{WFP},T_{j}^{WFP})$. Since wavefronts progress both clockwise and counter-clockwise around the ring manifold, the cascade size $q(t)$  will grow linearly at a rate $2j+2$.

%................................................
\subsection{Critical regimes for \emph{C. elegans}}\label{sec:criticalregimes}
%````````````````````````````````````````````````

Our bifurcation theory describes WFP and ANC on a 1D geometrical substrate is degree regular, so that STM cascade propagations occur identically for all boundary vertices. However, the empirical neuronal complex for \emph{C. elegans} is degree heterogeneous, and so we instead examine the median bifurcation curves that are associated with  median degrees, including  geometric degrees, non-geometric degrees, and 2-simplex. In principle, we could plot a different bifurcation curve for each vertex $v_i$ based on its unique degrees. (See Supplementary Figure 9 and related discussion on `Perturbed Bifurcation Results' in Supplementary Note 3 of  \cite{taylor2015topological}.) For simplicity, here we instead plot a single representative bifurcation curve for
\emph{C. elegans}  using the   median values  $d^{(G)} = 8$, $d^{(NG)} = 2$, and $d^{2} = 34$ to construct the bifurcation curves. Finally, we reiterate that the \emph{C. elegans} neuronal complex has a structure that is outside our assumed structure of a noisy ring complex, and so our  bifurcation theory should not be expected to be perfectly predictive. Our experiments highlight that these bifurcation curves are qualitatively predictive for the general effects of $T$ and $\Delta$.
%
%Notably, we also simulate STM cascades  on an undirected \emph{C. elegans} synapse network since our theory doesn't involve directed k-simplices. 

%................................................
\subsection{Simplicial cascade maps}\label{sec:CascadeMaps}
%````````````````````````````````````````````````

We introduce a notion of latent geometry for simplicial complexes called simplicial cascade maps in which the set $\mathcal{C}_0 = \{1,\dots, N\}$ of vertices is nonlinearly mapped as a set of  points (i.e., a `point cloud') in an $J$-dimensional Euclidean metric space $\mathbb{R}^J$. Simplicial cascade maps directly generalize contagion maps \cite{taylor2015topological}, which are   recovered under the choice of 1D STM cascades (and which do not utilize $k$-simplices for $k>1$).

We construct simplicial cascade maps using the activation times for STM cascades.
Given $J$ realizations of a STM cascade on a simplicial complex with different initial conditions with cluster seeding, the associated STM map is a map   $\{v_i\}   \mapsto \{\bm{ \tau}^{(i)}\} \in \mathbb{R}^J$ in which each vertex $v_i\in\mathcal{C}_0$ maps to a point $\bm{ \tau}^{(i)}= [\tau^{(i)}_1,\dots, \tau^{(i)}_J]^T$, where $\tau^{(i)}_j$ is the  activation time for vertex $v_i$ for the STM cascade with the $j$-th initial condition. See Supplementary Note  ``Visualizations of STM Cascade Maps'' for visualizations of these point clouds and further discussion.

In practice, we often let $J=N$ so that the $j$-th  initial condition corresponds to seed clustering at vertex $v_j$. However, extra attention is required for handling cascades that don't saturate the network, in which case   there would be $\tau^{(i)}_j$ values that are undefined. Herein, we choose to neglect such cascades. See  \cite{taylor2015topological} for alternative strategies in the context of cascades over graphs.

%................................................
\subsection{Data and code availability}\label{sec:code}
%````````````````````````````````````````````````

The authors declare that all data supporting the findings of this study are available within the paper. A codebase that implements STM cascades over noisy geometric complexes and reproduces our computational experiments can be found in a Python library \cite{neuronal_cascades}. Documentation on how to use this software is available at \cite{neuronal_cascades_doc}. The \emph{C. elegans} synapse network with physical vertex positions is publicly available and was downloaded from \cite{Kaiser-2011,data}.
\section*{Author Contributions}
Both authors developed the research plan and wrote the paper. BUK conducted the numerical experiments.

\section*{Competing Interests}
The authors declare that there are no competing interests.

\begin{acknowledgements}
BUK and DT were supported in part by the National Science Foundation (DMS-2052720) and the Simons Foundation (grant \#578333). Authors thank Sarah F. Muldoon for valuable discussions.
\end{acknowledgements}

%%%%%%%%%%%%%%%%%%%%%%%%%%%%%%%%%%%
%%%%%%%%%%%%%%%%%%%%%%%%%%%%%%%%%%%
\bibliographystyle{plain}
\bibliography{main.bib}

\end{document}

\title{Supplementary Material:\\
Simplicial cascades  are orchestrated by the multidimensional geometry of neuronal complexes}
\author{Bengier Ülgen Kılıç}\email{bengieru@buffalo.edu}
\author{Dane Taylor}\email{danet@buffalo.edu}
\affiliation{Department of Mathematics, University at Buffalo, State University of New York, Buffalo, NY 14260, USA}

\date{\today} 

\clearpage

%%%%%%%%%%%%%%%%%%%%%%%%%%%%%%%%%%%%%%
%%%%%%%%%%%%%%%%%%%%%%%%%%%%%%%%%%%%%%

\maketitle
\tableofcontents

~
\clearpage

%\linenumbers

%%%%%%%%%%%%%%%%%%%%%%%%%%%%%%%%%%%%%%
\section{Stochastic Simplicial Threshold Model} \label{stochastic}
%%%%%%%%%%%%%%%%%%%%%%%%%%%%%%%%%%%%%%

The simplicial threshold model (STM) for cascades that we studied in the main text is a nonlinear deterministic  process. We now develop a stochastic generalization of STM cascades and   show that its WFP and ANC behaviors  are qualitatively similar to those  for deterministic STM cascades, provided that the propagation mechanism is predominantly determined  by a threshold criterion and not randomness. In a practical setting, stochastic activations are important to consider because they can represent the effects of other influences that are not directly represented in our model, such as chemical signaling, {neuronal failures} and external stimuli \cite{shadlen1994noise,faisal2008noise,white2000channel,mcdonnell2011benefits}. 

Noting that the state $x_i^0(t) \in\{0,1\}$ of each vertex is a binary variable, we introduce stochasticity into STM cascades by allowing them to change as Bernoulli random variables, rather than as a deterministic function as defined by Eq.~(1) in the main text. We let $\text{Ber}(p)$ be the Bernoulli distribution with probability $p$ and denote  $e\sim \text{Ber}(p)$ as a Bernoulli random variable. We would like vertices with large exposure exposure $R_i(t)$ [see Eq.~(2) in the main text] to a cascade to become active with higher probability, and so we define the activation probabilities
\begin{equation}
    p_i(t) \equiv S(C(R_i(t)-T_i)),
\end{equation}
where $T_i$ is the threshold for vertex $i$ and 
$S:\mathbb{R}\to[0,1]$ is  a sigmoid activation function, $ S(y) \equiv \frac{1}{1+ e^{-y}}$. We refer to scalar $C>0$  as a stochasticity parameter, since we will use it  to control whether vertex activations occur mainly due to randomness or due to the threshold criterion $R_i(t)>T_i$ (or in this case, $R_i(t)-T_i>0$). Because of the sigmoid function, a probability $p_i(t)$ will be closer to $0$ when $R_i(t)-T_i<0$ and   closer to $1$ when $R_i(t)-T_i>0$.  For the deterministic STM cascades,   $p_i(t)$ would necessarily be exactly 0 and 1 under these two conditions. Thus motivated, we similarly have that $p_i(t)$ will converge to 0 and 1 under these respective conditions in the limit $C\to \infty$. Thus,  the stochastic dynamics will approach a similar deterministic dynamics as $C$ is increased. In contrast, as  $C$ is decreased, cascade propagations are increasingly determined by randomness rather than the nonlinear thresholding criterion. 

We define stochastic STM cascades identical to that in the main text, except that the vertices' states evolve instead   following the stochastic nonlinear equation

\begin{equation}\label{eq:stooc}
    x_{i}^0({t+1}) = \left\{ \begin{array}{rcl}
         1, &  &\text{ if either } x_{i}^0({t})=1 \text{ or } e_i(t)=1, \text{ where } e_i(t) \sim \text{Ber}(p_i(t)),\\
         0, &  &\text{ if } x_{i}^0({t})=0 \text{ and } e_i(t)=0.
    \end{array}\right.
\end{equation}
 
As before, once they become active,  vertices and $k$-simplices remain active for all later times. However, vertices' transitions  from inactive to active are now stochastic events that depend on the difference $R_i(t)-T_i$ and parameter $C$. 
Importantly,  stochastic and deterministic STM cascades significantly differ  when the stochasticity parameter $C$ is small or moderately valued, but they are  nearly identical  in the limit $C\to\infty$. The only slight difference occurs when $R_i(t)=T_i$, in which case a vertex remains inactive for deterministic STM cascades but becomes active with probability $0.5$ according to Eq.~\eqref{eq:stooc}. (This occurs due to the symmetry and continuity of the sigmoid function.)
 
In Supplementary Figures~\ref{fig9} and~\ref{fig10}, we numerically study stochastic versions of 2D STM cascades over the  \emph{C. elegans} neuronal complex from  the text. These two figures extend and generalize our findings that were shown in Figs.~6 and 4, respectively, in the main text. As before, we study different choices for the threshold $T\in[0,1]$ (we again focus on  $T_i=T~\forall~i$) and the 2-simplex influence parameter $\Delta\in[0,1]$. However, we now study the effect of stochasticity by showing results for three different choices of the $C$. In both figures, the left, center and right columns reflect when $C=1280$, $30$, and $5$, respectively. 

In Supplementary Figure~\ref{fig9}a), we illustrate the shapes of sigmoidal activation functions $S(y) \equiv \frac{1}{1+ e^{-y}}$, where $y = C(R_{i}(t)-T_{i})$, for the different choices of $C$. These values were selected to explore when the thresholding criterion has a greater effect than randomness. Specifically, the sigmoid function approaches a step function for large $C$ (left column), which coincides with a deterministic model. For small $C$ (right column), stochastic responses are less nonlinear and more stochastic. For intermediate values of $C$ (center column), STM cascades behave differently from the limiting cases of very large and very small   $C$.

In Supplementary Figures~\ref{fig9}b), \ref{fig9}c) and \ref{fig9}d), we extend our findings from Fig.~6 in the main text by studying the effects of stochasticity on WFP and ANC properties for STM cascades over the \emph{C. elegans} neuronal complex. Recall that we construct the \emph{C. elegans} neuronal complex by creating the clique complex for an empirically measured synapse network \cite{Choe-2004-connectivity,Kaiser-2006-placement}. In Supplementary Figures~\ref{fig9}b) and \ref{fig9}c), we use color to depict the average rate of change $dq/dt$  for cascade size $q(t)$ and average number  $C(t)$ of clusters, respectively. We choose $t=5$, which is an early time in which these values provide empirical quantitative measures for WFP and ANC, respectively. (At larger   $t$, it is difficult to distinguish WFP and ANC propagations since the cascades are so large.) Recall that $dq/dt$ and  $C(t)$ are empirical measures for WFP speed and ANC rate, respectively. We compare these empirical measurements with the same bifurcation lines $T_j^{WFP}$ and $T_j^{ANC}$ from Fig.~6.

In Supplementary Figure~\ref{fig9}d), we depict the correlation coefficient $\rho$ that quantifies the extent to which the geometry of a simplicial cascade map (see Methods Section `Simplicial cascade maps') appropriately recovers the geometry of the \emph{C. elegans} neuronal complex, which is an elongated 1D manifold. Recall that larger values of $\rho$ indicate better manifold recovery, which corresponds to when cascades exhibit a prevalence of WFP rather than ANC. (Smaller $\rho$ values indicate the opposite.)

\begin{figure*}[t]
	\centering
	\includegraphics[width= 1\linewidth]{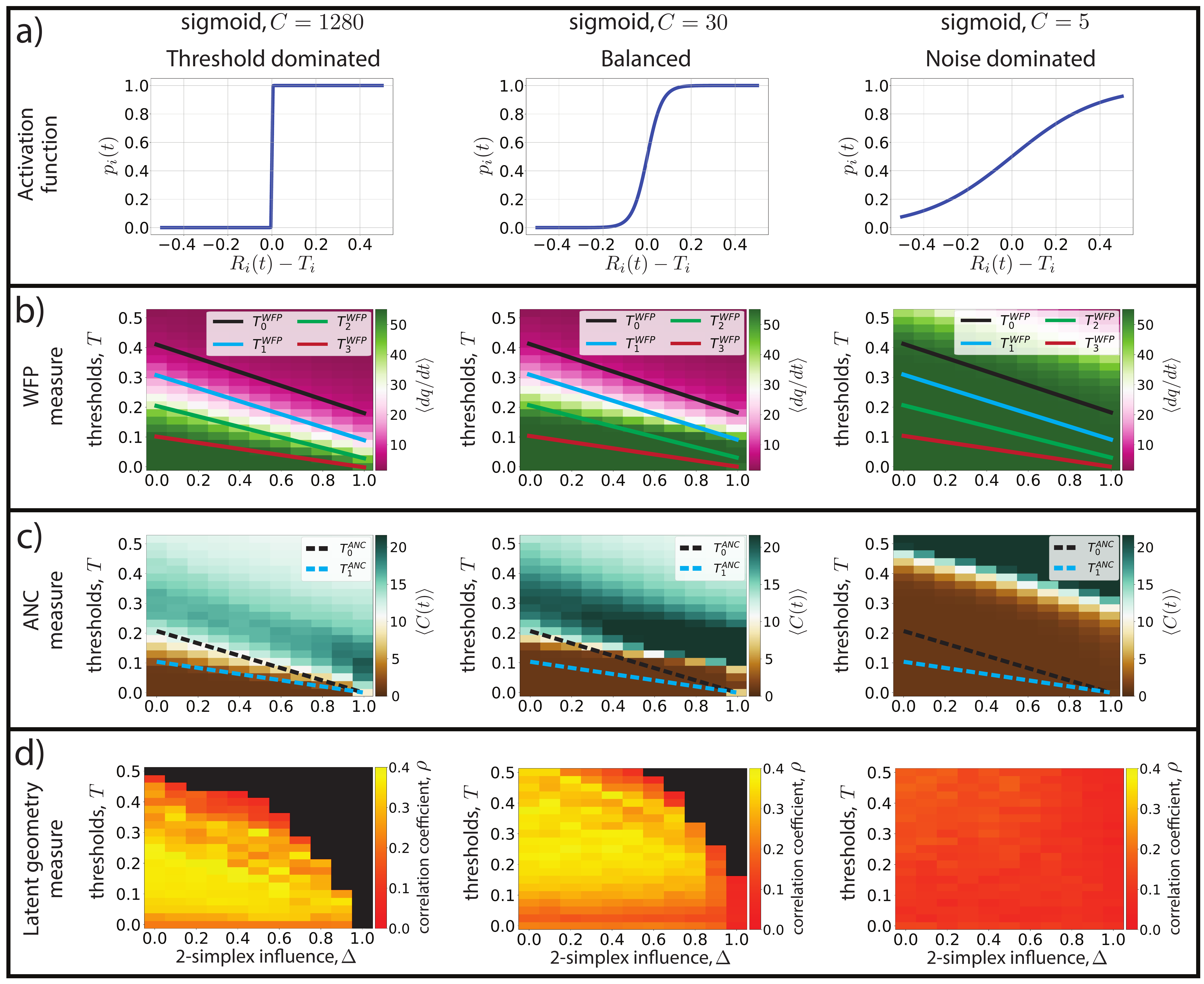}
	\caption{
		{\bf Effects of  stochasticity on WFP and ANC for 2D STM cascades on a \emph{C. elegans} neuronal complex}.
		[See Fig.~6 in the main text for similar experiments for deterministic STM cascades.]
		{\bf a)}~
		Each vertex $v_i$ becomes active with a probability $p_i(t)$ that is determined by a sigmoidal activation function, $p_i(t)  \equiv1/{(1+ e^{-C(R_i(t) - T_i)})}$, where $C$ is a stochasticity parameter, $T_i$ is $v_i$'s threshold, and $R_i(t)$ is $v_i$'s simplicial exposure to the cascade at time $t$. For large $C$, activations are nearly deterministic with $p_i(t) \approx 0$  when $R_i(t)-T<0$ and $p_i(t) \approx 1$  when $R_i(t)-T>0$. As $C$ decreases, vertices' responses becomes less nonlinear and more stochastic.
		{\bf b)}~
		We display  WFP speeds $dq/dt$ for different choices of $T$ and $\Delta$, which we average across all initial conditions with cluster seeding and compute at $t = 5$. We also  plot the critical threshold lines given by Eq.~(3) in the main text. 
		{\bf c)}~
		We display cascade sizes $C(t)$ at $t=5$, which we average across all initial conditions with cluster seeding. We also plot the critical threshold lines given by Eq.~(4) in the main text.
		{\bf d)}~
		We display Pearson correlation coefficients $\rho$ for point clouds resulting from STM cascade maps. $\rho$ quantifies the extent to which  stochastic STM cascades predominantly propagation by WFP along the 1D manifold of the nematode  \emph{C. elegans}. Black regions indicate parameter values in which no initial conditions yield cascades that saturate the network, and so we do not compute $\rho$   for those values of $(T,\Delta)$. 
	}
	\label{fig9}
\end{figure*}

\begin{figure*}[t]
	\centering
	\includegraphics[width= 1\linewidth]{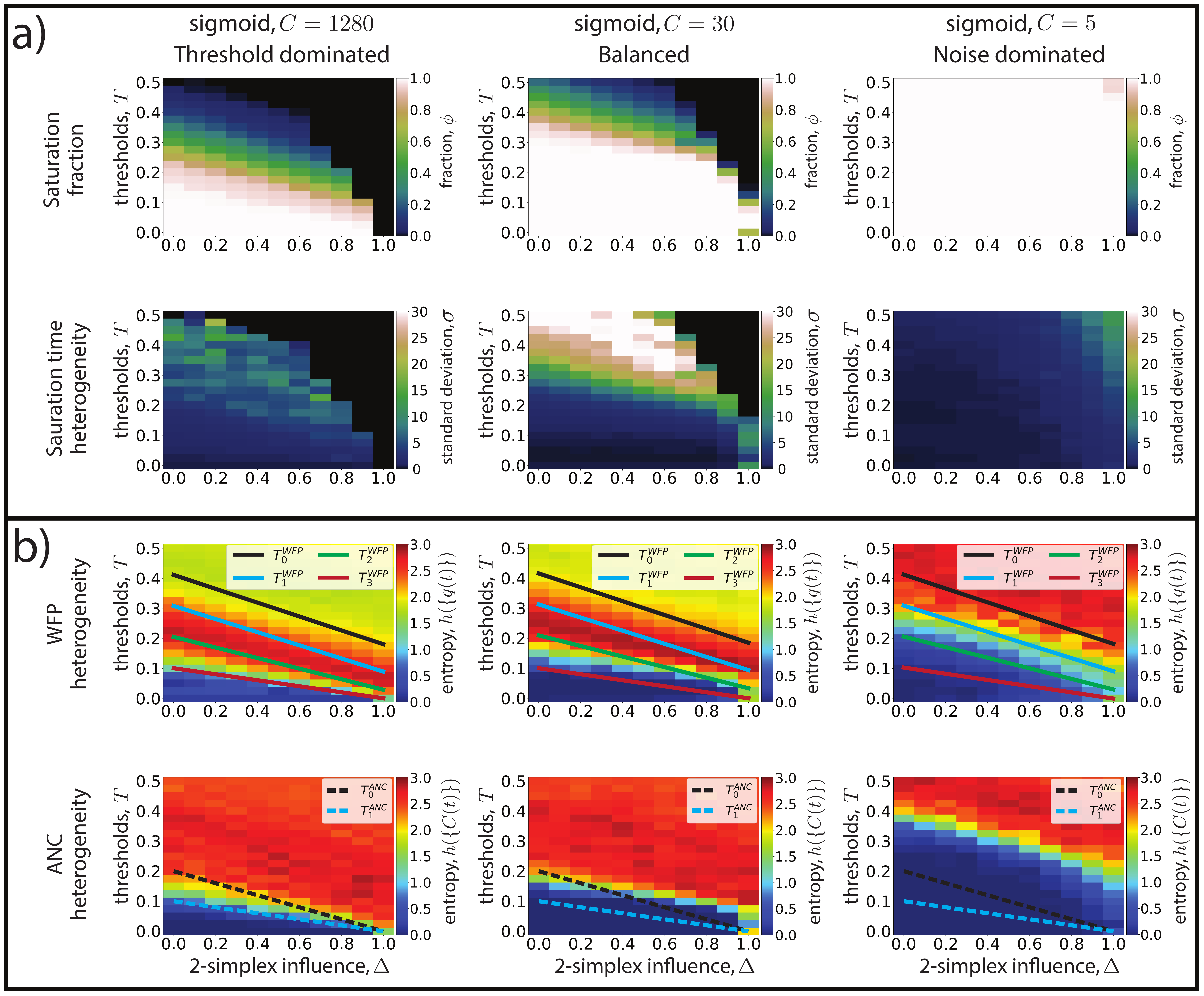}
	\caption{
		{\bf Effects of   stochasticity on cascade heterogeneity for 2D STM cascades on a \emph{C. elegans} neuronal complex}.
		[See Fig.~4 in the main text for similar experiments for deterministic STM cascades.]
		We study a stochastic variant of 2D STM cascades over the \emph{C. elegans} neuronal complex with three choices for a stochasticity parameter $C$. 
		{\bf a)}~
		We display the saturation fractions $\phi$ (top row) and the standard deviation $\sigma$ for the saturation times  (bottom row) for the parameter space $(T,\Delta)$. (Recall that a saturation occurs when all vertices become active.) Black regions indicate where no initial conditions lead to a saturation across the network.
		{\bf b)}~
		We measure the heterogeneity in the cascade sizes (top row) and number of clusters (bottom row) at $t = 5$ for different initial conditions. See Methods Section `Entropy Calculation' in the main text for the definitions of these entropic measures for heterogeneity. 
	}
	\label{fig10}
\end{figure*}

First, observe in the first column of Supplementary Figures~\ref{fig9}b), \ref{fig9}c) and \ref{fig9}d) that the results for stochastic STM cascades with large $C$ are nearly identical to those for deterministic STM cascades (i.e., compare to Fig.~6(D)--(F) in the main text for the \emph{C. elegans} neuronal complex). Second, observe in the second column that for moderate $C$, the WFP, ANC and latent geometry exhibit the same qualitative behaviors across the $(T,\Delta)$ parameter space as those shown in the first column. The main difference is that as $C$ decreases, the transitions between WFP/ANC regimes move  upward, i.e., they occur at larger values for the threshold $T$. Third, observe in the third column that when stochasticity is very high (i.e., $C=5$), then these qualitative regimes are significantly different and the bifurcation theory is no longer relevant. As a result, our bifurcation theory fails to predict dynamical transitions for WFP and ANC when cascade propagations occur predominantly due to stochastic noise rather than  a nonlinear thresholding mechanism. The right-most subpanel in Supplementary Figure~\ref{fig9}d) highlights that the ring manifold structure is not recovered by simplicial cascade maps when the noise is too high, regardless of $T$ or $\Delta$.

In Supplementary Figure~\ref{fig10}, we extend our findings from Fig.~4 in the main text by studying the effects of stochasticity on heterogeneity properties for STM cascades over the \emph{C. elegans} neuronal complex.  In Supplementary Figure~\ref{fig10}a), we study the heterogeneity of STM cascades across different initial conditions (i.e., cluster seeding starting at different vertices) for different choices of $T$, $\Delta$ and $C$. (Recall Methods Section `Cluster Seeding' in the main text.) We  plot the fraction $\phi$ of cascades that saturate the network (top row)  and the standard deviation $\sigma$ for the times at which saturations occur (bottom row). Recall that saturation occurs at the time in which all vertices become active. The black-colored regions highlight that no STM cascades saturate the network if $T$ and/or $\Delta$ are too large. Comparing the center and left columns for the entire $(T,\Delta)$ parameter space, observe that as $C$ decreases, the regimes of similar saturation fractions shift upward. That is, with increasing stochasticity, STM cascades are more likely to saturate a network (i.e., to prevent saturation, the threshold $T$ must be larger).
Observe in the right column for small $C$ that the cascades almost always saturate the network, and   there is little heterogeneity with respect to times at which saturations occur for stochastic-dominated STM cascades. Interestingly, observe in the lower center panel for $C=30$ that the  standard deviation $\sigma$ exhibits a sharp increase at approximately $T\approx 0.45$ and $\Delta\le 0.6$. This peak in saturation heterogeneity was not observed for the deterministic case [see Fig.~4(B) in the main text] or when $C=1280$. This suggests that a small amount of stochasticity can potentially aid in promoting pattern diversity (and subsequently, memory capacity \cite{shew2011information,shew2013functional} in the context of neuronal activity cascades).

In Supplementary Figure~\ref{fig10}b), we  study heterogeneity for the cascade size $q(t)$ and number  $C(t)$ of cascade clusters when $t=5$ for different $C$. We quantify heterogeneity by computing the Shannon entropy of $h(\{q(t)\})$ (top row) and $h(\{C(t)\})$ (bottom row) across the  different initial conditions. See Methods Section `Entropy Calculation' for details. Observe that the general effect is that as $C$ decreases (i.e., when there is more stochasticity), the heterogeneity measures'  transitions tend to occur for larger $T$ values. Moreover, recall for deterministic STM cascades that  $h(\{q(t)\})$ is the largest when $T$ and $\Delta$ are neither too small or too large. This remains true when $C$ is large (see the red-colored region in the left column, upper row of Supplementary Figure~\ref{fig10}b)), and that this intermediate regime shifts toward larger $T$ as $C$ decreases (see the red-colored region in the center column, upper row). In contrast, observe in the right column of Supplementary Figure~\ref{fig10}b) that this intermediate peak does not occur under large stochasticity (i.e., small $C$), and in that case, $h(\{q(t)\})$ monotonically increases with   $T$ and $\Delta$. Thus, for noise-dominated STM cascades, increasing $T$ and/or $\Delta$ increases the heterogeneity of spatiotemporal patterns (i.e., memory capacity \cite{shew2011information,shew2013functional}).

%%%%%%%%%%%%%%%%%%%%%%%%%%%%%%%%%%%%%%%%%%%%%%%%%%%%%%%%%%%%
%%%%%%%%%%%%%%%%%%%%%%%%%%%%%%%%%%%%%%%%%%%%%%%%%%%%%%%%%%%%
%%%%%%%%%%%%%%%%%%%%%%%%%%%%%%%%%%%%%%%%%%%%%%%%%%%%%%%%%%%%
%%%%%%%%%%%%%%%%%%%%%%%%%%%%%%%%%%%%%%%%%%%%%%%%%%%%%%%%%%%%

\clearpage

\section{Visualizations of STM Cascade Maps} \label{geometry}

In Methods Section `Simplicial Cascade Maps' we introduce   STM cascade maps that define a latent-space geometry for a simplicial complex. Such maps  can be used to embed the simplicial complex (i.e., thereby generalizing the pursuit of graph embedding \cite{cai2018comprehensive}) or for nonlinear dimension reduction \cite{tenenbaum2000global,coifman2006diffusion} (i.e., also generalizing this pursuit to the setting of simplicial complexes). STM cascade maps generalize contagion maps \cite{taylor2015topological,mahler2021analysis} (which are restricted to thresholded cascades over graphs), and we use them to quantify the extent to which STM cascades propagate predominantly due to WFP or  ANC. We presented these in Results section `Latent geometry of simplicial cascades quantifies WFP vs ANC' in the main text.

Here, we expand this study and provide visualizations for STM cascade maps.  We construct STM cascade maps $\{v_i\}\mapsto \{\bm{\tau }^{(i)}\}$ for different $\Delta$ and $T$ values, and we visualize these point clouds in Supplementary Figures~\ref{PCA2d} and \ref{PCA3d}. Importantly, if one uses $J$ different cascades to construct an STM cascade map (each using a different initial condition with cluster seeding), then it yields a $J$-dimensional point cloud $  \{\bm{\tau }^{(i)}\}\in\mathbb{R}^J$, which is inherently difficult to visualize for large $J$. Therefore, here we   visualize each point cloud using a low-dimensional projection that we construct using  principal component analysis (PCA)--- particularly, classical multidimensional scaling \cite{cox2008multidimensional}.

In Supplementary Figures~\ref{PCA2d} and \ref{PCA3d}, we visualize two-dimensional and three-dimensional projections, respectively, of point clouds resulting from STM cascade maps for noisy ring complexes with $N=1000$ vertices. We choose the same values of $T$ and $\Delta$ as were previously studied in Fig.~5 in the main text (i.e., where we presented our bifurcation theory for noisy ring complexes.). Observe for several choices of $\Delta$ and $T$ that the resulting point clouds approximately lie on a ring manifold, which implies that the geometry of spreading for STM cascades agrees with the geometry of the generative model for the noisy ring complex (i.e., the manifold is `learned' using the cascade dynamics). We quantify this similarity by computing the Pearson correlation coefficient $\rho$ for the   distances between pairs of points in a cascade map and pairs of points along the ring manifold. Therefore, $\rho\approx1$ when  the point clouds'  geometries match (e.g., a ring arises in the STM cascade map), and $\rho\approx0$ when they are very different. Intuitively, $\rho$ will be larger when STM cascades predominantly spread by WFP along a geometrical substrate that discretizes the manifold.

\begin{figure*}[h!]
	\centering
	\includegraphics[width= 1\linewidth]{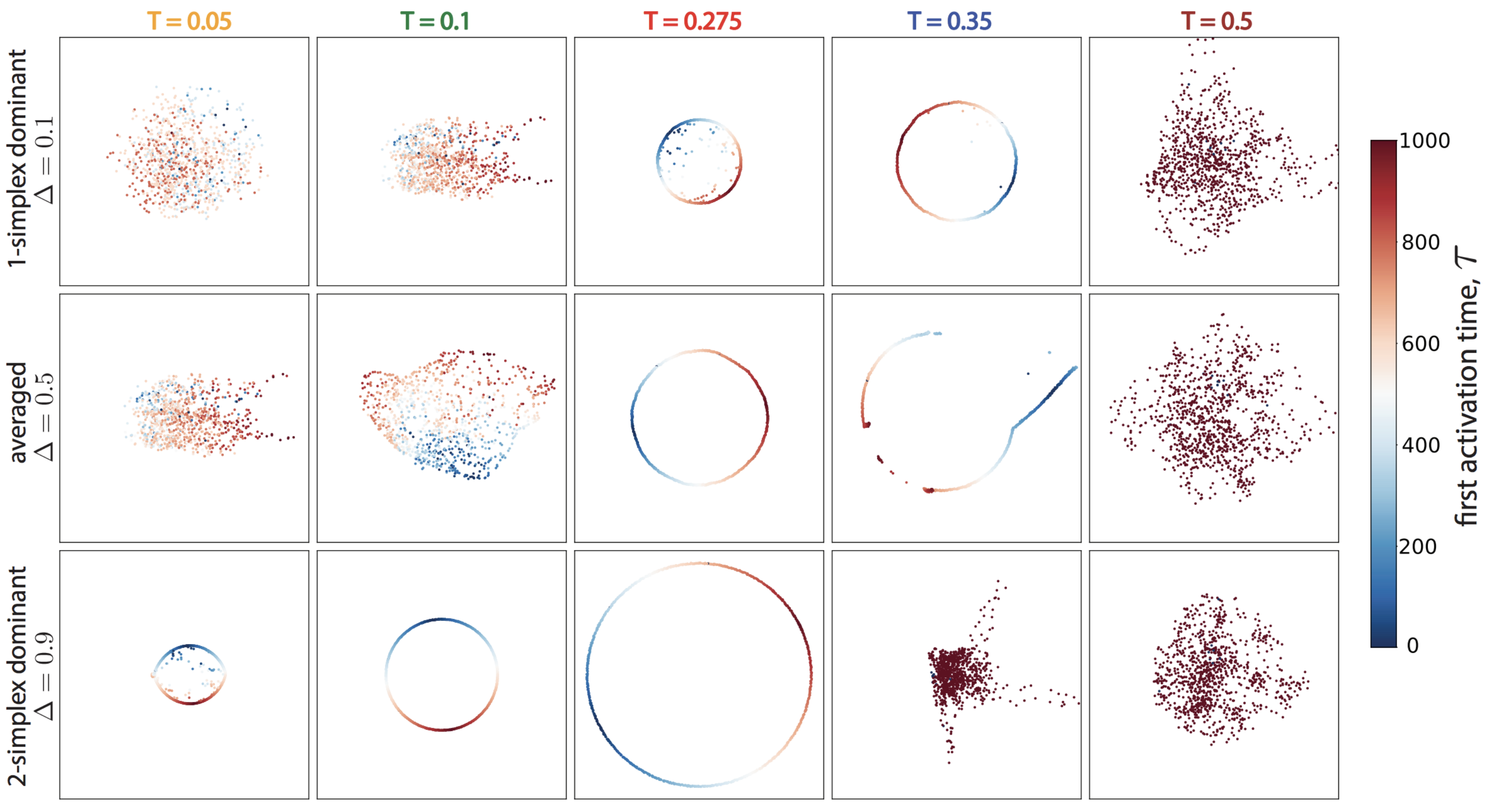}
    \caption{
    {\bf 2D visualizations of simplicial cascade maps for noisy ring complexes.}
		We visualize $N$-dimensional point clouds $\{v_i\}\mapsto \{\bm{\tau }^{(i)}\}\in\mathbb{R}^N$ resulting STM cascade maps by projecting them onto $\mathbb{R}^2$ using classical multidimensional scaling. Rows and columns  reflect the parameter choices $\Delta\in\{0.1,0.5,0.9\}$ and  $T\in\{0.05,0.1,0.275,0.35,0.5\}$, respectively. Cascades spread predominantly by WFP versus ANC when the ring manifold is recovered in the point clouds resulting from STM cascade maps. Observe that the point clouds' best recover the ring manifold when $\Delta$ is larger and $T$ is neither too small or larger.
		}
	\label{PCA2d}
\end{figure*}

\begin{figure*}[t!]
	\centering

	\includegraphics[width= 1\linewidth]{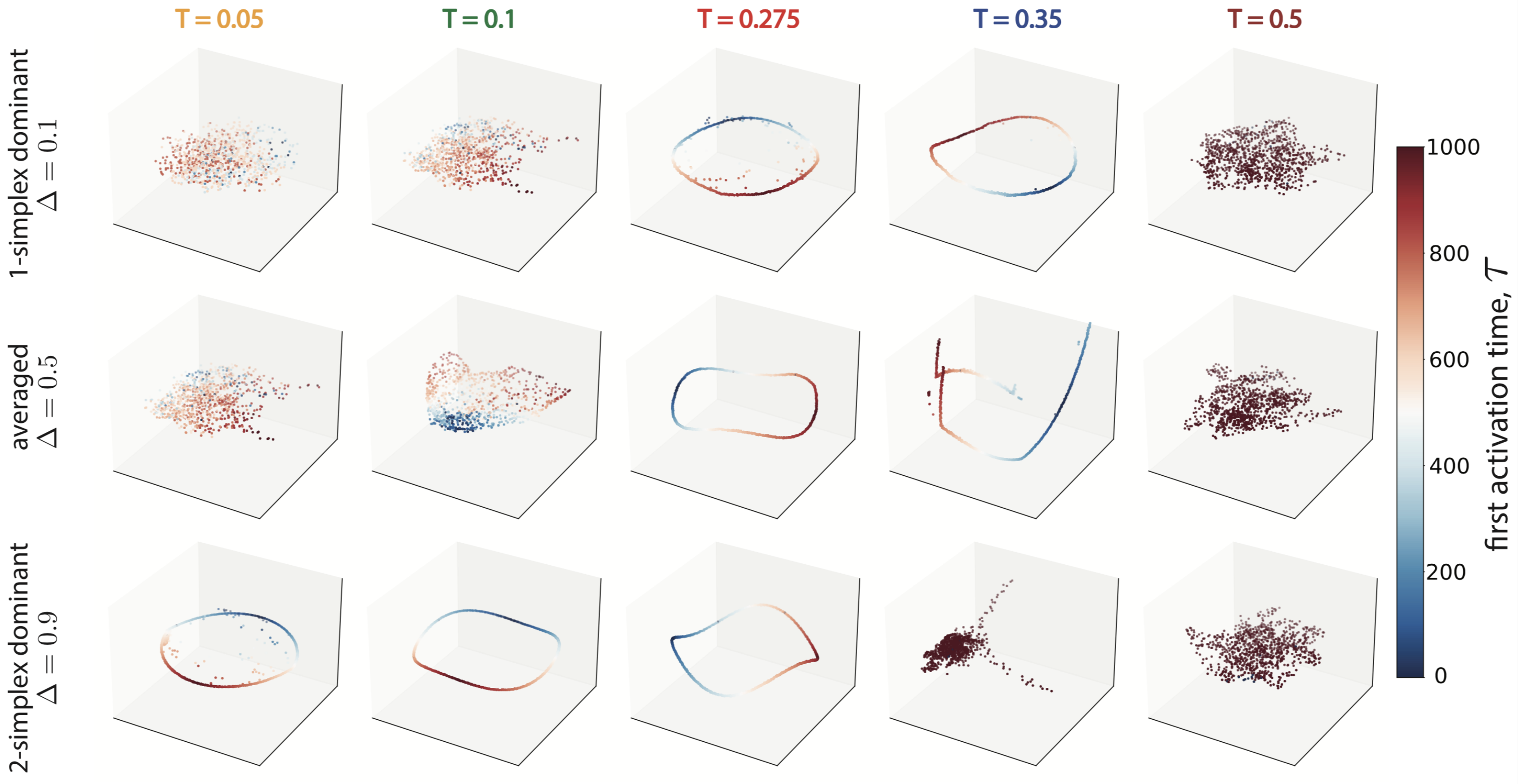}
    \caption{
    {\bf 3D visualizations of simplicial cascade maps for noisy ring complexes.}
        Here, we show the same information as Supplementary Figure \ref{PCA2d}, except that we show 3D projections. 
		}
	\label{PCA3d}
\end{figure*}

%%%%%%%%%%%%%%%%%%%%%%%%%%%%%%%%%%%%%%%%%%%%%%%%%%%%%%%%%%%%
%%%%%%%%%%%%%%%%%%%%%%%%%%%%%%%%%%%%%%%%%%%%%%%%%%%%%%%%%%%%
%%%%%%%%%%%%%%%%%%%%%%%%%%%%%%%%%%%%%%%%%%%%%%%%%%%%%%%%%%%%
%%%%%%%%%%%%%%%%%%%%%%%%%%%%%%%%%%%%%%%%%%%%%%%%%%%%%%%%%%%%

\clearpage

\section{Effects of 1-Simplex Degree Heterogeneity on STM Cascades} \label{networkfams}

In the main text, we studied  a family of noisy geometric complexes over ring manifolds that we described in the Methods Section `Generative model for noisy ring complexes' (see also Family I below). One reason that we focused on this model  was because it lacks degree heterogeneity---that is, each vertex $v_i$ has exactly $d_i^{1,G}=d^{(G)}$ geometric edges and $d_i^{1,NG}=d^{(NG)}$ non-geometric edges. Here, we now expand this study to investigate the effects of 1-simplex degree heterogeneity on WFP and ANC   for four   stochastic generative models for noisy ring complexes.  

\begin{description}
    \item[$\bullet$ Family I] See  Methods Section `Generative model for noisy ring complexes'. To review, we place $N$ vertices $v_i$ at angles $\theta_i = 2\pi (i/N)$ for $i\in\{1,\dots, N\}$ so that each vertex is embedded in $\mathbb{R}^{2}$ with coordinates $[\cos(\theta_{i}), \sin(\theta_{i})]$. We then add   geometric edges by connecting each vertex to its $d^{(G)}$ nearest neighbors. We assume $d^{(G)}$ to be an even number so that ${d^{(G)}}/{2}$ edges go in either direction along the 1D manifold. Next, we add non-geometric edges uniformly at random between the vertices so that each vertex has exactly $d^{(NG)}$ non-geometric edges. We create non-geometric edges using the configuration model, except that we introduce a re-sampling procedure to avoid creating a non-geometric edge that already exists as a geometric edge. The resulting graph is degree regular both in terms of geometric and non-geometric 1-simplex degrees. 
    
    \item[$\bullet$ Family II] We first allow heterogeneity for the non-geometric 1-simplex degrees $d_i^{1,NG}$. We construct geometric edges exactly as for Family I, but we randomly create  non-geometric edges using a different approach. Specifically, we create  $N d^{(NG)}/2$ edges  by selecting them uniformly at random from the $\dfrac{N(N-1-d^{(G)})}{2}$ possible edge locations that are unoccupied by geometric edges. The process for generating non-geometric edges is analogous to an Erd\H{o}s-R\'enyi $G_{NM}$ model, with the added constraint that a set of of edges (i.e., the geometric edges) are unavailable for creation. The   resulting distribution of non-geometric degrees $d_i^{1,NG}$ is  a binomial distribution that is centered at $d^{(NG)}$. 
    
    \item[$\bullet$ Family III] Next, we allow heterogeneity for the geometric 1-simplex degrees $d_i^{1,G}$. We create non-geometric edges exactly as for Family I, but we now create geometric edges using a different approach. We place the vertices non-uniformly on the ring manifold (i.e., unit circle). Given the vertices' angles $\theta_i = 2\pi (i/N)$ for $i\in\{1,\dots, N\}$, we define randomly adjusted angles $\theta_i \mapsto \theta_i + \delta \theta_{i}$, where each $\delta \theta_i$ is a Gaussian-distributed random variable $\delta \theta_{i} \sim \mathcal{N}(0,(s\frac{2\pi}{N})^2)$ with variance $(s\frac{2\pi}{N})^2$. We vary $s$ to adjust the amount of heterogeneity in vertex locations along the ring manifold. To generate geometric edges, we choose a parameter $\epsilon >0$ and place edges between all pairs of vertices $v_{i}$ and $v_{j}$ such that $|\theta_{i} - \theta_{j}| < \epsilon$. We choose   $\epsilon$   so that the median geometric degree is  unchanged, $\langle d_{i}^{1,G} \rangle = d^{(G)}$. 
    
    \item[$\bullet$ Family IV] Finally, we introduce heterogeneity for both   geometric and non-geometric 1-simplex degrees by creating non-geometric and geometric edges using the approaches for Family {II} and  III, respectively.
    
\end{description}

\begin{figure*}[t!]
	\centering
	\includegraphics[width= 1\linewidth]{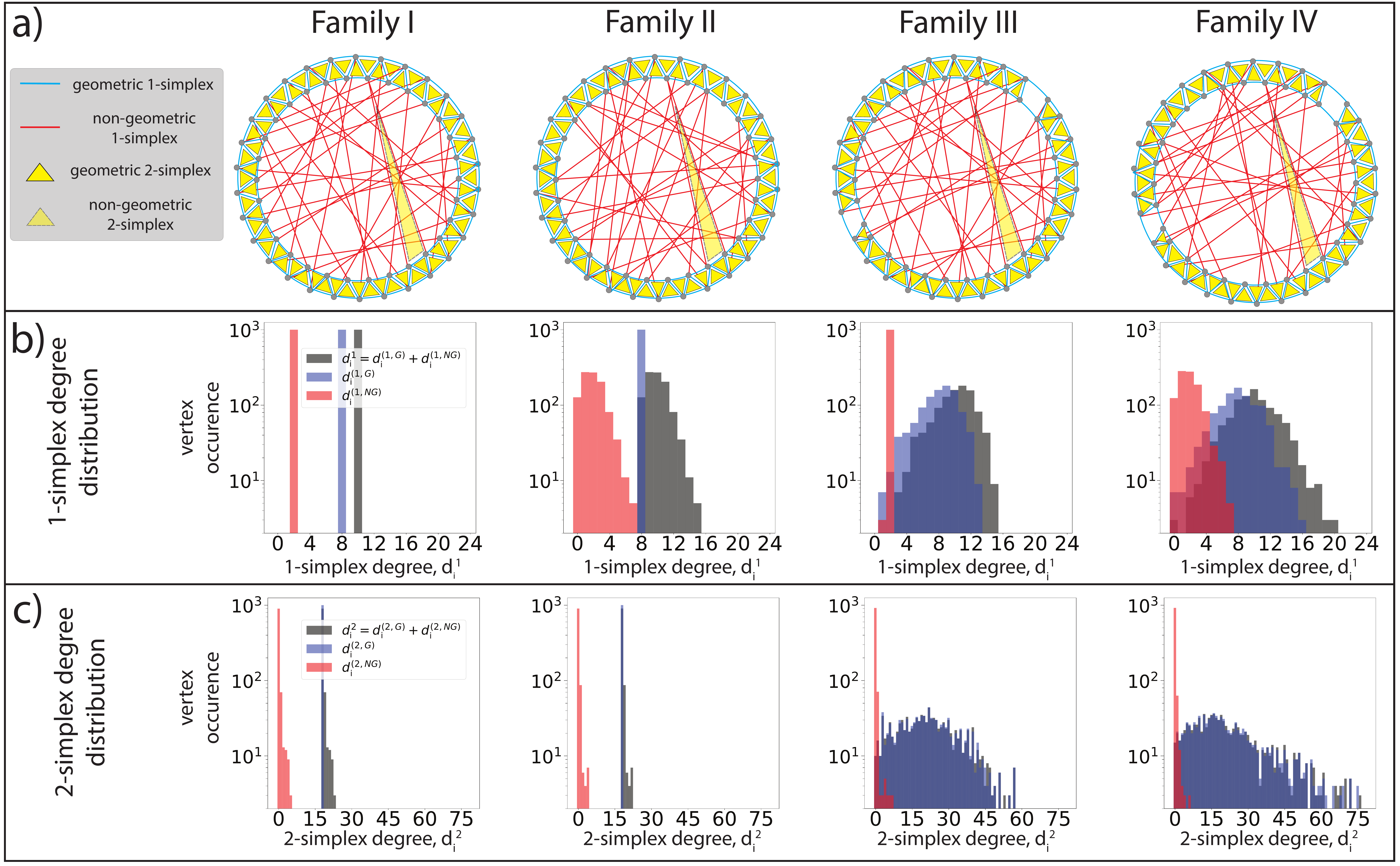}
	\caption{
	{\bf Four generative models for noisy ring complexes with varying 1-simplex degree heterogeneity.} 
	{\bf a)}~Examples that are sampled from  Families I--IV with $N=30$,   $d^{(G)} = 4$ and $d^{(NG)}=1$. These Families are defined so that Families II and IV have heterogeneity for non-geometric 1-simplex degrees, while Families III and IV have heterogeneity for geometric 1-simplex degrees.
    {\bf b)}~Histograms  depict the distributions of 1-simplex degrees $d_i^1$ (gray), geometric 1-simplex degrees $d_i^{1,G}$ (blue), and non-geometric 1-simplex degrees $d_i^{1,NG}$ (red). Observe that these distributions are homogeneous  for Family I,
	that   Families II and III introduce heterogeneity for $d_i^{1,NG}$  and $d_i^{1,G}$, respectively, and that both types of degree are heterogeneous for Family IV.
    {\bf c)}~Histograms depict the distributions of 2-simplex degrees $d_i^2$ (gray), geometric 2-simplex degrees $d_i^{2,G}$ (blue), and non-geometric 2-simplex degrees $d_i^{2,NG}$ (red). Observe that non-geometric 2-simplex  degrees are largely the same across all models (i.e., in all cases, there are very few non-geometric 2-simplices). In contrast, the introduction of heterogeneity for the geometric edges leads to heterogeneity in $d_i^{2,G}$ for Families III and IV.  Panels b) and c) use $N=1000$, $d^{(G)}=8$ and $d^{(NG)} = 2$.
	%\todo{does the last statement contradict the first?}
	}
	\label{fig12}
\end{figure*}

These four Families I--IV were previously defined  in the supplementary material of \cite{taylor2015topological} to generate and investigate different models of noisy ring lattices. Here, we now generalize these models by constructing   associated simplicial complexes---specifically their clique complexes---and by defining $k$ simplices has being geometric or non-geometric. That is, we first construct geometric and non-geometric edges. Then we create higher-order simplices by constructing the associated clique complexes.  In any case, a $k$-simplex is defined to be geometric if and only if all its associated edges are geometric.  A $k$-simplex is defined to be non-geometric if it involves  at least one non-geometric edge.

In Supplementary Figure~\ref{fig12}, we study example noisy ring complexes from these four models and discuss the associated heterogeneity for the geometric and/or non-geometric $k$-simplex degrees. We depict results for Family I, II, III and IV in the first, second, third, and fourth columns, respectively.  In Supplementary Figure~\ref{fig12}a), we illustrate examples with $d^{(G)}  = 4$, $d^{(NG)} = 1$, $N=30$,  $s=5$, and either $\epsilon=0.025$ for Families I and II and $\epsilon=0.032$ for Families III and IV. Observe that the vertices are uniformly spaced along the unit circle for Families I and II, but they are non-uniformly spaced for Families III and IV.  
%We indicate geometric $k$-simplices by blue and yellow circles along the ring manifold, and we do not draw them for clarity (i.e., there are many of them and they would all be overlapping if shown). See Fig.~2 in the main text for visualizations of such geometric $k$-simplices and the geometric subtrates/channels that they compose.

In Supplementary Figure~\ref{fig12}b), we visualize the distributions of 1-simplex degrees $d_i^1$ (gray), geometric 1-simplex degrees $d_i^{1,G}$ (blue), and non-geometric 1-simplex degrees $d_i^{1,NG}$ (red). Observe that the 1-simplex degrees are appropriately homogeneous, or heterogeneous, for the different models as described above. Notably, heterogeneity for either $d_i^{1,G}$ or $d_i^{1,NG}$ leads to heterogeneity for $d_i^{1} = d_i^{1,G}+d_i^{1,NG}$.

In Supplementary Figure~\ref{fig12}c),  we visualize the distributions of 2-simplex degrees $d_i^2$ (gray), geometric 2-simplex degrees $d_i^{2,G}$ (blue), and non-geometric 2-simplex degrees $d_i^{2,NG}$ (red). First, observe that the 2-simplex degrees can be heterogeneous even if the 1-simplex degrees are homogeneous. Second, observe that non-geometric 2-simplex  degrees are all almost identical across the different models; in all cases, there are very few non-geometric 2-simplices, which is an observation that our bifurcation theory relied on (see Methods Section `Combinatorial analysis for bifurcation theory') in the main text. Thirdly, observe that the introduction of heterogeneity for the geometric edges significantly increases the heterogeneity of $d_i^{2,G}$ for Families III and IV. Because heterogeneity added to the creation of geometric edges causes heterogeneity for the associated geometric $2$-simplices (and  higher-dimensional geometric simplices), we can deduce that the associated geometric substrate (e.g., channel) can be significantly affected by introducing heterogeneity for the geometric edges.

\begin{figure*}[t!]
	\centering
	\includegraphics[width= 1\linewidth]{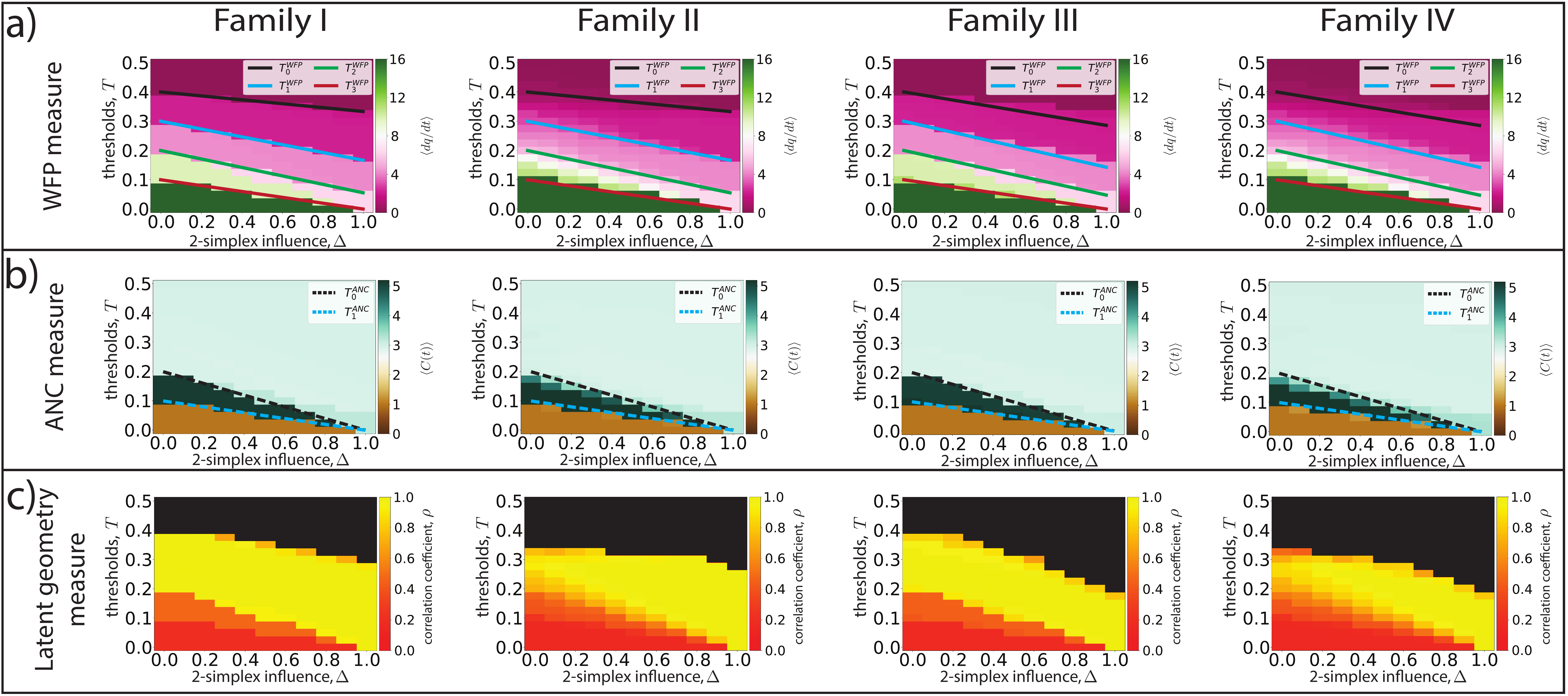}
	\caption{
	    {\bf Effects of 1-simplex  degree heterogeneity on WFP and ANC for noisy ring complexes.}
		[See Fig.~6 in the main text for related results.] 
        {\bf a)}~We display an WFP measure $dq/dt$, which we average across all initial conditions with cluster seeding and compute at $t = 5$ for a range of $T$ and $\Delta$ values and plot the critical threshold lines given by Eq. 3 in the main text. 
		{\bf b)}~We display an ANC measure, $C(t)$, which we average across all initial conditions with cluster seeding and compute at $t = 5$ on the $(T,\Delta)$ parameter space and plot the critical threshold lines given by Eq. 4 in the main text.
		{\bf c)}~We display a Pearson correlation coefficient, $\rho$, which quantifies the extent to which    STM cascades predominantly follow along the ring manifold via WFP. Black regions indicate no saturation for any seed, so $\rho$ is not computed for those values of $(T,\Delta)$. 
		Results were computed with $N=1000$, $d^{(G)}=8$ and $d^{(NG)} = 2$.
	}
	\label{fig13}
\end{figure*}

In Supplementary Figures~\ref{fig13} and~\ref{fig14}, we numerically study the effects of 1-simplex degree heterogeneity on 2D STM cascades spreading over noisy ring complexes. These two figures extend and generalize our findings that were shown in Figs.~6 and 4, respectively, of the main text. (Note that Fig.~4 in the main text examined the \emph{C. elegans} neuronal complex, whereas here we study noisy ring complexes.) As before, we study different choices for the threshold $T\in[0,1]$ (we again focus on  $T_i=T$ for all $v_i$) and the 2-simplex influence parameter $\Delta\in[0,1]$. However, we now study the effect of 1-simplex degree heterogeneity by studying noisy ring complexes that are sampled from the four generative models. In both figures, the four columns correspond to   Families I--IV, respectively.

In Supplementary Figure~\ref{fig13}, we extend our findings from Fig.~6 in the main text by studying the effects of 1-simplex  degree heterogeneity on WFP and ANC properties for STM cascades over noisy ring complexes. In Supplementary Figures~\ref{fig13}a) and~\ref{fig13}b), we use color to depict the average rate of change $dq/dt$  for cascade size $q(t)$ and average number  $C(t)$ of clusters, respectively, with  $t=5$. Recall that $dq/dt$ and  $C(t)$ are empirical measures for WFP speed and ANC rate, respectively. We compare these empirical measurements with   bifurcation lines $T_j^{WFP}$ and $T_j^{ANC}$. (See Methods Section `Combinatorial analysis for bifurcation theory' in the main text for details.) The bifurcation lines for Family I are identical to those shown in Figs.~6(A) and 6(B) in the main text. For Families II-IV, we present   critical thresholds that we obtain using empirically computed means of the  $k$-simplex degrees. In Supplementary Figure~\ref{fig13}c), we depict the correlation coefficient $\rho$ that quantifies the extent to which the geometry of a simplicial cascade map (see Methods Section `Simplicial cascade maps') appropriately recovers the geometry of the original noisy ring complex. Recall that larger values of $\rho$ indicate better manifold recovery, which corresponds to when cascades exhibit an increased amount of WFP versus ANC. 

By comparing the different columns of Supplementary Figure~\ref{fig13}, one can observe that the introduction of heterogeneity for the geometric and non-geometric 1-simplex degrees has surprisingly little effect on the overall WFP and ANC behaviors for these experiments in which  $N=1000$, $d^{(G)}=8$ and $d^{(NG)} = 2$. This most likely occurs because even though we introduced heterogeneity, we have not introduced significant heterogeneity in which one has heavy-tailed (e.g., power law) degree distributions. Here, we  highlight two subtle changes that occur when there is slight-to-moderate degree heterogeneity. First, observe in  Supplementary Figure~\ref{fig13}a) that the transitions that occur at bifurcation lines are smoother (i.e., less abrupt) for Families II and IV,  which are the generative models in which the non-geometric degrees are heterogeneous. Second,   consider  Supplementary Figure~\ref{fig13}c)  when $\Delta$ is very small (e.g., $\Delta=0.05$). For Families I and III, there is a large range of $T$ for which $\rho\approx 1$ (i.e., $T\in[0.2,38]$); however, this range decreases for Families II and IV (i.e., $\rho\approx 1$ only when $T\in[0.26,29]$ and $\Delta = 0.05$). That is, for cascades dominated by dyadic interactions, the introduction of heterogeneity for non-geometric degrees decreases the range of  $T$ in which STM cascades exhibit a prevalence of WFP over ANC. Interestingly,   increasing the    2-simplex influence (i.e., larger $\Delta$) counter balances this effect---i.e., higher-order interactions  thereby help  overcome the heterogeneity for non-geometric degrees to allow STM cascades to more strongly exhibit WFP along a geometric substrate composed of geometric $k$-simplices.

\begin{figure*}[t!]
	\centering
	\includegraphics[width= 1\linewidth]{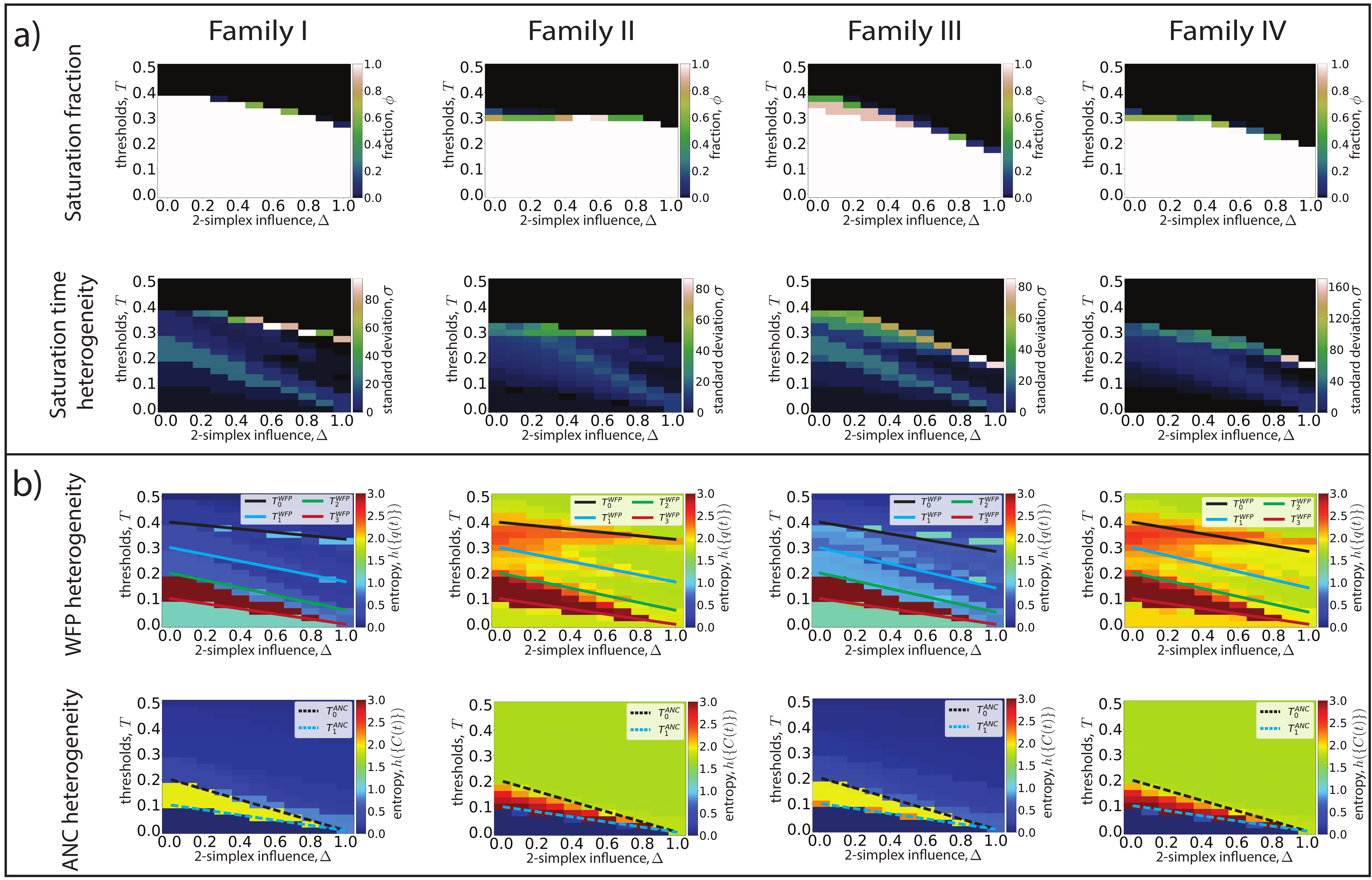}
	\caption{
		{\bf Effects of 1-simplex degree heterogeneity on cascade heterogeneity for noisy ring complexes.}
        We study  2D STM cascades over the noisy ring complexes generated using the four different models Families I-IV. See Fig.~4 in the main text for similar experiments for the \emph{C. elegans} neuronal complex.
		{\bf a)}~We display the saturation fractions $\phi$ (top row) and the standard deviation $\sigma$ for the saturation times  (bottom row) for the parameter space $(T,\Delta)$. Black regions indicate where no initial conditions lead to a saturation across the network.
		{\bf b)}~We measure the heterogeneity in the cascade sizes (top row) and number of clusters (bottom row) at $t = 5$ for different initial conditions. See Methods Section `Entropy Calculation' in the main text for the definitions of these entropic measures for heterogeneity. The   only difference is that we now use 80 bins, whereas   20 bins were  used for our experiments with \emph{C. elegans}.
	}
	\label{fig14}
\end{figure*}

In Supplementary Fig.~\ref{fig14}, we study how 1-simplex degree heterogeneity  affects the heterogeneity of spatio-temporal patterns for 2D STM cascades over noisy ring complexes. Our results in Supplementary Figure~\ref{fig14}a) and Supplementary Figure~\ref{fig14}b) are similar to those shown in Fig.~4 in the main text, except that we now consider the four Families I--IV. In Supplementary Figure~\ref{fig14}a), we study the heterogeneity of STM cascades across different initial conditions  for different   $T$ and $\Delta$. We  plot the fraction $\phi$ of cascades that saturate the network (top row)  and the standard deviation $\sigma$ for the times at which saturations occur (bottom row).  Interestingly, for all four families, STM cascades with different initial conditions  either all saturate the network (white colored pixels) or none do (black colored pixels). That is, there is a very small intermediate regime in which some initial conditions yield saturations and some do not. In comparison, one can observe in Fig. 4 in the main text and Supplementary Figure 2(A) that the intermediate regime is large for the \emph{C. elegans} neuronal complex. Moreover, observe for Families II and IV that the transition from saturation to no saturation is shifted downward (i.e., smaller $T$) when  heterogeneity is introduced for the non-geometric edges shifts, but that shift only occurs when $\Delta $ is small. In contrast, a similar shift occurs for Families III and IV, which have added heterogeneity for the geometric edges, but those shifts only occur for larger $\Delta$ values. Thus, both effects impact this transition for Family IV.

In Supplementary Figure~\ref{fig14}b), we  study heterogeneity for the cascade size $q(t)$ and number  $C(t)$ of cascade clusters when $t=5$, which are quantitative measures for WFP and ANC, respectively. We quantify heterogeneity by computing the Shannon entropy of $h(\{q(t)\})$ (top row) and $h(\{C(t)\})$ (bottom row) across the  different initial conditions as described in  Methods Section `Entropy Calculation' for details. The only difference is that we now use 80 histogram bins to compute entropies (whereas we used 20 for the \emph{C. elegans} neuronal complex). First, we find for Family I that there is an intermediate regime of $T$ and $\Delta$ where different initial conditions yield diverse WFP and ANC phenomena---see red-colored pixels in the top row  and yellow pixels in the bottom row    in Supplementary Figure~\ref{fig14}b). (Recall that a similar regime was observed for the \emph{C. elegans} neuronal complex.) Second, we find  that heterogeneity for non-geometric edges  has a greater effect on WFP and ANC heterogeneity across initial conditions than heterogeneity for geometric edges. That is, by comparing the results for Families II and IV to those for  Family I, observe  that when heterogeneity  is added to non-geometric edges, it amplifies the diversity of   WFP and ANC across the  different initial conditions. In contrast, observe for Family III  that adding  heterogeneity to the geometric edges has little effect on these measures for heterogeneity for WFP and ANC.

%%%%%%%%%%%%%%%%%%%%%%%%%%%%%%%%%%%%%%%%%%%%%%%%%%%%%%%%%%%%
%%%%%%%%%%%%%%%%%%%%%%%%%%%%%%%%%%%%%%%%%%%%%%%%%%%%%%%%%%%%
%%%%%%%%%%%%%%%%%%%%%%%%%%%%%%%%%%%%%%%%%%%%%%%%%%%%%%%%%%%%
%%%%%%%%%%%%%%%%%%%%%%%%%%%%%%%%%%%%%%%%%%%%%%%%%%%%%%%%%%%%

\clearpage

\section{Effects of 2-Simplex Degree Heterogeneity on STM Cascades} \label{networkfams2}

In the previous Supplementary Note, we studied how  the introduction of heterogeneity for the vertices' 1-simplex degrees affects the properties of WFP and ANC for STM cascades over noisy ring lattices. Here, we extend these findings by  studying the effects on WFP and ANC due to the introduction of heterogeneity for vertices' 2-simplex degrees. To this end, we now introduce three additional families of noisy ring complexes.

\begin{description}
    \item[$\bullet$ Family I] This family is identical to that which was described in Supplementary Note 3 as well as the Methods Section `Generative model for noisy ring complexes' in the main text.

    \item[$\bullet$ Family V] We create geometric 1- and 2-simplices in the same way as for Family I; however, we now construct non-geometric 1-simplices and 2-simplices in a way that leads to many more non-geometric 2-simplices. Specifically, for each vertex $v_i$, we create a non-geometric 2-simplex by selecting two other vertices without replacement uniformly at random. Note that each non-geometric 2-simplex that is created also requires the addition of  two 1-simplices to that vertex. Therefore, in order to control the distribution of 1-simplices in the simplicial complex,  we repeat drawing vertices at random if the corresponding 1-simplices are already existent in the complex, i.e., we don't add a 2-simplex if any of its faces are present in the simplicial complex. It follows that  $d_i^{2,NG}\ge1$ for each vertex $v_i$, and $d_i^{1,NG} =d^{(NG)} = 2$, which makes the comparison to Family I easier since their 1-simplex distributions are exactly the same.  

    \item[$\bullet$ Family VI]  We create non-geometric 1- and 2-simplices in the same way as for Family I; however, we now introduce irregularity in the geometric substrate by imposing heterogeneity for the geometric 1- and 2-simplices. Specifically, for each vertex $v_i$ we delete $1$ of its geometric edges, which is selected  uniformly at random. The removal of geometric 1-simplices also leads to the removal of the associated  2-simplices. After applying this processes to all $N$ vertices, the geometric 1-simplex degree $d_i^{1,G}$ of each node $v_i$ will decrease by an average amount of two (i.e., since each edge removal decreases the 1-simplex degree for to vertices).   
    
    \item[$\bullet$ Family VII] Finally, we introduce heterogeneity for both the geometric and non-geometric 2-simplex degrees by removing geometric 1- and 2-simplices using the approach for Family VI and by creating non-geometric 1- and 2-simplices using the same procedure as for Family  V.
    
\end{description}

\begin{figure*}[t!]
	\centering
	\includegraphics[width= 1\linewidth]{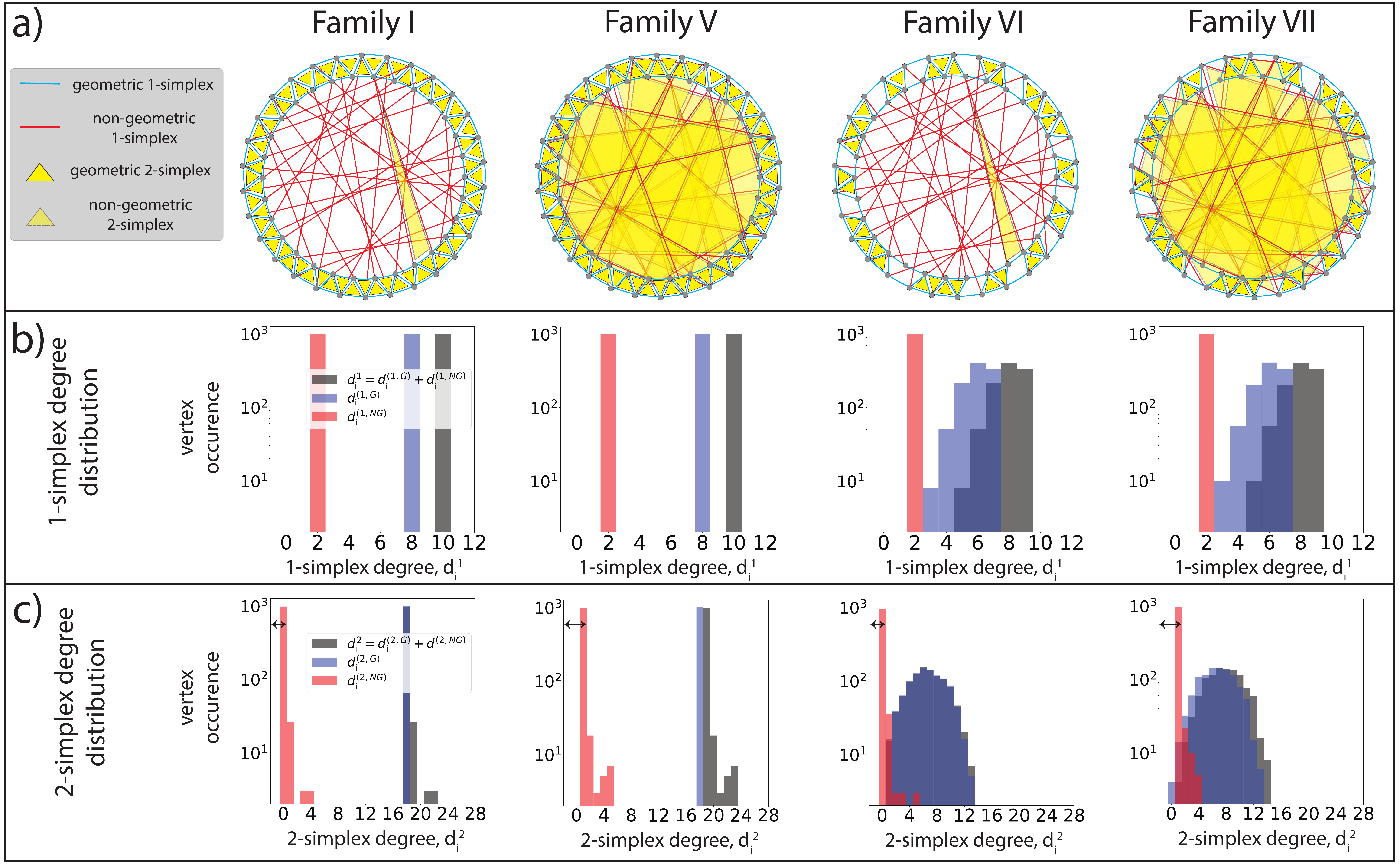}
	\caption{
	{\bf Four generative models for noisy ring complexes with varying 2-simplex degree heterogeneity.} 
	{\bf a)}~Examples that are sampled from  Families I, V, VI and VII with $N=30$, $d^{(G)}=4$ and $d^{(NG)}=1$ for Families I and VI and $d^{(2,NG)}=1$ for Families V and VII. These Families are defined so that some geometric 2-simplices are missing from Families VI and VII, and there are many non-geometric 2-simplices for Families V and VII.
	{\bf b)}~Histograms   depict the distributions of 1-simplex degrees $d_i^1$ (gray), geometric 1-simplex degrees $d_i^{1,G}$ (blue), and non-geometric 1-simplex degrees $d_i^{(1,NG)}$ (red). Observe that these distributions are homogeneous  for Family I and Family V, and
	that Families VI and VII introduce heterogeneity for $d_i^{1,G}$ whilst keeping $d_i^{1,NG}$ homogeneous.
    {\bf c)}~Histograms depict the distributions of 2-simplex degrees $d_i^2$ (gray), geometric 2-simplex degrees $d_i^{2,G}$ (blue), and non-geometric 2-simplex degrees $d_i^{2,NG}$ (red). Observe for Families I and VI that $d_i^{2,NG}=0$ for most vertices $v_i$, whereas $d_i^{2,NG}=1$ for most vertices $v_i$ under Families V and VII. Also note that Families VI and VII have significantly fewer geometric 2-simplicies as compared to that for Families I and V. Panels b) and c) use $N=1000$, $d^{(G)}=8$ and $d^{(NG)} = 2$.}
	\label{fig15}
\end{figure*}

Our motivation for studying STM cascades of Families V, VI and VII is that we'd like to gain a better understanding for how geometric and non-geometric 2-simplices can affect WFP and ANC for STM cascades. 
We study Families VI and VII to gain insight into the effects of irregularity/heterogeneity for the geometric substrate. Recall that Families III and IV in Supplementary Note 3 also introduced such  irregularity for both geometric 1- and 2-simplices. The main difference here is the geometric $k$-simplex degrees are decreased for all vertices under Families VI and VII, whereas these degrees could either increase or decrease under Families III and IV. Thus, the geometric substrate is more greatly affected by the perturbations introduced by Families VI and VII as compared to that for Families III and IV.
At the same time, we study Families V and VII to investigate  when there exist many non-geometric 2-simplices. While some non-geometric 2-simplices are randomly created in Families I--IV, there are very few of them. In contrast, each vertex has at least one non-geometric 2-simplex under Families V and VII.

\begin{figure*}[t!]
	\centering
	\includegraphics[width= 1\linewidth]{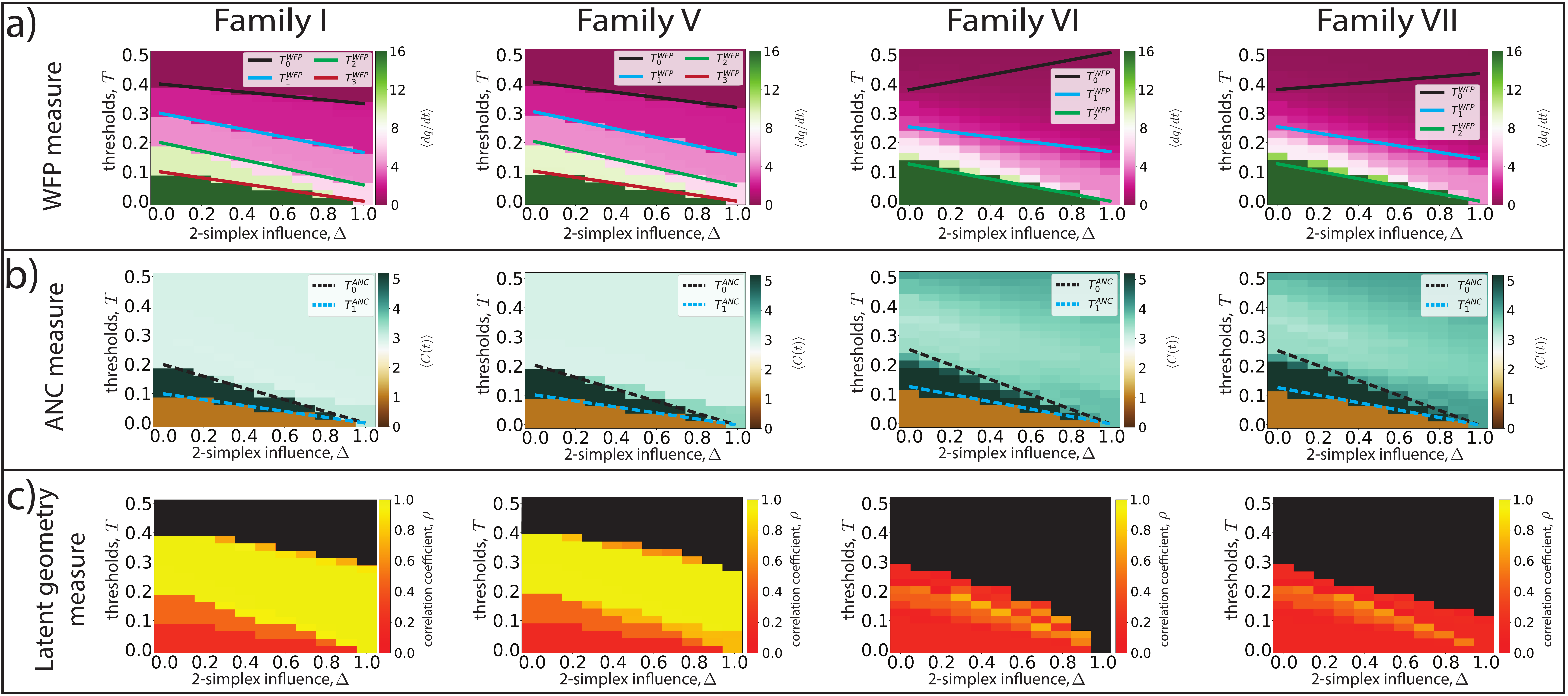}
	\caption{
	    {\bf Effects of 2-simplex  degree heterogeneity on WFP and ANC for noisy ring complexes.}
		[See Fig.~6 in the main text as well as Supplementary Figure~\ref{fig13} for related plots.] 
        {\bf a)}~We display an WFP measure $dq/dt$, which we average across all initial conditions with cluster seeding and compute at $t = 5$ for a range of $T$ and $\Delta$ values and plot the critical threshold lines given by Eq. 3 in the main text. 
        {\bf b)}~We display an ANC measure, $C(t)$, which we average across all initial conditions with cluster seeding and compute at $t = 5$ on the $(T,\Delta)$ parameter space and plot the critical threshold lines given by Eq. 4 in the main text.
		{\bf c)}~We display a Pearson correlation coefficient, $\rho$, which quantifies the extent to which STM cascades predominantly follow along the ring manifold via WFP. Black regions indicate no saturation for any seed, so $\rho$ is not computed for those values of $(T,\Delta)$. 
		Results were computed with $N=1000$,  $d^{(NG)} = 2$, and $d^{(G)}=8$. Note, however, that the removal of geometric edges for  Families VI and VII decrease the expected geometric 1-simplex degrees by two on average, and so we depict bifurcation lines for these families using $\langle d_i^{1,G}\rangle=6$ as opposed to 8. 
	}
	\label{fig16}
\end{figure*}

In Supplementary Figure~\ref{fig15}, we study example noisy ring complexes from the Families I, V, VI, VII and discuss the associated heterogeneity for the geometric and/or non-geometric $k$-simplex degrees. We depict results for Family I, V, VI and VII in the first, second, third, and fourth columns, respectively.  In Supplementary Figure~\ref{fig15}a), we illustrate examples with $ N=30$, $d^{(G)}  = 4$, and either $d^{(NG)} = 1$ for Families I and VI or  $d^{(NG)}=2$ for Families V and VII.  First, observe that many geometric $k$-simplices missing for Families VI and VII, which imposes irregularity into the geometric substrate (i.e., channel).  Second, observe that there are far more non-geometric 2-simplicies in Families V and VII as compared to Families  I and VI.

In Supplementary Figure~\ref{fig15}b), we visualize the distributions of 1-simplex degrees $d_i^1$ (gray), geometric 1-simplex degrees $d_i^{1,G}$ (blue), and non-geometric 1-simplex degrees $d_i^{1,NG}$ (red). Observe that the 1-simplex distributions of Family I and V are identical. Also, note that all non-geometric 1-simplex degrees are homogeneous across all four Families. However, geometric 1-simplex degrees are heterogeneous for Families VI and VII due to the deletion of geometric edges. Notably, heterogeneity for either $d_i^{1,G}$ or $d_i^{1,NG}$ leads to heterogeneity for $d_i^{1} = d_i^{1,G}+d_i^{1,NG}$.

In Supplementary Figure~\ref{fig15}c),  we visualize the distributions of 2-simplex degrees $d_i^2$ (gray), geometric 2-simplex degrees $d_i^{2,G}$ (blue), and non-geometric 2-simplex degrees $d_i^{2,NG}$ (red). First, observe that Families I and VI don't contain many non-geometric 2-simplices, whereas these are abundant in Families V and VII. Second, note that although Families I and V have identical 1-simplex distributions, their 2-simplex distributions are different, since we strategically create many non-geometric 2-simplices  for Families V and VII. Third, vertices in Families VI and VII have far fewer geometric 2-simplices as compared to those of Families I and V.

In Supplementary Figures~\ref{fig16} and~\ref{fig17}, we numerically study the effects of 2-simplex degree heterogeneity on 2D STM cascades spreading over noisy ring complexes. These two figures extend and generalize our findings that were shown in Supplementary Figures~\ref{fig13} and~\ref{fig14} as well as Figs.~6 and 4 of the main text. As before, we study different choices for the threshold $T\in[0,1]$ (we again focus on  $T_i=T$ for all $v_i$) and the 2-simplex influence parameter $\Delta\in[0,1]$. However, we now study the effect of 2-simplex degree heterogeneity by studying noisy ring complexes that are sampled from the four generative models described in this section. In both figures, the four columns correspond to Families I, V, VI and VII, respectively.

In Supplementary Figure~\ref{fig16}, we extend our findings of Supplementary Figure~\ref{fig13} and Fig.~6 in the main text by studying the effects of 2-simplex  degree heterogeneity on WFP and ANC properties for STM cascades over noisy ring complexes. In Supplementary Figures~\ref{fig16}a) and~\ref{fig16}b), we use color to depict the average rate of change $dq/dt$  for cascade size $q(t)$ and average number  $C(t)$ of clusters, respectively, with  $t=5$. We compare these empirical measurements with   bifurcation lines $T_j^{WFP}$ and $T_j^{ANC}$ as described in Methods Section `Combinatorial analysis for bifurcation theory' in the main text. The bifurcation lines for Family I are identical to those shown in Figs.~6(A) and 6(B) in the main text. For Family V, we present   critical thresholds that we obtain using empirically computed means of the  $k$-simplex degrees. For   Families I and V, we plot the bifurcation lines for $d^{(G)}=8$, whereas we instead plot bifurcation lines using $\langle d^{1,G}_i\rangle =6$ for Families VI and VII since the removal of geometric $k$-simplices decreases  each $d^{1,G}_i$ by two on average.
In Supplementary Figure~\ref{fig16}c), we depict the correlation coefficient $\rho$ that quantifies the extent to which the geometry of a simplicial cascade map (recall Methods Section `Simplicial cascade maps') appropriately recovers the geometry of the original noisy ring complex. Recall that larger values of $\rho$ indicate better manifold recovery, which corresponds to when cascades exhibit an increased amount of WFP versus ANC.

By comparing the different columns of Supplementary Figure~\ref{fig16}, observe that the plots for Families I and V are very similar and that these are very different from the plots for Families VI and VII (which are also similar to one another). One main reason is that the deletion of geometric 1-simplices decreases each $d_i^{1,G}$ by two on average, which is why we present two different sets of bifurcation lines. In addition, observe that one bifurcation line $T_{0}^{WFP}$ for Families VI and VII is tilted upwards as $\Delta$ increases. This occurs since the deletion of geometric 1-simplicies decreases each $d_i^{2,G}$ by more than half (see Supplementary Figure~\ref{fig15}c) for Families VI and VII), on average, and  the denominator in the second term in the right-hand-side of Eq.~3 in the main text becomes small.

Comparing the panels for Families I and V in Supplementary Figure~\ref{fig16}, we observe  abrupt transitions for the critical regimes at the same parameter values, highlighting that the introduction of non-geometric 2-simplicies doesn't affect  the $\langle dq/dt \rangle$, $\langle C(t) \rangle$ and $\rho$ at $t=5$. That said, we also find that the introduction of non-geometric 2-simplices can lead to higher rates of ANC, but only after an STM cascade has grown large. In contrast, our bifurcation theory and this experiment focus on early times $t$ in which the cascade size is small.
%
%Although we report that long-range 2-simplicies contribute to ANC events for high $\Delta$ values, these events occur at later times during the propagation of a cascade. Notably, our theory and this experiment aim to study STM cascades early on during a cascade, and so these later effects are not seen here.
%is valid for early stages of the cascade, and WFP, ANC and latent geometry measures do not capture this influence. 
Turning our attention to
%However, transitions between regimes are not as abrupt for 
Families VI and VII, we find that the introduction of irregularity in the geometric substrate makes the bifurcations less distinct. Or in other words, %the bifurcation lines are ``blurred''.
%in comparison with the former two families because of the irregularity of the geometric substrate. Instead, these changes are reflected as a blur 
we observe in the last two columns of Supplementary Figures~\ref{fig16}a) and b) that the transitions between regimes appear ``blurred''.
Lastly, by comparing $\rho$    for Families I and V with Families VI and VII,   observe in Supplementary Figure~\ref{fig16}c)  that $\rho$ is significantly decreased for Families VI and VII. In particular, there is no longer a regime of $(T,\Delta)$ parameter values with $\rho\approx1$ (i.e., the regime in which WFP is more prevalent than ANC). 

\begin{figure*}[t!]
	\centering
	\includegraphics[width= 1\linewidth]{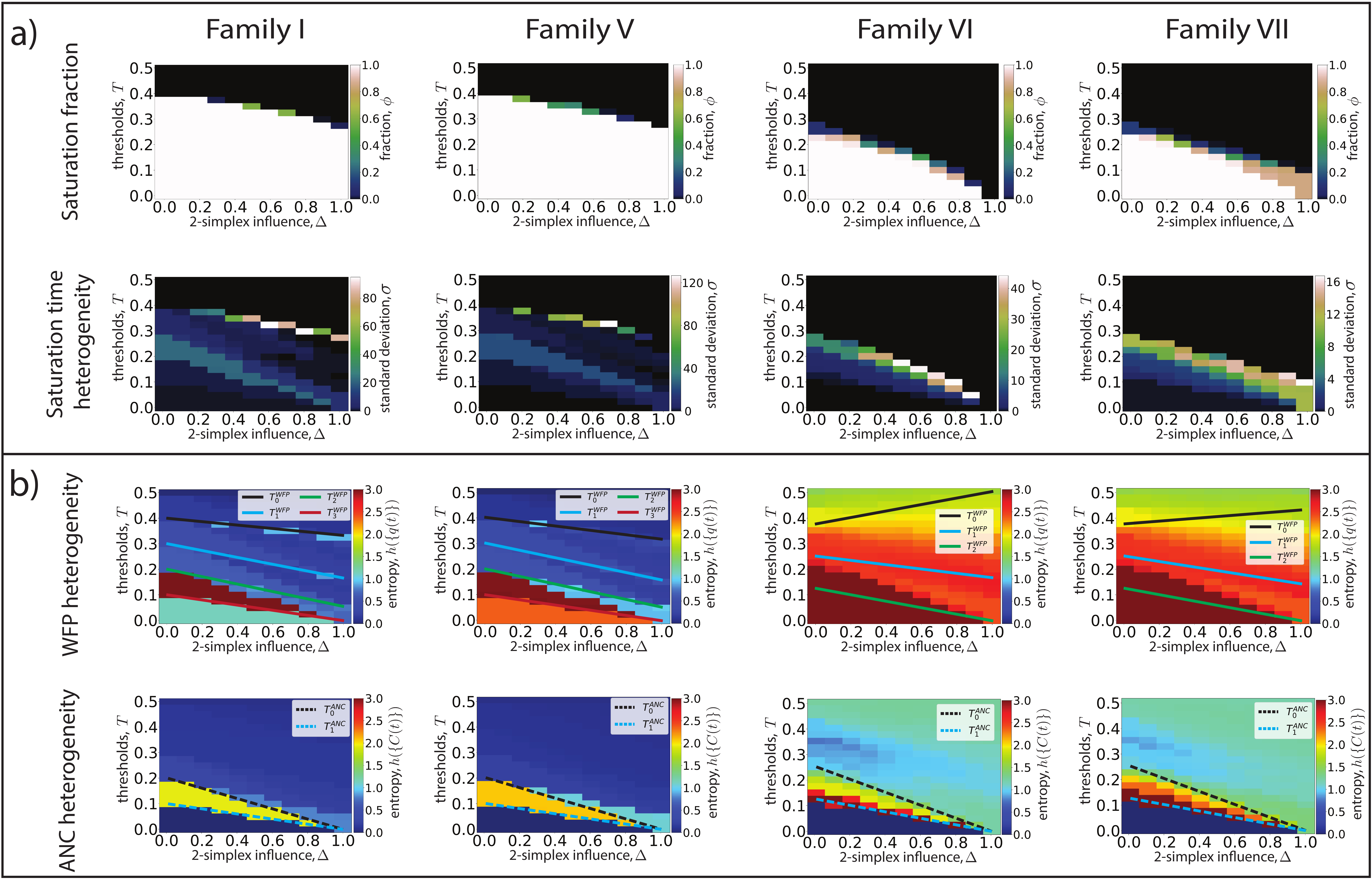}
	\caption{
		{\bf Effects of 2-simplex degree heterogeneity on cascade heterogeneity for noisy ring complexes.}
        We study  2D STM cascades over   noisy ring complexes generated from models Families I, V, VI and VII. See Supplementary Figure \ref{fig14} for similar experiments for the other Families that were defined in Supplementary Note 3.
		{\bf a)}~We display the saturation fractions $\phi$ (top row) and the standard deviation $\sigma$ for the saturation times  (bottom row) for the parameter space $(T,\Delta)$. Black regions indicate where no initial conditions lead to  saturation.
		{\bf b)}~We measure the heterogeneity in the cascade sizes (top row) and number of clusters (bottom row) at $t = 5$ for different initial conditions. See Methods Section `Entropy Calculation' and for details.As before, results were computed with $N=1000$,  $d^{(NG)} = 2$, and $d^{(G)}=8$, and we similarly  depict bifurcation lines for Families VI and VII  using $\langle d_i^{1,G}\rangle=6$ as opposed to 8. 
	}
	\label{fig17}
\end{figure*}

In Supplementary Fig.~\ref{fig17}, we study how 2-simplex degree heterogeneity  affects the heterogeneity of spatio-temporal patterns for 2D STM cascades over noisy ring complexes. Our results in Supplementary Figure~\ref{fig17}a) and Supplementary Figure~\ref{fig17}b) are similar to those shown in Supplementary Figure~\ref{fig14}a) and Supplementary Figure~\ref{fig14}b), respectively,  except that we now consider   Families I, V, VI and VII. In Supplementary Figure~\ref{fig17}a), we study the heterogeneity of STM cascades across different initial conditions  for different   $T$ and $\Delta$. We  plot the fraction $\phi$ of cascades that saturate the network (top row)  and the standard deviation $\sigma$ for the times at which saturations occur (bottom row).  Interestingly, for Families I, V and VI, STM cascades with different initial conditions  either all saturate the network (white colored pixels) or none do (black colored pixels). That is, there is only a very small intermediate regime in which some initial conditions yield saturations and some do not. In contrast, for Family VII there is a larger parameter regime  in which cascades do not fully saturate [i.e., see the bottom-right corner of the top-right panel in Supplementary Figure~\ref{fig17}a).] In comparison, one can observe in Fig.~4 in the main text and Supplementary Figure~\ref{fig14}a) that the intermediate regime is also   larger for the \emph{C. elegans} neuronal complex. This suggests that the heterogeneity in 2-simplex degrees significantly contributes to the heterogeneity of cascade saturations across the different cascade seeds.

In Supplementary Figure~\ref{fig17}b), we  study heterogeneity for the cascade size $q(t)$ and number  $C(t)$ of cascade clusters when $t=5$. As before, we quantify heterogeneity by computing the Shannon entropy of $h(\{q(t)\})$ (top row) and $h(\{C(t)\})$ (bottom row) across the  different initial conditions as described in  Methods Section `Entropy Calculation' for details. The only difference is that we now use 80 histogram bins to compute entropies (whereas we used 20 for the \emph{C. elegans} neuronal complex). 
Our first observation is that the results for Family V are nearly identical to those for Family I, implying that the addition of many non-geometric 2-simplices has little effect on these measures for heterogeneity. In contrast, by comparing the results for Families VI and VII to those for  Families I and V, one can observe that the introduction of irregularity into the geometric substrate  has a significant effect. It amplifies the overall diversity of WFP and ANC across the  different initial conditions.

\bibliographystyle{plain}
\bibliography{supp.bib}